\documentclass{article}

\usepackage{epsfig}
\usepackage{mysprocl}
\usepackage{latexsym}
\begin{document}
  
\newcommand  {\new}  {\newcommand}
\new {\ri} {{\rm i}}
\new {\rd} {{\rm d}}
\new {\re} {{\rm e}}
\new {\s} {$\,$}

\makeatletter 
\def\subsubsection{\@startsection{subsubsection}{3}{\z@}
   {-1ex plus -1ex minus -0.5ex}{0.1ex plus 0.1ex}{\it}}
\makeatother

\renewcommand{\thefootnote}{\fnsymbol{footnote}}

\begin{flushright}
FAU-TP3-00/9\\
hep-th/0008175
\end{flushright}
\hspace{0.5cm}

\title{2D MODEL FIELD THEORIES AT FINITE TEMPERATURE AND DENSITY
  \footnote{Contribution to the Festschrift in honor of Boris Ioffe,
    edited by M.~Shifman.}}

\author{V. SCH\"ON and M. THIES}
\address{Institute for Theoretical Physics III,
University of Erlangen-N\"urnberg, \\
Staudtstr. 7, 91058 Erlangen, Germany}

\maketitle

\abstracts{
In certain 1+1 dimensional field theoretic toy models, one can go  
all the way from microscopic quarks via the hadron spectrum to the
properties of
hot and dense baryonic matter in an essentially analytic way.
This ``miracle"
is illustrated through case studies of two popular large $N$
models, the
Gross-Neveu and the 't Hooft model --- caricatures of the
Nambu--Jona-Lasinio
model and real QCD, respectively. The main emphasis will
be on aspects  
related to spontaneous symmetry breaking (discrete or continuous
chiral symmetry,
translational invariance) and confinement. 
 }

\tableofcontents

\section{Introduction}
 
With the shift in emphasis in field theory from perturbative
to non-per\-tur\-ba\-tive
phenomena during the last decades, exactly soluble models have
acquired a
quite respectable status. The limited number of analytical tools
available at strong
coupling, so painfully felt in quantum chromodynamics (QCD),
makes it tempting to look for 
other sources of inspiration.
Except for the celebrated case of supersymmetric $N=2$ Yang-Mills
theory\s\cite{Seiberg94}, ``soluble model" in quantum field 
theory still means ``1+1 dimensional model". This makes it
clear at once that 
we should not expect such toy models to be in any sense
realistic; they can teach us only  
little (if anything) of direct relevance to the real world.
Nevertheless, we find it 
quite appropriate to include them in a book devoted to analytic
approaches to QCD. 
One can perhaps picture the relationship between QCD and the soluble
toy models 
as follows: If QCD is the sun, then the 1+1 dimensional models are
like planets which are gravitating around it, shining in its light.
Their distance to the sun may vary and each of them develops a life of
its own, although powered by the sun. Only then can one understand why 
a theory like the Schwinger model\s\cite{Schwinger62} describing massless
electrons which only move along a line, have a linear
Coulomb potential and
form non-in\-ter\-ac\-ting, massive bosons can accumulate close to 
1000 citations in the hep-archive up to now.
This striking success is shared by the Gross-Neveu
(GN) model\s\cite{Gross74} (massless fermions
with a point interaction) and the 't Hooft model\s\cite{tHooft74b}
(QCD$_2$ in the large $N$ limit). 
Whether one likes these models
or not --- they have become an integral part of the intellectual endeavor
to understand the strong interactions and their traces pervade the
QCD literature. 

There is another aspect which we should like to mention. A soluble
model in quantum
mechanics is typically solved once and for all, then put into the
textbooks where
it ceases to play any scientific role. By contrast, a soluble
field theory is a laboratory which
can be used over and over again for new types of questions ---
this is the reason
why it is so much richer. When browsing through the numerous
citations of models
like the GN model or the 't Hooft model, 
one finds the whole history of QCD mapped out -- ranging from
questions of asymptotic freedom, quark-hadron duality, spectroscopy,
scattering, over issues of confinement, covariance, vacuum structure,
chiral symmetry breaking, up to sum rules, supersymmetric extensions, 
``stringy" aspects, heavy quarks, and finally hot and dense baryonic
matter, our main
topic here.
Also some technical developments like light-cone quantization
would be unthinkable
without 2d models as firm testing ground. Whether in the end
these models will
help to solve the real hard problems in QCD is still an open
question; they certainly
help to prepare us intellectually and widen our horizon about the
fascinating world of non-perturbative physics and non-Abelian gauge theories.

As will be seen later on, one of the main themes of this article
is chiral symmetry and its spontaneous breakdown (SSB). We are
aware that studying
SSB in low dimensions is like dancing on thin ice --- there
is the constant danger
of falling into the trap of some no-go theorem and
making a fool of oneself.
Without claiming mathematical rigor, we believe that the large $N$ limit
provides a reasonable protection against these perils.
Thereby, it significantly
enlarges the spectrum of
questions which can be addressed within lower dimensional soluble models. 

In the spirit of what has been said so far, we have attempted
to write this article in a
rather self-contained and hopefully pedagogical way. The literature
about two-dimensional field theory is
not always easy to access for the non-expert, mainly because a number of
different, highly specialized techniques are employed by different
sub-communities
(e.g. bosonization, light-cone quantization, functional methods, conformal
and topological field theory). We have refrained from using genuine
2d-technology as
much as possible
and instead based our presentation upon a simple yet powerful technique
familiar to every physicist, the Hartree-Fock (HF)
approximation.\cite{Hartree28,Fock30}
For the large $N$
models which we will consider here, one can actually go quite
far with a relativistic
extension of this basic tool from many-body theory.
This should also
make the article more readable for those who want to learn
field theory while working
on QCD related problems (e.g. students or researchers with a nuclear physics
background like ourselves). 

Finally we should apologize for the perhaps not always balanced choice of
citations: It is strongly biased towards work which fits into our
approach and may
not do justice to other important activities in the field. 
Our article is not intended as a comprehensive review article
of the usual type,
but rather an attempt to convey some of the intellectual pleasure of
working on QCD motivated, analytically soluble problems, using as example
of topical interest hot and dense matter in two-dimensional
large $N$ models. This is presumably more adequate for 
a contribution to a Festschrift like the present one in honor
of Boris Ioffe.

\section{Basics about 1+1 dimensions}\label{sec:basics}

The transition from the real world to 1+1 dimensions has drastic consequences.
These consequences are responsible for tremendous technical simplifications
of quantum field theory, but also for many unrealistic features
and even pathologies which have to be kept in mind. In the present
section, we have collected the main kinematical and geometrical
facts which are independent of the specific dynamics, as well as some
characteristic aspects of two-dimensional gauge theories. The selection of topics
is guided
by the applications which we have in mind in later sections. Therefore,
we leave out a large number of important topics which are closely
associated with low dimensions but less relevant for
us.\cite{Blau93,Abdalla96}
These include boson-fermion mappings, existence of anyons
with arbitrary statistics, conformal
invariance, sigma models, integrability questions as well as
relations to matrix models, string theories, and gravity.
We will also make no use of
the technique of light-cone quantization which has become a powerful tool
for dealing with two-dimensional model field theories.\cite{Brodsky98}
It is not clear to us whether it is of any help
in the study of field theory at finite temperature and chemical potential.

\subsection{Kinematics}\label{subsec:kinem}

We use the metric and antisymmetric tensors
\begin{equation}
g^{\mu \nu} = \left( \begin{array}{rr} 1 & 0 \\ 0 & -1 \end{array}
\right) \ , \qquad
\epsilon^{\mu \nu} = \left( \begin{array}{rr} 0 & -1 \\ 1 & 0 \end{array}
\right)
\end{equation}
in 1+1 dimensions. Clearly, the
Lorentz group SO(1,1) consists only of boosts --- there are no rotations
(and hence there is no spin). The inhomogeneous Lorentz group or
Poincar\'{e} group ISO(1,1) has three generators
(boost $K$, momentum $P$, Hamiltonian $H$) which satisfy
the Poincar\'{e} algebra
\begin{equation}
[H,P]  =  0 \ , \qquad [K,P]={\rm i}H \ , \qquad [K,H]={\rm i}P \ .
\end{equation}
States are characterized by the Casimir operator
\begin{equation}
P_{\mu}P^{\mu} = m^2 \ ,
\end{equation}
the dispersion relation has the standard form $E=\sqrt{p^2+m^2}$.
A left-over type of helicity can be defined
via Lorentz-boosts
\begin{equation}
\left( \begin{array}{c} x^0 \\ x^1 \end{array} \right)' =
\left( \begin{array}{cc} \cosh \xi & \sinh \xi \\ \sinh \xi
& \cosh \xi \end{array} \right)
\left( \begin{array}{c} x^0 \\ x^1 \end{array} \right)
\label{eq:boost}
\end{equation}
with rapidity
\begin{equation}
\xi = {\rm artanh}\, v \ .
\end{equation}
States are said to have helicity $s$ if they transform according to
\begin{equation}
|s \rangle ' = {\rm e}^{\xi s}|s \rangle \ .
\end{equation}
Lorentz scalars, vectors
and tensors can be constructed in the usual way.
If $V^{\mu}$ is a two-vector, its light-cone
components transform irreducibly according to helicity $\pm 1$,
\begin{equation}
(V^0 \pm V^1 ) ' = {\rm e}^{\pm \xi} (V^0 \pm V^1) \ .
\end{equation}
The trace of a tensor $T^{\mu}_{\mu}$ is Lorentz scalar,
as is an antisymmetric tensor (e.g. the electric field strength).
A symmetric, traceless tensor has helicity 2,
\begin{equation}
(T^{00} \pm T^{01} )'={\rm e}^{\pm 2 \xi} (T^{00} \pm T^{01}) \ .
\end{equation}
One peculiar feature of massless states in two dimensions
is the following: Any number of massless non-interacting
particles moving in the same direction with momenta $p_i$
behave kinematically like a single massless particle of momentum $\sum_ip_i$.
This is partly responsible for the proliferation of massless
modes in certain models.

\subsection{Dirac fermions}\label{subsec:dirac}

We start our survey of various aspects of two-dimensional field theory
with Dirac fermions which will be the main actors in the following.
Canonically, fermions are distinguished from bosons by equal-time
anticommutation relations,
\begin{equation}
\{ \psi_i(x) , \psi_j^{\dagger}(y)  \} = \delta_{ij}\delta(x-y) \ .
\end{equation}
Since there is no spin,
a Dirac fermion has only two components (par\-ti\-cle/an\-ti\-par\-ti\-cle
degrees of freedom). The free Dirac equation reads
\begin{equation}
({\rm i} \gamma^{\mu}\partial_{\mu}-m)\psi = 0
\end{equation}
where the $\gamma$-matrices are 2$\times$2 matrices satisfying
\begin{equation}
\{ \gamma^{\mu}, \gamma^{\nu} \} = 2 g^{\mu \nu} \ .
\end{equation}
Unless stated otherwise, they will be chosen as
\begin{equation}
\gamma^0 = \left( \begin{array}{rr} 0 & 1 \\ 1 & 0 \end{array} \right) \ ,
\qquad
\gamma^1 = \left( \begin{array}{rr} 0 & -1 \\ 1 & 0 \end{array} \right) \ .
\end{equation}
The free Dirac Hamiltonian (in first quantization) is
\begin{equation}
h = \gamma^5 p + \gamma^0 m
\label{eq:hDirac}
\end{equation}
with
\begin{equation}
\gamma^5 = \gamma^0 \gamma^1 = \left( \begin{array}{rr} 1 & 0 \\
0 & -1 \end{array} \right) \ .
\end{equation}
This representation is well suited
for the chiral limit $m\to 0$. Free spinors (eigenvalues $\pm E_p$ with
$E_p=\sqrt{m^2+p^2}$) are
\begin{eqnarray}
u(p) & = &
\left( \begin{array}{r} \cos \theta_p/2 \\ \sin \theta_p /2 \end{array} \right)
= \frac{1}{\sqrt{2 E_p}}
\left( \begin{array}{r} \sqrt{E_p+p} \\ \sqrt{E_p-p} \end{array} \right)\ ,
\nonumber \\
v(p) & = &
\left( \begin{array}{r} -\sin \theta_p/2 \\ \cos \theta_p /2 \end{array} \right)
= \frac{1}{\sqrt{2 E_p}}
\left( \begin{array}{r} -\sqrt{E_p-p} \\ \sqrt{E_p+p} \end{array} \right)\ .
\label{eq:spinors}
\end{eqnarray}
For later convenience, we have introduced the free Bogoliubov angle as
\begin{equation}
\theta_p = {\rm arccot} \left( \frac{p}{m} \right) \, \in \, [0,\pi]
\ .\label{freebog}
\end{equation}
It appears if one diagonalizes $h$ of Eq. (\ref{eq:hDirac}) by a canonical
transformation with
the unitary operator
\begin{equation}
U_p = \exp \left\{-
  \frac{\theta_p}{2} (a_p^{\dagger}b_p-b_p^{\dagger}a_p ) \right\}
\end{equation}
for each mode where $a_p$ and $b_p$ are second quantized operators for
right- and lefthanded fermions with momentum $p$.
In the massless limit, right- and left-handed fermions are projected
out by
\begin{equation}
P_{R,L} = \frac{1 \pm \gamma^5}{2} \ .
\end{equation}
Chiral transformations are defined as
\begin{equation}
\psi \to {\rm e}^{{\rm i}\alpha \gamma^5} \psi \ ,
\end{equation}
and represent a symmetry of the massless theory
(note that $\gamma^5$ can be defined in an odd number of space
dimensions only).
To see the meaning of right/left-handedness in 1+1 dimensions, we
specialize the
above spinors to $m=0$, where
\begin{equation}
\theta_p = \Theta(-p) \pi
\end{equation}
and hence
\begin{equation}
u(p) = \left( \begin{array}{c} \Theta(p) \\ \Theta(-p) \end{array} \right)
\end{equation}
($\Theta(p)$ denotes the Heaviside step function).
Thus handedness in two dimensions is correlated with the direction of motion
(``right- and left movers"). The
Lorentz transformation is consistent with the above definition
of kinematic helicity $s=\pm 1/2$ for right- and left-handed
fermions, respectively,
\begin{equation}
\psi(x) \to {\rm e}^{\xi \gamma^5/2} \psi(x') \ ,
\end{equation}
with $x'$ as given in Eq.~(\ref{eq:boost}).
Another important peculiarity of fermions in two dimensions concerns
the vector and axial currents,
\begin{equation}
j_V^{\mu} = \bar{\psi}\gamma^{\mu} \psi \ , \quad j_A^{\mu} =
\bar{\psi} \gamma^{\mu} \gamma^5 \psi
\end{equation}
and their (partial) conservation laws,
\begin{equation}
\partial_{\mu} j_V^{\mu} = 0 \ , \quad \partial_{\mu}j_A^{\mu}
= 2 {\rm i} m \bar{\psi} \gamma^5 \psi \ \label{partialcons}.
\end{equation}
Due to the severe restrictions for $\gamma$-matrices
in two dimensions, these
currents have actually only two independent components,
\begin{equation}
j_V^0 = j_A^1 \ , \quad j_V^1 = j_A^0\ ,
\end{equation}
or
\begin{equation}
j_A^{\mu} =  \epsilon^{\mu \nu} j_{V,\nu} \ .
\end{equation}
This intimate relationship between vector and axial currents
is the key for understanding the existence of massless
mesons and baryons (see Sec.~\ref{subsec:chiral}).

\subsection{Scalar bosons}\label{subsec:scalar}

Bosons are quantized canonically with equal-time commutation relations
\begin{equation}
\left[ \pi(x) , \phi(y) \right] = -{\rm i} \delta(x-y) \ .
\end{equation}
Whereas the free massive scalar field has no special features as compared to
3+1 dimensions and satisfies the Klein-Gordon equation
\begin{equation}
(\partial_{\mu} \partial^{\mu} + m^2 )\phi = 0 \ ,
\end{equation}
massless bosons are quite delicate. The
simplest object in 3+1 dimensional quantum field theory, the massless
scalar field, becomes the most subtle one in 1+1 dimensions due to
severe infrared divergences.\cite{Nakanishi} As pointed out by
Coleman,\cite{Coleman73} formally,
the integral appearing in the free two-point function
\begin{equation}
\langle 0 | \phi(x) \phi(0) |0\rangle = \int \frac{\rd^2k}{(2\pi)}
{\rm e}^{{\rm i}kx} \delta(k^2) \Theta(k_0)
\label{eq:2point}
\end{equation}
is IR divergent, as can be seen by performing the
$k^0$-integration,
\begin{equation}
\int_0^{\infty} \frac{\rd k^1}{2\pi k^1} \cos(k^1x^1) {\rm e}^{{\rm i} k^1 x^0} \ .
\end{equation}
Since this is also the simplest example of a Goldstone boson in
3+1 dimensions (broken symmetry: $\phi \to \phi + c$),\cite{Itzykson}
one expects
trouble with the Goldstone theorem in two dimensions.
This is indeed what happens:
There is a rigorous no-go theorem which forbids SSB
of a continuous symmetry in two dimensions
(Coleman theorem\s\cite{Coleman73}).
A similar phenomenon is known in statistical physics: there is no
long range order in two-dimensional systems (at least with short range
interactions\s\cite{Mermin66}).
The strong infrared fluctuations destroy the order, for
instance in a crystal, a magnet or a superfluid.
There is a different type of long range order though which
we discuss in the next subsection.

\subsection{Long range order}\label{subsec:long}

The
absence of long range order
in 1+1 dimensional QFT or two-dimensional statistical physics
could be a fatal blow to
our investigation --- what interesting phase structure can
possibly be left? To understand the way out, let us first come back to
the essence of two-dimensionality alluded to in the preceding subsection:
Long range order cannot be maintained over arbitrarily large
distances due to strong fluctuations of the Goldstone mode; there is no SSB.
A nice example for this effect are atoms in a plane which want to
form a crystal.\cite{Huang}
The amplitude of fluctuations of each atom
around its equilibrium position grows logarithmically
with $L$, if $L$ is the extension of the crystal.
The calculation
involves a similar infrared divergence as in the
2-point function (\ref{eq:2point}),
\begin{equation}
\langle \vec{u}^2 \rangle \sim \int \frac{\rd^2 k}{(2\pi)^2}
\frac{1}{\omega^2(k)} \sim \int_{2\pi/L}^{\Lambda} \frac{\rd k}{k} \ ,
\end{equation}
($\Lambda = 1/a$, inverse lattice spacing).
In order to circumvent this no-go theorem one can go to the 
large $N$ limit, invoking a diverging number of degrees of freedom
at each point. This was demonstrated
by Witten\s\cite{Witten78,Affleck86} for
the chiral GN
model, a case very pertinent to the present study. After integrating out
the fermions with the help of a complex scalar auxiliary field
$\Phi=\Phi_0 {\rm e}^{{\rm i}\theta}$, one gets the following
effective low-energy action for the Goldstone mode $\theta$,
\begin{equation}
{\cal L}_{\rm eff} = \frac{N}{4\pi} (\partial_{\mu}\theta)^2 \ .
\end{equation}
Chiral symmetry breaking can be probed by the correlator
\begin{equation}
\langle \bar{\psi}\psi(x) \bar{\psi}\psi(y) \rangle \to \langle \Phi^*(x)
\Phi(y) \rangle \ .
\end{equation}
It
approaches a constant $|\Phi_0|^2$ for $|x-y| \to \infty$ if the
symmetry is broken.
Without SSB, one expects an exponential decay
\begin{equation}
\langle \Phi^*(x) \Phi(y) \rangle \sim {\rm e}^{-M|x-y|} \ .
\end{equation}
Here, a different behavior is found,
\begin{equation}
\langle \Phi^*(x) \Phi(y) \rangle \sim \langle {\rm e}^{-{\rm i} \theta(x)}
{\rm e}^{{\rm i} \theta(y)} \rangle \sim {\rm e}^{-2\pi G(x-y)/N}
\sim \frac{1}{|x-y|^{1/N}}  \ .
\end{equation}
$G$ is the free massless boson propagator, logarithmic in two dimensions.
Exponentiation
of this propagator yields a power law behavior for the propagator.
Evidently, there is
no SSB at any finite $N$, in agreement with the no-go theorem, but there
is a loophole at $N=\infty$ since
\begin{equation}
\langle \Phi^*(x) \Phi(y) \rangle \sim 1 + \frac{1}{N} \ln |x-y| + O(1/N^2)\ .
\end{equation}
The correlator becomes more and more long-range until it cannot be
distinguished from a constant, in the limit $N\to \infty$.
A similar phenomenon is known from two-dimensional $xy$
spin-model\s\cite{Kosterlitz73,Kosterlitz74,Berezinsky71}
under the name of topological order.

\subsection{Gauge fields}\label{subsec:gauge}

Gauge symmetry implies redundant variables; this is nowhere as
clear as in a two dimensional world where, due to the absence of
transverse directions,
one might guess that all degrees of freedom of a gauge field
are ``pure gauge''.
This is not quite true --- there are left-over quantum mechanical
degrees of freedom stemming from gauge invariant zero-mode fields.
They can only be discussed reasonably well
on a finite interval of length $L$.
Evidently, pure gauge theories are a totally different story in 1+1
and 3+1 dimensions,
as are the corresponding interactions between static charges.

\subsubsection{Abelian gauge fields}\label{subsubsec:abelian}

For the U(1) gauge theory, Lagrangian and field strength tensor
look as usual,
\begin{equation}
{\cal L} = - \frac{1}{4} F_{\mu \nu} F^{\mu \nu} \ , \quad
F_{\mu \nu} = \partial_{\mu} A_{\nu} - \partial_{\nu} A_{\mu} \ .
\end{equation}
In two dimensions however, an antisymmetric tensor has only one independent
component, a Lorentz scalar
(the electric field); there is no magnetic field. Canonical quantization
in the Weyl gauge ($A_0=0)$ reveals that the
electric field $E=\partial_0 A^1$ is conjugate to (minus) the vector
potential $A^1$,
\begin{equation}
[E(x),A^1(y)] = {\rm i} \delta(x-y) \ .
\end{equation}
The Gauss law in this gauge has to be implemented as a
constraint on the physical states
(this is possible since $G(x)$ is conserved),
\begin{equation}
G(x)|\rangle = \partial_1 E(x)|\rangle = 0 \ .
\end{equation}
This leaves only the 0-mode of $A_1$, the spatially constant electric
field, as physical variable. Switching to finite $L$, we have
\begin{equation}
a = \frac{1}{L} \int_0^L {\rm d}x A^1(x)\ , \quad e =\int_0^L
{\rm d}x E(x)
\end{equation}
with
\begin{equation}
H=\frac{1}{2L} e^2 \ , \qquad [e,a] = {\rm i} \ .
\end{equation}
Since the integer part of $\frac{gLa}{2\pi}$, $g$ being the
electric charge of static sources, can also be gauged away by a
periodic gauge transformation
\begin{equation}
U={\rm e}^{{\rm i} 2\pi n x/L} \ ,
\end{equation}
pure QED reduces to quantum mechanics of a particle on
a circle.\cite{Manton85} If we
couple static charges to the gauge field, this changes Gauss's law into
$G=\partial_x E -e \rho$. The ``longitudinal" electric field
(i.e., the $x$-dependent part) is the Coulomb field of
the static charge, which
is linear in one space dimension. For comparison with
Yang-Mills theory below,
assume two charges $\pm g$ on a circle of length $L$ at points $x,y$;
then the change in energy is
\begin{equation}
\Delta E  = g^2 K(x-y)
\end{equation}
with the periodic Coulomb potential
\begin{eqnarray}
K(x-y) &=& \langle x | \frac{1}{\partial_x^2} |y \rangle =
-\frac{L}{4\pi^2} \sum_{n \neq 0} \frac{1}{n^2} {\rm e}^{{\rm i}
2\pi n (x-y)/L} \nonumber \\
& = & -\frac{L}{12} + \frac{1}{2}|x-y| -\frac{1}{2L}(x-y)^2 \ .
\label{eq:percoul}
\end{eqnarray}
Pure QED is confining in 1+1 dimensions for rather trivial reasons. In the
limit $L\to \infty$, the fact that the potential to order $g^2$ is
linear also follows on purely dimensional grounds, since
$[g]=L^{-1}$.
Finally, we note that one has to impose the residual Gauss law
($Q|\rangle = 0$); only the charge 0 sector survives on the circle.

\subsubsection{Non-Abelian gauge fields}\label{subsubsec:non-abelian}
Consider the Lagrangian of
Yang-Mills theory with a SU(N) gauge group on
a circle,\cite{Rajeev88,Hetrick89,Hetrick93}
\begin{equation}
{\cal L} = - \frac{1}{4} F_{\mu \nu}^a F^{a \mu \nu} \ ,
\end{equation}
and
\begin{equation}
F_{\mu \nu}^a = \partial_{\mu} A^a_{\nu} - \partial_{\nu} A^a_{\mu}
- g f^{abc} A_{\mu}^b A_{\nu}^c \ .
\end{equation}
Whereas
canonical quantization in the Weyl gauge is as straightforward
as in the Abelian case,
the resolution of Gauss's law becomes more intricate due to its non-linearity
\begin{equation}
G^a(x) = (DE)^a(x)= \partial_x E^a -g f^{abc} A^{1,b} E^c \ .
\end{equation}
Generically, the covariant derivative has $N-1$ zero modes; the
projections of $E$ onto these are the physical (quantum
mechanical) variables $e^p$.\cite{Lenz94} They are non-hermitean due to the
projection onto
$A^1$-dependent basis vectors.
The corresponding gauge invariant coordinates
are the eigenphases of the path ordered integrals (untraced
Polyakov loops) around the compact space direction,
\begin{equation}
{\cal P} = P {\rm e}^{{\rm i}g \int_0^L {\rm d}x A^1(x)}
= V {\rm e}^{{\rm i} g a L} V^{\dagger} \ , \qquad a = a^p t^p \ .
\end{equation}
One finds
\begin{equation}
[e^p, a^q] = {\rm i} \delta_{pq}
\end{equation}
and the Hamiltonian
\begin{equation}
H_a = \frac{1}{2L} e^{p \dagger} e^p = - \frac{1}{2L}
\frac{1}{{\cal J}(a)} \frac{\partial}{\partial a^p} {\cal J}(a)
\frac{\partial}{\partial a^p} \ .
\end{equation}
This
form makes explicit the fact that the variables are curvilinear
coordinates --- $H_a$ is just the Laplacian on the SU(N) group manifold,
the Jacobian ${\cal J}(a)$ being the reduced Haar measure
(Vandermonde determinant for unitary matrices)
\begin{equation}
{\cal J}(a) = \prod_{i>j} \sin^2 \left( \frac{1}{2} gL(a_{ii}-a_{jj})
\right) \ .
\end{equation}
Convenient angular variables are
\begin{equation}
\varphi_i = g L a_{ii}
\end{equation}
with the constraint $\sum_i \varphi_i = 0$ for SU(N).
If one introduces the analogue of a ``radial wavefunction"
\begin{equation}
\Phi(\varphi) = \frac{u(\varphi)}{\sqrt{{\cal J}(\varphi)}} \ ,
\end{equation}
the Laplacian is reduced to cartesian form (except for a
center of mass correction in the SU(N) case, as opposed to U(N)),
and a constant effective potential appears,
\begin{equation}
H= - \frac{g^2 L}{4} \left( \sum_{i=1}^N \frac{\partial^2}
{\partial \varphi_i^2}
- \frac{1}{N}\left( \sum_{i=1}^N\frac{\partial}
{\partial \varphi_i} \right)^2 \right)
- \frac{Ng^2L(N^2-1)}{48} \ .
\end{equation}
The reduced wave function $u(\varphi)$
now has to satisfy the boundary conditions
\begin{equation}
u(\varphi)=0 \ \  {\rm if} \ \ \varphi_i = \varphi_j \ .
\label{eq:bound}
\end{equation}
The ground state is simply $\Phi_0 = $ const.,
normalizable owing to the compactness of the variables, with $E_0=0$.
To get the excited states for any $N$, it is
advantageous to reinterpret this problem in terms of free, non-relativistic
fermions on a circle\s\cite{Minahan93a} (the Pauli principle
is enforced by Eq.~(\ref{eq:bound}))
with periodic (antiperiodic) boundary
conditions for odd (even) $N$. Then the wavefunctions are
Slater determinants of single particle wave functions
\begin{equation}
\psi_m(\varphi) = \frac{1}{\sqrt{2\pi}} {\rm e}^{{\rm i}m \varphi}
\end{equation}
with $m$ integer for odd, half integer for even $N$.
The energy is
\begin{equation}
E = \frac{g^2 L}{4} \left( \sum_m^{{\rm occ}}m^2 - \frac{1}{N}
\left( \sum_m^{{\rm occ}}m \right)^2 \right)
 - \frac{Ng^2L(N^2-1)}{48}      \ ,
\end{equation}
where the sums run over all occupied orbits.
By filling the lowest $N$ single particle states,
one can easily ascertain that
the ground state wavefunction is constant and that the ground state energy
vanishes.
The configuration space is determined by the fundamental domain,
the smallest region bounded by zeros of the Jacobian. We can choose
\begin{equation}
\varphi_1 < \varphi_2 < ... < \varphi_N
\end{equation}
and the center of mass constraint
$\sum_i \varphi_i=0$  for $SU(N)$ (this represents
a more complete gauge fixing). Since the ordering is on a circle
rather than a line,
it still
leaves a residual gauge freedom, namely cyclic permutations of
color labels. This residual $Z_N$ symmetry is expected on
topological grounds due to
\begin{equation}
\Pi_1(SU(N)/Z_N)=Z_N.
\end{equation}
This is the well-known center symmetry of pure Yang Mills theory which plays an
important role at finite temperature.\cite{Svetitsky86}
By way of illustration, take SU(2): There is only one independent
angular variable,
$\varphi = (\varphi_1-\varphi_2)/2 \in [0,\pi]$. Then, YM$_2$ reduces
to quantum mechanics
of a particle in an infinite square well with
\begin{equation}
H = -\frac{g^2 L}{8} \frac{\partial^2}{\partial \varphi^2}-
\frac{g^2 L}{8}
\end{equation}
and
\begin{equation}
u_n(\varphi) = \sqrt{\frac{2}{\pi}}\sin n\varphi \ , \ \ \
{\cal J}(\varphi) = \sin^2 \varphi \ ,
\ \ \ E_n= \frac{g^2 L}{8}(n^2-1) \ .
\end{equation}
The Z$_2$ center symmetry is just parity. A detailed discussion of the
SU(3) case with two independent variables can be found in
Ref. 27.
At very large $N$, since the $\varphi_i$ repel
each other (cf. Eq.~\ref{eq:bound}))
and live on a circle,
they will tend to distribute themselves along the circle
like a pearl necklace, with fluctuations which are $1/N$ suppressed in
amplitude. This observation will become important later on
for the 't Hooft model at finite temperature.

\begin{figure}[t]
  \begin{center}
  \epsfig{file=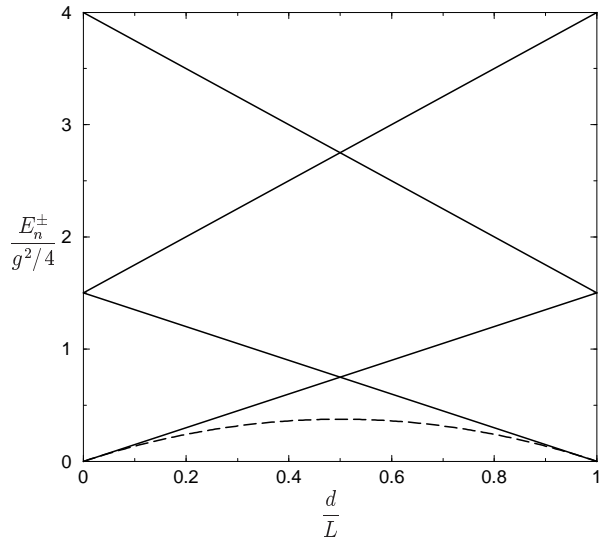,width=8cm}
  \caption{Interaction energy of static quarks at distance $d$
    on a circle of length $L$.
    Dashed line: QED$_2$, cf. Eq. (\ref{eq:percoul}); solid lines: SU(2) YM$_2$
    (cf. Refs. 28, 30).}
  \label{fig:1}
  \end{center}
\end{figure}

The existence of
zero mode gluons also complicates the interaction between static
charges, at least for finite $L$. If one puts a static $q\bar{q}$-pair on
a circle,
the Coulomb potential evidently involves $D^{-2}$ (covariant derivatives)
rather than
$\partial^{-2}$, so that the Coulomb interaction gets modified by the
gauge field remnants. The Coulomb potential
can be transformed to coordinate space,\cite{Engelhardt95b,Dhar94a}
\begin{eqnarray}
K_{ij}(z) & = & - L\sum_n \frac{1}{(2\pi n-\varphi_j+\varphi_i)^2}
{\rm e}^{{\rm i}2\pi nz/L}
\\
& = &-\frac{L}{4} {\rm e}^{{\rm i}(\varphi_j-\varphi_i)z/L}
\left( \frac{1}{\sin^2(\varphi_i-\varphi_j)/2} - \frac{2}{L}|z|
+ {\rm i} \frac{2}{L} z \cot (\varphi_i-\varphi_j)/2\right)\nonumber
\end{eqnarray}
for $i \neq j$, whereas it is identical to the Abelian one,
Eq.~(\ref{eq:percoul}), for $i=j$.
Since the $\varphi_i$ are dynamical variables, one is not yet through
but still has to solve a fairly complicated quantum mechanical
$(N-1)$-body problem
with a two-body potential defined by the Coulomb interaction.
This reflects the fact that
due to color spin dynamics, charges are never static in the
non-Abelian case, even if they are nailed down in space.
A full analytic solution is available for
the SU(2) case,\cite{Engelhardt95b} where a
dynamical quantum mechanical supersymmetry is
at work.\cite{Seeger98} It explains why the potential is strictly linear
(at all $L$, not just in the limit $L\to \infty$), and why
-- for vanishing separation of the charges -- all excited
states are doubly degenerate whereas the ground state is
non-degenerate and has strictly zero energy,
see Fig.~\ref{fig:1}.
In the limit $L \to
\infty$, all excited states move up to infinite energy, the gauge
field dynamics get frozen, and the
difference between QED$_2$ and YM$_2$ disappears.

The discussion about the zero mode gauge fields may seem a bit pedantic:
Why bother about finite $L$ at all? The reason is of course that we
are interested among other things in the finite temperature problem.
The physics relevance of the Polyakov loop is beyond doubt there.

Finally, the fact that
the left-over gluon variables are eigenphases of path-ordered integrals
has been used in the literature to emphasize the string picture:
Closed strings that wrap around the compact
direction in pure Yang-Mills theory, open strings with quarks at the
ends if one includes matter fields in the fundamental
representation.\cite{Dhar94a,Dhar94b,Dhar94c}
We shall not exploit this picture here any further.

\subsection{Renormalization}\label{subsec:renormalization}

From the kinetic term in the Lagrangian in two dimensions,
one infers that Dirac
fermion, scalar boson
and vector fields have dimensions
\begin{equation}
[\psi_i]= L^{-1/2} \ , \quad [\phi] = [A^{\mu}] = L^0 \ .
\end{equation}
In self-in\-ter\-ac\-ting fermion theories, a quartic
self-in\-ter\-ac\-tion has
dimensionless coupling constant and is therefore perturbatively
re\-nor\-ma\-li\-zable (GN model) unlike in four dimensions
(NJL model).
Self-in\-ter\-ac\-ting
bosons can have interactions of arbitrary high order with couplings
of positive mass dimension and are all super
renormalizable (e.g. sine-Gor\-don model). The
gauge coupling has dimension of mass, therefore gauge theories like
the Schwinger model or QCD$_2$ are also su\-per-re\-nor\-ma\-li\-zable.

In the U(1) theory, an axial anomaly occurs, as is well known from the
Schwinger model. In one-flavor SU(N) QCD$_2$, there is no axial
anomaly.

Pure fermionic theories of GN type are asymptotically free;
they share this property with QCD and are rather exceptional in this respect.
The $\beta$-function is negative (see Sec.~\ref{subsubsec:vacuum}).
If the bare mass is put equal to zero (as in the original GN model),
dimensional
transmutation occurs like in pure Yang-Mills theory
in four dimensions: A massless
theory generates mass dynamically. All dimensionless ratios can be
predicted without free parameter, only the physical fermion mass
is needed as an input.\cite{Gross74}

In the
massive GN model, there are two distinct parameters of dimension mass
(matching the two bare parameters $Ng^2, m_0$):
The physical fermion mass $m$ and the ratio $m_0/(Ng^2)$. The first
parameter is directly observable via the spectrum. The second
parameter is also observable: In the chiral GN model, it
determines
the mass of the would-be Goldstone boson, cf. Sec.~\ref{subsec:chiral}.
In the non-chiral GN model, it can be measured via the
$\bar{q}q$ scattering length which is proportional
to $N g^2/m_0$.
In the limit $m_0 \to 0$, the pion mass goes to zero and the scattering
length in the non-chiral model diverges. In this way, the second
parameter disappears and all observables can be expressed in terms of
$m$ only.

QCD$_2$ is trivially asymptotically free, being
su\-per-re\-nor\-ma\-li\-zable.
The perturbative (short distance) interaction is determined by
dimensional considerations.

\subsection{Soluble models of interest for QCD}\label{subsec:soluble}

The first generation of soluble models like the Thirring
model\s\cite{Thirring58} or the
massless Schwinger model\s\cite{Schwinger62} from $\pm$1960 were
not yet rich enough for
the questions we are interested in. Although they can be solved exactly
and have certain interesting non-perturbative aspects, they essentially
describe free theories with a trivial $S$-matrix.
If one tries to make
them non-trivial by adding e.g. a mass term or flavor, one loses solvability.
The Schwinger model has confinement, but provides no model for hadrons.

From this point of view, 't~Hooft's idea of the $1/N$
expansion\s\cite{tHooft74a}
is very nice indeed.
By invoking a large number of fermion species
and using $1/N$ as a small parameter, one finds highly non-trivial
results already to leading order. Semiclassical methods become
applicable and yield systematic, reliable results. All symmetries
like Poincar\'{e} invariance, chiral symmetry, gauge invariance are
preserved.
The phenomenological arguments in favor of the large $N$ limit for QCD
(Zweig rule etc.\cite{Coleman79,Witten79,Manohar98}) naturally play
no role in two dimensions, where it is
a purely theoretical tool.

Models based on the large $N$ expansion can be considered as second
generation field theory models, also historically.
The two models we are particularly
interested in came up in 1974, the GN\s\cite{Gross74} and
the 't Hooft model.\cite{tHooft74b}
The GN model generalizes the Thirring model to $N$ flavors,
while the 't Hooft model generalizes the Schwinger model to the
non-Abelian SU(N) gauge group with a large number of colors.
They acquire a rich physical content, as we hope to convince the reader
in the remainder of this article. Here, we only
mention the following basic facts: The GN models describe
self-interacting massless fermions with a
point-like four-fermion interaction.
Depending on the particular variant, they can have
either a discrete or a continuous chiral symmetry; in the latter case,
the GN model may be thought of as the NJL model\s\cite{Nambu60}
in two dimensions.
Mass terms which violate chiral symmetry explictly
can be included as well without loss of solvability
(at large $N$).
The 't Hooft model is QCD$_2$ with SU(N) color in the limit $N \to \infty$,
with massless or massive
quarks in the fundamental representation. As mentioned above, the
model on the infinite line has no gluon degrees of freedom,
but is also a self-interacting fermion theory, here with a linear
Coulomb potential characteristic for the two dimensional world
(in an appropriate gauge).
Unlike the GN models, this model shows confinement,
which makes it particularly interesting from the
point of view of QCD. The fact that
also in real life (in four dimensions),
the NJL model and QCD are used in parallel
nowadays for certain questions of strong
interaction physics makes this pair of models a good match and
immediately suggests many questions to be asked.
Throughout this work,
we do not consider any embellishments or generalizations
(e.g. flavor, more complicated interactions, different matter fields)
of these models which
have occasionally been invoked.
The nice thing about the original models is just the contrast between
the simplicity of
the Lagrangian and
the complexity of the phenomena it produces, and we do not want
to spoil this property.

Finally, we should point out that the main goal of the present study are
questions of hot and dense matter in these toy models.
For this purpose, the large $N$ limit is not
only a technical trick leading to solvability of the models, but
is in fact instrumental for enabling chiral symmetry breakdown in two
dimensions. This is the reason why another aspect of the GN
model will play no role here, its
integrability.\cite{Zamolodchikov78,Andrei79,Andrei80}
A lot of progress
has been made towards solving the finite $N$ model exactly. Here,
we shall restrict ourselves to the large $N$ limit so that the
corresponding difference between the non-integrable 't Hooft model
and the integrable GN models will not play any role,
but a common approach for both
models will be used throughout.

\section{Two dimensional models at zero temperature and chemical potential}
\label{sec:models}

\subsection{Gross-Neveu models}\label{subsec:GN-Models}

We cannot discuss finite temperature or density without first reviewing
the particle content of the theory.
Originally, the GN model has been solved with the help of semi-classical
functional methods,\cite{Gross74} supplemented by inverse
scattering techniques from soliton theory
for the baryons.\cite{Dashen75b} This approach can be rigorously
justified in the large $N$ limit.
Typically, one introduces
auxiliary bosonic
fields and integrates out the fermions (Gaussian integral
over Grassmann variables), then applies a
saddle point approximation to the remaining bosonic
functional integral.
More recently, an alternative to the inverse scattering method
has been proposed, based on the Gel'fand-Dikii equation.\cite{Feinberg94}
Here, in order
to have a coherent scheme for all questions that interest us as well as
for pedagogical reasons, we shall
make use of time honored concepts well known from atomic and nuclear physics,
such as the HF and random phase approximation (RPA), although
in a fully relativistic setting.\cite{Salcedo91,Lenz91,Pausch91} In
contrast to the functional integral methods,
we thereby handle fermions more
directly and, since we work canonically, can also more readily address
questions of the vacuum structure (Dirac sea). Needless to say,
the equations one eventually solves are always the same
(HF or Schwinger-Dyson, RPA or Bethe-Salpeter equations).
More than anything else the choice is a question of language.

\subsubsection{Vacuum, physical fermion}\label{subsubsec:vacuum}

The Lagrangian of the GN model family, including a bare fermion
mass term and suppressing color labels, reads\s\cite{Gross74}
\begin{equation}
{\cal L} = \bar{q} \, \ri \gamma^{\mu} \partial_{\mu} q + \frac{1}{2}
g^{2} \left[ \left( \bar{q}q\right)^{2}-\lambda \left( \bar{q}\gamma^5
q\right)^2 \right] - m_0 \bar{q}q  \ .
\end{equation}
The original model with discrete chiral symmetry $q\to \gamma^5 q$ is
recovered for
$\lambda=0, m_0=0$; the choice $\lambda=1, m_0=0$ corresponds
to the NJL-type model
with continuous chiral symmetry; the $m_0$ term breaks both chiral symmetries
explicitly.
To leading order in $1/N$, the Hartree approximation can be used, replacing
$\bar{q}q$ in the Euler-Lagrange equation
\begin{equation}
 \ri \gamma^{\mu} \partial_{\mu} q +
g^{2} \left( \bar{q}q -\lambda (\bar{q}\gamma^5 q)\gamma^5\right)
q  -  m_0 q = 0
\label{eq:q_eom}
\end{equation}
by its vacuum expectation value (the Fock term is $1/N$ suppressed).
The non-zero value of the order parameter for chiral symmetry
signals SSB. We first assume
$\langle \bar{q} \gamma^5 q \rangle = 0$,
\begin{equation}
\left\{ \ri \gamma^{\mu} \partial_{\mu} - m_0  +
N g^{2} \rho_s \right\} q(x) = 0   \ .
\end{equation}
A physical fermion mass appears,
\begin{equation}
m=m_0-Ng^2\rho_s \ , \quad \rho_s =\frac{1}{N} \langle \bar{q}q\rangle \ ,
\label{eq:ferm_mass}
\end{equation}
as dictated by covariance.
The essence of the Hartree approximation is to impose a self-consistency
condition,
here on the scalar density $\rho_s$ of the vacuum.
Using the free fermion
spinors of Eq.~(\ref{eq:spinors}) above we get
\begin{eqnarray}
\rho_{s}=-\frac{(m-m_0)}{Ng^{2}} & \stackrel{!}{=} & \int_{-\Lambda/2}^{+\Lambda/2}
\frac{\rd k}{2\pi} \, \bar{v}(k)v(k)   \nonumber \\
& = & - \int_{-\Lambda/2}^{+\Lambda/2} \frac{\rd k}{2\pi} \frac{m}{\sqrt{k^{2}
+m^{2}}}    \simeq  - \frac{m}{2\pi} \log \frac{\Lambda^{2}}{m^{2}} \  .
\label{eq:gapint}
\end{eqnarray}
As usual, this is also a variational solution obtained by minimizing
the vacuum energy in the space of Slater determinants. The
Hartree- or gap equation,
\begin{equation}
m= m_0 + m \frac{Ng^2}{2\pi}\ln \frac{\Lambda^2}{m^2}  \ ,
\label{eq:gap}
\end{equation}
contains much information. For $m_0=0$ (discrete or continuous model),
it has two solutions, either $m=0$ or
\begin{equation}
1= \frac{Ng^2}{2\pi}\ln \frac{\Lambda^2}{m^2}  \ .
\end{equation}
The broken symmetry solution is always lower in energy density by
\begin{equation}
{\cal E}_{\rm vac} - {\cal E}_{\rm vac}^{\rm free} = - \frac{Nm^2}{4\pi} \ .
\end{equation}
The gap equation for the broken phase does not determine $m$, as is clear
from the fact
that the theory contains no
dimensionful parameter. If $m$ is adjusted (to the ``observed" fermion mass,
whatever that means in two dimensions), then the
gap equation is a renormalization condition which tells us how
$Ng^2$ (the bare coupling constant) depends on the cutoff $\Lambda$.
Correspondingly, the
physical mass has the dependence
\begin{equation}
m= \Lambda \exp \left \{-\frac{\pi}{Ng^2}\right \}
\label{eq:mass}
\end{equation}
familiar from similar dependences in real QCD and in line with the
renormalization group. All physical quantities with
dimension of mass have to depend in the same way on $g, \Lambda$. The
essential
singularity in $g$ at $g=0$ underlines the power of the $1/N$ expansion
as compared to ordinary perturbation theory in $g$. Dimensionless
quantities can be predicted without any parameter, at $m_0=0$.
Apart from this phenomenon of
dimensional transmutation, the other remarkable feature which we find here
is asymptotic freedom (AF).\cite{Gross73,Politzer73}
One can determine the $\beta$-function of the GN model in an
elementary way as follows: The physical fermion mass
$m$ should be cut-off independent, therefore
\begin{equation}
\Lambda\frac{{\rm d}}{{\rm d}\Lambda} m(\Lambda, g(\Lambda))=
\left( \Lambda \frac{\partial}{\partial \Lambda} + \Lambda
\frac{\rd g}{\rd \Lambda}
\frac{\partial}{\partial g}\right) m = 0 \ .
\end{equation}
Set
\begin{equation}
\Lambda \frac{\rd g}{\rd \Lambda} = \beta(g)
\end{equation}
and find, by inserting $m$ from Eq.~(\ref{eq:mass}),
\begin{equation}
\beta(g) = - \frac{N g^3}{2\pi} < 0
\end{equation}
or AF (to this order, the result is independent of the
renormalization scheme). More elaborate computations of the
higher loop beta-functions\s\cite{Wetzel85,Luperini91}
do not play a role in the large $N$ limit.
We shall return to the issue of AF in connection with $q
\bar{q}$ scattering, where a running coupling constant will be defined.

For $m_0=0$ and the case of continuous chiral symmetry, the assumption
$\langle \bar{q}
\gamma^5 q \rangle=0$  is too restrictive; one gets a continuum of
vacua related by U(1) transformations and labeled by a
chiral angle $\varphi$,
\begin{equation}
\langle \bar{q} q \rangle = -\frac{m}{g^2}\cos \varphi \ , \quad
\langle \bar{q} \, {\rm i}\gamma^5 q \rangle
= \frac{m}{g^2}  \sin \varphi \ .
\end{equation}
For non-vanishing bare mass $m_0$, the gap equation (\ref{eq:gap}) signals
the appearance of a second physical
parameter $\frac{m_0}{Ng^2}$, in addition to the overall
scale set by $m$. Its relevance for scattering and bound state observables
will become clear later on.

\subsubsection{Baryons}\label{subsubsec:baryons}

It would be premature to conclude from the preceding section that the GN model
at $N\to \infty$ reduces to a free, massive fermion theory. As observed by
Witten,\cite{Witten79}
baryons have to be considered at large $N$
and can be treated in HF approximation, like the vacuum. They correspond
to a different kind of HF solution which breaks translational invariance.
Their mass diverges like
$N$, the baryon-baryon interaction also scales like $N$, and these
effects have to be taken into account in leading order.
Baryons
were found originally with functional
techniques and inverse scattering methods.\cite{Dashen75b}
Since we now know the scalar potential, we can use it to verify Witten's
picture of baryons as relativistic HF solutions --- this is what we propose
to do here.\cite{Pausch91}
We consider the discrete chiral model first,
since the continuous model has in addition
totally different, light baryons, see Sec. \ref{subsec:chiral} below.

\begin{figure}[t]
  \begin{center}
    \epsfig{file=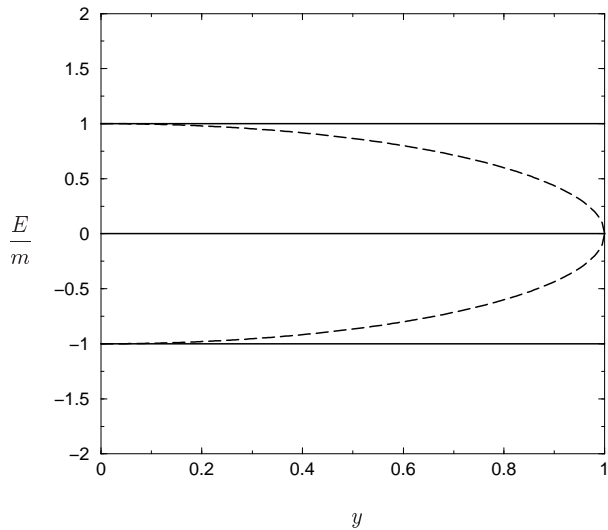,width=8cm}
    \caption{Single particle spectrum belonging to the baryon
      in the GN model. Continuum states fill 
      the outer region $|E/m|>1$; dashed curves inside the gap
      are valence levels;
      $y$ measures the occupation number of the positive energy
      valence state, cf. Eq. (\ref{upsilon}).}
    \label{fig:2}
  \end{center}
\end{figure}

If we restrict ourselves to baryons made up of $n \leq N$ quarks, it is
sufficient to take into account one positive energy ``valence" level
filled with $n$ fermions, together with the completely filled Dirac sea.
The Hartree equation without invoking translational
invariance becomes
\begin{equation}
\left\{ \ri \gamma^{\mu} \partial_{\mu} + N g^{2}
\rho_{s}(x)\right\} q(x) = 0
\label{eq:Hart}
\end{equation}
where the scalar density now refers to the baryon state $|B\rangle$\, ,
\begin{equation}
\rho_{s}(x) = \frac{1}{N} \sum_{i=1}^{N} \langle B|\overline{q}_{i}(x)q_{i}(x)|B\rangle  \ .
\end{equation}
We introduce the single particle energies and eigenfunctions of the
Dirac Hamiltonian corresponding to Eq.~(\ref{eq:Hart}),
\begin{equation}
\left\{ \gamma^{5} \frac{1}{\ri} \frac{\partial}{\partial x} -
\gamma^{0} N g^{2} \rho_{s}(x) \right\} \psi_{\alpha}^{(\pm)}(x)
= \pm \epsilon_{\alpha}\psi_{\alpha}^{(\pm)}(x) \ .
\end{equation}
Then $\rho_s(x)$ is given by
\begin{equation}
\rho_{s}(x) =
\frac{n}{N}\overline{\psi}_{0}^{\,(+)}(x)\psi_{0}^{(+)}(x)+
\sum_{\alpha} \overline{\psi}_{\alpha}^{\,(-)}(x)\psi_{\alpha}^{(-)}(x)
\end{equation}
where $\psi_0^{(+)}$ is the valence state.
Dashen {\em et al.}\s\cite{Dashen75b} have found two different
types of solutions for
the scalar density, the ``kink" and the ``double kink" (or rather
kink-antikink).
Let us start with the less exotic double kink,
characterized by the following scalar
potential:
\begin{equation}
S(x) \equiv - Ng^{2}\rho_{s}(x) = m \left\{1+y\left(\tanh \xi_{-}-\tanh\xi_{+}\right)
\right\}
\label{eq:scalar}
\end{equation}
with the definitions
\begin{equation}
y=\sin \theta  \ , \ \ \ \ \theta=
\frac{\pi}{2}\frac{n}{N}  \label{upsilon}\ ,
\end{equation}
\begin{equation}
\xi_{\pm} = ymx \pm \frac{1}{2}\mbox{artanh}\, y   \ .
\end{equation}
$S(x)$ has the form of a potential well and approaches for $x\to \pm \infty$
the asymptotic value $m$, i.e., the physical fermion mass.
Varying $y$ or equivalently the degree of occupation
$n/N$, $S(x)$ changes from a $1/\cosh^{2}(ymx)$ shape at small $y$
to a Woods-Saxon like shape
at $y \simeq 1$,
\begin{equation}
S(x) \sim \left(1+e^{(|x|-R)/c}\right)^{-1}   \ ,
\end{equation}
with skin thickness $c$ and radius $R$ given by
\begin{equation}
c=\frac{1}{2m}  \ , \ \ \ \ R=-\frac{1}{2m} \log \frac{\cos \theta}{2}
\ \ \ \ \ (y \to 1)  \ ,
\end{equation}
respectively.
We proceed
to verify the self-consistency of this scalar potential.
After a straightforward solution of the Dirac equation (\ref{eq:Hart})
with potential (\ref{eq:scalar}), one finds the following results: The normalized
wavefunctions of the
discrete (valence) states are given by (in a basis where $\gamma^0=-\sigma_1,
\gamma^1={\rm i}\sigma_3$)
\begin{equation}
\psi_{0}^{(\pm)}(x) = \frac{\sqrt{ym}}{2} \left(
\begin{array}{r} \frac{1}{\cosh \xi_{-}} \\ \mp \frac{1}{\cosh \xi_{+}}
\end{array} \right)
\label{eq:psi0}
\end{equation}
with eigenvalues
\begin{equation}
   E_{0}^{(\pm)} = \pm m \sqrt{1-y^{2}}    \ .
\label{eq:E0}
\end{equation}
The continuum states are
\begin{equation}
\psi_{k}^{(\pm)}(x) = \frac{1}{\sqrt{2}E(k)(\ri k+ym)}
\left( \begin{array}{r} (\ri k-m)(\ri k-ym\tanh\xi_{-}) \\
\pm E(k) (\ri k-ym\tanh \xi_{+}) \end{array}\right)
\re^{\ri kx}  \ ,
\label{eq:psik}
\end{equation}
with
\begin{equation}
E(k) = \sqrt{k^{2}+m^{2}}  \ .
\end{equation}
They are reflectionless, a characteristic feature of time
independent solutions.\cite{Shei76}
\cite{Feinberg96a,Feinberg96b} The
spectrum of the Dirac Hamiltonian therefore consists of the usual
positive and negative energy continua starting from $\pm m$
and a pair of discrete states inside the mass gap at energies
given by Eq.~(\ref{eq:E0}), see Fig.~\ref{fig:2}.
These latter states are of course a prerequisite for
obtaining bound baryons. The discrete negative energy state has
to be filled completely ($N$ fermions), its positive energy
partner with $n \leq N$ fermions (remember that, unlike in QCD, there is no
restriction to color singlets).
Reinserting Eqs.~(\ref{eq:psi0},\ref{eq:psik}) into (\ref{eq:scalar}), one
can then verify the
self-consistency of the solution. This requires the use of the vacuum
gap equation (\ref{eq:gap}) to
eliminate infinities.\cite{Pausch91}

Finally, one has to compute the mass of the baryon.
This is somewhat delicate because of the vacuum subtraction.
The single particle energies of two systems must be subtracted:
One of the systems has only
continuum states, whereas the other system has one extra pair of discrete
states. The easiest way of doing this is to enclose the system
in a finite box, thereby discretizing all states, and to use the
well-known relation between phase shifts and
density of states.
In this way, one recovers
the result of Dashen {\em et al.}\s\cite{Dashen75b} for the baryon mass,
\begin{equation}
M_{B} = \frac{2Nym}{\pi} = nm \left( \frac{\sin \theta}{\theta} \right)  \ .
\end{equation}

Let us briefly compare this exact result
to the non-relativistic approach (in the
present model, the number of valence quarks
governs the degree of relativity).
In the non-relativistic Hartree approximation, we start from the
Schr\"{o}dinger equation for a particle of mass $m$, interacting
with the other particles through a $\delta$\,-function potential.
The Hartree equation then reduces to the non-linear Schr\"odinger
equation,\cite{Scott73}
\begin{equation}
\left\{ -\frac{1}{2m}\frac{\partial^{2}}{\partial x^{2}}
- ng^{2} |\phi_{0}|^{2}
\right\} \phi_{0}(x) = E_{0} \phi_{0}(x)  \ ,
\end{equation}
with the following bound state solution,
\begin{equation}
\phi_{0}(x) = \sqrt{\frac{\kappa}{2}} \frac{1}{\cosh \kappa x}  \ ,
\end{equation}
\begin{equation}
\kappa= \frac{1}{2} ng^{2}m  \ , \ \ \ \ E_{0}=-\frac{\kappa^{2}}{2m}  \ .
\end{equation}
The baryon mass turns out to have the value
\begin{equation}
M_{B}= nm\left(1-\frac{1}{6}\left( \frac{ng^{2}}{2} \right)^{2}\right)  \ .
\end{equation}
If we choose
\begin{equation}
g^{2} = \frac{\pi}{N} \  ,
\end{equation}
both $M_B$ and the scalar density in the full calculation and
in the non-relativistic
limit agree up to O($y^2$).
The reason why the GN model corresponds to a specific
choice of the coupling constant is that it involves only one
parameter, whereas the non-relativistic model has two independent parameters,
the fermion mass and the coupling constant.

\begin{figure}[t]
    \parbox[t]{3.6cm}{
      \epsfig{file=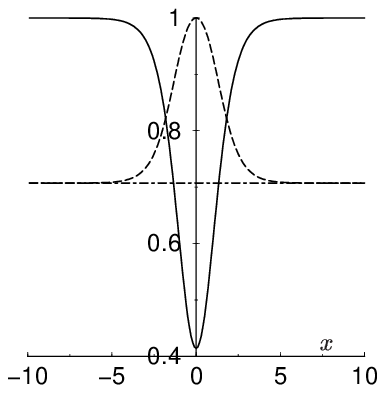,width=3.6cm}
      }\hfill
    \parbox[t]{3.6cm}{
      \epsfig{file=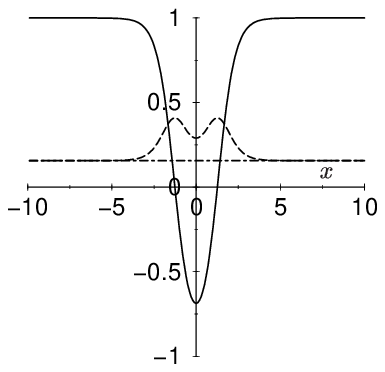,width=3.6cm}
      }\hfill
    \parbox[t]{3.6cm}{
      \epsfig{file=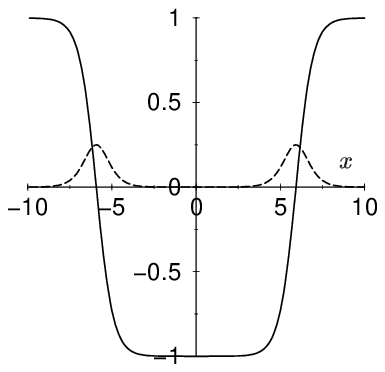,width=3.6cm}
      }\hfill
    \caption{Scalar potential $S$ (solid), valence fermion
      density $\rho_f^0$ (dashed)
and valence energy $E_0$ (dash-dotted lines) for GN model 
baryon and three different occupation fractions, $n/N=0.5$ (left), 0.9 (middle),
and 0.99999 (right). In units of $m$.}
\label{fig:3}
\end{figure}

Now turn to the opposite, extreme relativistic limit.
As $\theta$ approaches $\pi/2$,
the effects of the Dirac sea become
overwhelming. Eventually, the baryon goes
over into a well separated kink anti-kink
pair. Although the scalar density becomes flat,
the fermion density for the discrete state is
concentrated near the surface, since
\begin{equation}
\rho_f^0 = \psi_{0}^{\dagger}(x) \psi_{0}(x) = \frac{ym}{4}
\left( \frac{1}{\cosh^{2}\xi_{-}}
+ \frac{1}{\cosh^{2}\xi_{+}}\right)  \ .
\end{equation}
In the limit $y \to 1$, the two kinks are completely
decoupled, see Fig.~\ref{fig:3}.

If one looks at a single kink now, one finds
another type of baryon of the GN model, referred
to as Callan-Coleman-Gross-Zee kink in the literature.\cite{CCGZ}
It has the scalar potential
\begin{equation}
S(x) = m \tanh\left(xm\right)  \ .
\end{equation}
This solution has many unusual features:
$S(x)$ interpolates between the two different vacua related
by the discrete $\gamma^5$-trans\-for\-ma\-tion. Unlike the double kink baryon
which is only stabilized by the binding of fermions, the single kink
is topologically stable.
The fact that a kink type potential can bind particles is a well-known, purely
relativistic effect which has no analogue in  non-relativistic
quantum mechanics.
The discrete state corresponding to Eq.~(\ref{eq:psi0})
becomes
\begin{equation}
\psi_{0}(x) = \sqrt{\frac{m}{2}} \left(
\begin{array}{c}    \frac{1}{\cosh mx} \\ 0 \end{array}
\right)  \ ,
\end{equation}
with eigenvalue $E_{0} = 0$, while
the continuum wavefunctions replacing Eq.~(\ref{eq:psik}) are
\begin{equation}
\psi_{k}^{(\pm)}(x) = \frac{1}{\sqrt{2}E(k)} \left(
\begin{array}{c} \ri k-m \tanh mx \\ \pm E(k) \end{array}
\right) \re^{\ri kx} \ .
\end{equation}
The valence state has evidently vanishing scalar density so that
there is no feedback at all to the Hartree potential.
This is consistent with the fact that the mass
of the kink baryon is independent
of the number $n$ of valence fermions.
The self-consistency for the scalar
potential can again be verified using the vacuum Hartree condition
(\ref{eq:gap}).
As expected intuitively, the mass is $1/2$ of the double kink
mass in the limit
$y\to 1$
\begin{equation}
\left. M_{B}\right|_{\rm kink} = \frac{Nm}{\pi}  \ ,
\end{equation}
in agreement with Dashen {\em et al}.\s\cite{Dashen75b}

\subsubsection{Mesons}\label{subsubsec:mesons}

In this section, we shall rederive the known fermion-antifermion
bound state (meson) of
the GN model, using the relativistic RPA.
This meson has zero binding energy in the large $N$ limit, i.e., its mass
is just twice the physical fermion mass.\cite{Gross74}
We discuss only the discrete chiral symmetry case here, without bare mass; the
NJL type model with its massless
Goldstone mode will be treated in Sec.~\ref{subsec:chiral}.
We derive the RPA equations using an equations of
motion approach\s\cite{Salcedo91,Lenz91} and solve them
analytically for mesons with arbitrary momentum, thereby
demonstrating explicitly the covariance of the spectrum.\cite{Pausch91}

\begin{figure}[t]
  \begin{center}
\epsfig{file=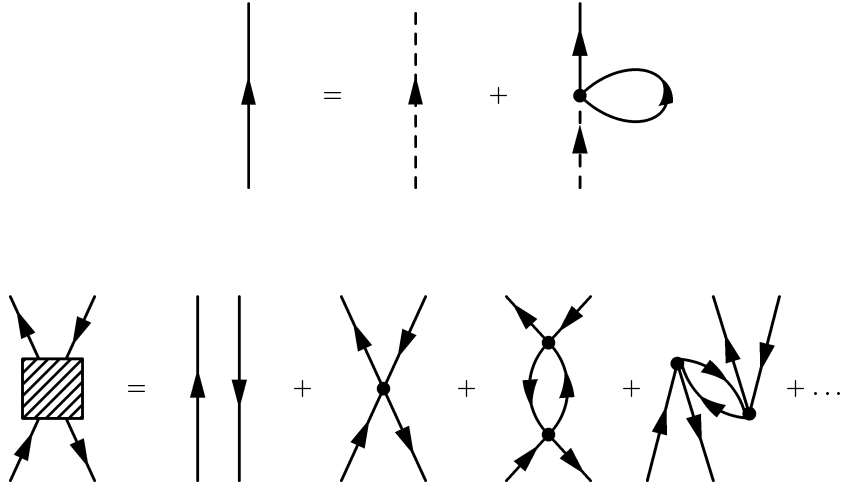}
    \caption{Diagrammatic illustration of HF (upper part) and RPA (lower part)
for the GN model. Dashed lines: bare (massless) quark propagator,
solid lines: dressed (massive) quark propagator.}
    \label{fig:4}
    \end{center}
\end{figure}

In RPA, the meson is described in terms of particle-hole excitations
on top of the HF vacuum. We therefore start from the equation
of motion of the following operator
bilinear in the fermion fields,
\begin{equation}
Q_{\alpha \beta}(x,y) = \frac{1}{N} \sum_{i} q_{i\beta}^{\dagger}(y)q_{i\alpha}(x)  \ .
\end{equation}
Using the equation of motion for $q$, Eq.~(\ref{eq:q_eom}), and neglecting
$1/N$ suppressed terms,
we find
\begin{eqnarray}
\ri\frac{\partial}{\partial t} Q(x,y) & = & -\ri \left\{ \frac{\partial}{\partial y}
Q(x,y) \gamma^{5} + \frac{\partial}{\partial x} \gamma^{5} Q(x,y) \right\}
\label{eq:Q_eom}\\
& &- Ng^{2} \left\{ \mbox{tr} \left(\gamma^{0} Q(x,x) \right) \gamma^{0}
Q(x,y)
- Q(x,y) \gamma^{0} \mbox{tr} \left( \gamma^{0} Q(y,y)\right)
\right\}\nonumber
\end{eqnarray}
Furthermore, we
expand the operator $Q(x,y)$ around its vacuum expectation value
\begin{equation}
Q(x,y) = \,\langle 0|Q(x,y)|0\rangle  + \,\frac{1}{\sqrt{N}} \widetilde{Q}(x,y) + ...
\label{eq:expandQ}
\end{equation}
where the $c$-number part corresponds to the density matrix of the vacuum,
\begin{equation}
\langle 0|Q(x,y)|0\rangle  \, = \rho(x-y)  \ .
\end{equation}
Inserting (\ref{eq:expandQ}) into (\ref{eq:Q_eom}) and equating terms
with the same
power in $1/\sqrt{N}$, we get to zeroth order the following alternative
formulation of the Hartree equation,
\begin{equation}
\ri\frac{\partial}{\partial t} \rho(x) =
-\ri \frac{\partial}{\partial x} [ \gamma^{5},\rho(x)]
- Ng^{2} \mbox{tr} \left( \gamma^{0} \rho(0)\right) [ \gamma^{0},
\rho(x)] = 0
\end{equation}
and to order $1/\sqrt{N}$ the linearized equation of motion for
the fluctuation $\widetilde{Q}$,
\begin{eqnarray}
\ri\frac{\partial}{\partial t} \widetilde{Q}(x,y)
&=& -\ri \left\{ \frac{\partial}{\partial y}
\widetilde{Q}(x,y) \gamma^{5} + \frac{\partial}{\partial x}
\gamma^{5} \widetilde{Q}(x,y) \right\}\label{eq:Qtilde_eom}\\
&&\hspace{-2cm}-Ng^{2} \left\{ \mbox{tr}
  \left(\gamma^{0} \rho(0) \right) \gamma^{0}
\widetilde{Q}(x,y) - \widetilde{Q}(x,y) \gamma^{0} \mbox{tr}
\left( \gamma^{0} \rho(0)\right) \right.
\nonumber \\
&&\hspace{-2cm}+\left.  \mbox{tr} \left(\gamma^{0}
    \widetilde{Q}(x,x) \right) \gamma^{0}
\rho(x-y) - \rho(x-y) \gamma^{0} \mbox{tr} \left( \gamma^{0}
  \widetilde{Q}(y,y)\right) \right\}  \ \nonumber.
\end{eqnarray}
Switching to momentum variables,
\begin{eqnarray}
\rho(k) & = & \int \rd x \, \re^{-\ri kx} \rho(x)
\nonumber \\
\widetilde{Q}(k',k) & = & \int \rd x
\int \rd y \, \re^{-\ri k'x+\ri ky} \widetilde{Q}(x,y)  \ ,
\end{eqnarray}
the Hartree equation simplifies to
\begin{equation}
    [h(k),\rho(k)] = [k\gamma^{5}+m\gamma^{0},\rho(k)]=0
\label{eq:HF_mom}
\end{equation}
where $h(k)$ denotes the single particle Dirac Hamiltonian
and where we have again introduced the physical fermion mass,
Eq.~(\ref{eq:ferm_mass}).
The Hartree solution of Sec.~\ref{subsubsec:vacuum} corresponds to
\begin{equation}
\rho(k) = v(k)v^{\dagger}(k) = \frac{1}{2} - \frac{m}{2E(k)} \gamma^{0}
-\frac{k}{2E(k)} \gamma^{5}
\end{equation}
and obviously satisfies Eq.~(\ref{eq:HF_mom}).
In order to convert Eq.~(\ref{eq:Qtilde_eom}) into the canonical form
of the RPA, we first have to project the operator $\widetilde{Q}$
onto self-consistent Hartree spinors. In the present case this simply
amounts
to a change of basis from massless to massive free spinors.
We define the ``particle-hole" and ``hole-particle" components
of the operator $\widetilde{Q}$ as
\begin{eqnarray}
\widetilde{Q}_{12}(k',k) & = & u^{\dagger}(k') \widetilde{Q}(k',k)
v(k)
\nonumber \\
\widetilde{Q}_{21}(k',k) & = & v^{\dagger}(k') \widetilde{Q}(k',k)
u(k) \label{hole-part}           \ .
\end{eqnarray}
Using the commutation relations, the
corresponding ``par\-ti\-cle-par\-ti\-cle" and ``hole-hole"
operators $\widetilde{Q}_{11}$ and $\widetilde{Q}_{22}$ can be
shown to be of higher order in $1/\sqrt{N}$ and can therefore
be neglected. The resulting system of coupled equations
is sandwiched between
the vacuum $|0\rangle $ and a one meson state $\langle n,P|$,
where $P$ denotes the total momentum and $n$ the type of meson.
We define the transition matrix-elements
\begin{eqnarray}
\langle n,P|\widetilde{Q}_{21}(k',k)|0\rangle  & = & 2\pi \delta(P-k+k')  X_{n}(P,k)  \ ,
\nonumber \\
\langle n,P|\widetilde{Q}_{12}(k',k)|0\rangle  & = & 2\pi \delta(P-k+k')  Y_{n}(P,k)  \label{transMel}\ .
\end{eqnarray}
Then, using
the Heisenberg equation of motion
in order to replace time derivatives by energies,
we obtain
\begin{eqnarray}
{\cal E}_{n}(P)  X_{n}(P,k) & = & E(k-P,k)  X_{n}(P,k)\\
& &\hspace*{-3.0cm}- Ng^{2} \overline{v}(k-P)u(k)
\int \frac{\rd k'}{2\pi} \left\{
\overline{v}(k')u(k'-P)Y_{n}(P,k')\right.\nonumber\\
&&\hspace{1.8cm}\left.+\overline{u}(k')v(k'-P)X_{n}(P,k')\right\}
\nonumber \\
{\cal E}_{n}(P)  Y_{n}(P,k) & = & -E(k-P,k)  Y_{n}(P,k)
 \\
& & \hspace*{-3.0cm} + Ng^{2} \overline{u}(k-P)v(k)
\int \frac{\rd k'}{2\pi} \left\{
\overline{v}(k')u(k'-P) Y_{n}(P,k')\right.\nonumber\\
& &\hspace{1.8cm}\left.+\overline{u}(k')v(k'-P) X_{n}(P,k')\right\}
\nonumber
\end{eqnarray}
with
\begin{equation}
E(k',k) = E(k') + E(k)   \ .
\end{equation}
These are the final RPA equations.
Taking into account the symmetry relation
\begin{equation}
\overline{u}(k)v(k') = - \overline{v}(k)u(k')
\end{equation}
which holds for the spinors (\ref{eq:spinors}), they can be cast into the
familiar matrix form of the RPA,
\begin{equation}
\left( \begin{array}{rr} A & B \\ -B & -A \end{array} \right)
\left( \begin{array}{c} X  \\ Y \end{array} \right)
= {\cal E}
\left( \begin{array}{c} X  \\ Y \end{array} \right)
\label{matrixRPA}\ .
\end{equation}
It is illustrated graphically in Fig.~\ref{fig:4}.
Characteristic of the RPA
(as opposed to the Tamm-Dancoff approximation (TDA)) are the
backward going bubbles (corresponding to $Y$-components) which
account for ground
state correlations.
Since the integral kernels
are separable, one can solve these equations algebraically.
After a few straightforward steps\s\cite{Pausch91} one
arrives at the eigenvalue equation
\begin{equation}
1= \frac{Ng^{2}}{2} \int \frac{\rd k}{2\pi} \left( \frac{1}{E(k-P)}
+ \frac{1}{E(k)} \right) \frac{4m^{2}+P^{2}-E^{2}(k-P,k)}{{\cal E}_{n}^{2}
(P)-E^{2}(k-P,k)}  \ .
\end{equation}
Clearly, the choice
\begin{equation}
{\cal E}_{0}^{2}(P) = 4m^{2} + P^{2}
\end{equation}
converts this into the vacuum Hartree equation. This shows that
there is a marginally bound meson with mass $2m$, and
that the spectrum is covariant.

It is instructive to compute also the $q\bar{q}$ scattering matrix in this
approach\s\cite{Thies93}. The scattering matrix can easily be found, since
we are dealing with a separable potential. With the Mandelstam variable
$s=2P{\cal E}$, find
\begin{equation}
-\frac{1}{\tau(s)} = - \frac{\sqrt{2}}{\pi m^2} \sqrt{1-\eta}
\left( \log \frac{1-\sqrt{1-\eta}}{1+\sqrt{1-\eta}}+{\rm i}\pi
\right)
\end{equation}
in the scattering regime
with $\eta=4m^2/s<1 $. ($\tau$ is normalized such that
$\tau=-Ng^2m^2/\sqrt{8}$ in tree approximation).
$\tau(s)$ has the expected pole at $s=4m^2$ corresponding to
the threshold
bound state. By going to large $s$, we can check asymptotic freedom in
a very physical manner,
\begin{equation}
\tau(s) \approx \frac{\pi m^2}{\sqrt{2}} \frac{1}{\log (m^2/s)} \ .
\end{equation}
The interaction shows the
logarithmic decrease in the UV characteristic for asymptotically
free theories.
A running coupling constant can be defined
by comparison with the tree approximation ($s=\mu^2$),
\begin{equation}
Ng_{\rm eff}^2(\mu) = \frac{2\pi}{\log \mu^2/m^2} \ .
\end{equation}
This illustrates once again the usefulness of the GN model as
toy model for QCD.

\subsection{'t~Hooft model}\label{subsec:thooft}

The 't~Hooft model\s\cite{tHooft74b} is defined as the large $N$ limit of
1+1 dimensional
SU($N$) gauge theory with quarks in the fundamental representation,
\begin{equation}
{\cal L} = \bar{q} \, {\rm i} D \!\!\!\!/ \, q - \frac{1}{2}{\rm tr}
F_{\mu \nu} F^{\mu \nu} \ , \qquad D \!\!\!\!/ =
\gamma^{\mu}(\partial_{\mu} + {\rm i}g A_{\mu})
\ .
\end{equation}
In the axial gauge, the gluons are gauged away, leaving
behind a theory of fermions interacting via a Coulomb
potential which is linearly rising in one space dimension.
't~Hooft originally solved this model for the meson spectrum, using
light-cone coordinates and the light-cone gauge.
This was the first practical demonstration of the power of the $1/N$ expansion
for gauge theories.\cite{tHooft74a}
Working diagrammatically, he identified planar diagrams without quark loops
as being of leading order in $1/N$ (nested rainbow graphs and ladders) and
summed them up by solving (light-cone) Schwinger-Dyson and
Bethe-Salpeter equations. He thus found an infinite
tower of mesons lying on a kind of Regge trajectory. Early applications
of this model dealt
with issues of confinement, asymptotic freedom, form factors,
scattering etc.\cite{Einhorn76,Callan76,Einhorn77,Ellis77,Weis78}
For some time, the precise character of the 't~Hooft limit was
somewhat disputed (strong vs. weak coupling, order of limits,
't~Hooft boundary condition, IR regularization of the quark
self-energy, Wu's alternative meson equation\s\cite{Wu77}).
Studies in ordinary coordinates followed where the vacuum and
baryons can be dealt with in a more direct
way;\s\cite{Bars78,Li86,Li87,Salcedo91,Lenz91} they
fully confirmed the original light-cone results for the meson spectrum.
During the last few years, various extensions of the 't~Hooft model
were found useful as laboratory for QCD related questions, such as
adjoint quarks\s\cite{Dalley93,Bhanot93}
(a toy model for gluons and strings), heavy quarks,
operator product expansion, duality etc.\cite{Bigi99,Grinstein99,Burkardt00}
In line with our main subject, we propose to go through the same
three topics as in Sec. \ref{subsec:GN-Models}
for the GN model --- the vacuum, baryons and mesons.
The light sector (mesons and baryons which become massless in
the chiral limit)
will be deferred to Sec. \ref{subsec:chiral}
where we will cover this aspect in a unified way for both models.

\subsubsection{Vacuum, confinement}\label{subsubsec:vacuumth}

\begin{figure}[t]
  \begin{center}
   \epsfig{file=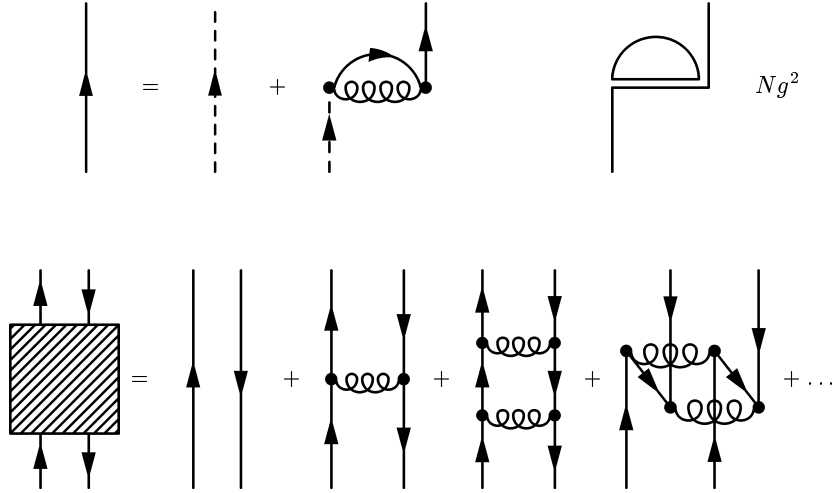,width=11cm}
    \caption{Same as Fig. \ref{fig:4}, but for 't Hooft model in axial gauge.
Curly lines: Static Coulomb
potential. The extra diagram in 't Hooft's double line representation 
(upper right) explains the dominance of the Fock term at large $N$.}
    \label{fig:5}
  \end{center}
\end{figure}

\begin{figure}[t]
  \begin{center}
    \epsfig{file=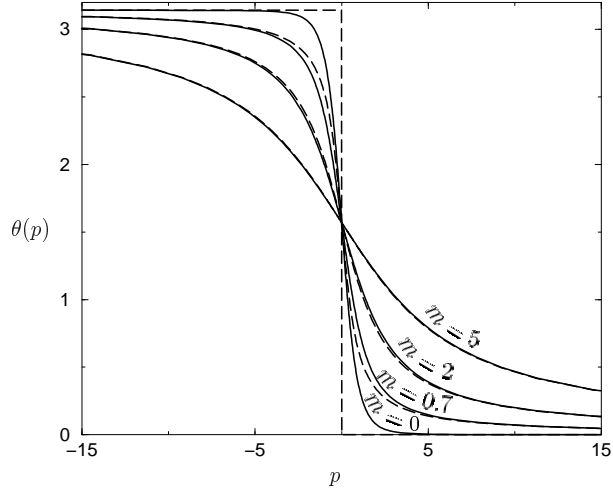,width=8cm}
    \caption{Bogoliubov angles for QCD$_2$ (solid curves) compared to free
theory (dashed curves) as a function of momentum, for several bare
quark masses. Units are such that $Ng^2=2\pi$.}
    \label{fig:6}
  \end{center}
\end{figure}

We start with a brief reminder of the vacuum, chiral symmetry breaking and the
role of confinement in the 't~Hooft model.\cite{Salcedo91,Lenz91}
Since this latter aspect is
intimately related to infrared singularities (absent in
the GN model), we work in a finite interval of length $L$, taking the
limit $L\to \infty$ at the end whenever possible.
Fermions are assumed to obey antiperiodic boundary conditions.
We introduce a bilinear operator which describes color
singlet dynamics to leading order in large $N$ (cf. GN model in
Sec. \ref{subsubsec:mesons}),
\begin{equation}
Q(p',p) = \frac{1}{N} \sum_i \left( \begin{array}{cc}
a_i^{\dagger}(p)a_i(p') & b_i^{\dagger}(p)a_i(p') \\
a_i^{\dagger}(p)b_i(p') & b_i^{\dagger}(p)b_i(p') \end{array} \right) \ .
\label{Q}
\end{equation}
The $a_i, b_i$
are annihilation operators for right- and left-han\-ded
quarks.
The Hamiltonian for the 't~Hooft model in the axial gauge
has the form
\begin{eqnarray}
H &=& \sum_{p,i} \frac{2\pi}{L}(p+1/2)\left( a_i^{\dagger}(p)a_i(p)
-b_i^{\dagger}(p)b_i(p) \right)\label{H_tHooft}\\
& & +\frac{g^2 L}{16 \pi^2}
\sum_{ij,n\neq 0} \frac{j_{ij}(n)j_{ji}(-n)}{n^2}
+ m \sum_{p,i} \left( a_i^{\dagger}(p)b_i(p)+b_i^{\dagger}(p)
a_i(p) \right) \nonumber \ .
\end{eqnarray}
The currents $j_{ij}(n)$ can be taken in the U($N$) form at large $N$,
\begin{equation}
j_{ij}(n) = \sum_p \left( a_j^{\dagger}(p)a_i(p+n)+b_j^{\dagger}(p)
b_i(p+n) \right) \ .
\end{equation}
$H$ is expressed in terms of $Q$ as
\begin{eqnarray}
H & = & \frac{2\pi}{L} N \sum_p \left(p+\frac{1}{2} \right)
{\rm tr} \gamma^5 Q(p,p) + m
N \sum_p {\rm tr} \gamma^0 Q(p,p)\\
& & + \frac{N^2 g^2L}{16\pi^2} \sum_{n \neq 0,rs}
\frac{1}{n^2} {\rm tr} \{ \delta_{rs}
Q(r,r) - Q(s,r)Q(r-n,s-n)\} \nonumber\ ,
\end{eqnarray}
whereas the equation of motion for $Q$ becomes
\begin{eqnarray}
{\rm i}\partial_t Q(p^\prime,p) & = & \frac{2\pi}{L} \left( (p'+1/2) \gamma^5
Q(p',p)-(p+1/2) Q(p',p)\gamma^5 \right)\label{Q_eom}\\
&&\hspace{-0.3cm}+ m [\gamma^0,Q(p',p)]
+ \frac{Ng^2L}{8\pi^2} \sum_{n \neq 0, p''} \frac{1}{n^2}
\{ Q(p',p'')Q(p''-n,p-n)\nonumber\\
&&\hspace{4.8cm}
-Q(p'-n,p''-n)Q(p'',p) \} \ .\nonumber
\end{eqnarray}
Expansion of $Q$ around its $c$-number part (vacuum expectation value)
\begin{eqnarray}
Q(p',p) &=& \langle 0 | Q(p',p) | 0 \rangle + \frac{1}{\sqrt{N}} \tilde{Q}
(p',p) + ... \nonumber \\
& = & \delta_{p'p} \rho(p) +  \frac{1}{\sqrt{N}} \tilde{Q}
(p',p) + ...
\label{expandQ}
\end{eqnarray}
yields, to leading order, the HF equation
\begin{equation}
[h_{HF},\rho(p)]=0
\end{equation}
with the single particle HF Hamiltonian
\begin{equation}
h_{HF}= \frac{2\pi}{L} (p+1/2)\gamma^5 + m \gamma^0 - \frac{Ng^2L}{8\pi^2}
\sum_{n \neq 0} \frac{1}{n^2} (\rho(p-n)-1/2) \ .
\label{h_HF}
\end{equation}
In contrast to the GN model where the Hartree term dominates,
here only the Fock term
survives at large $N$,
see Fig. \ref{fig:5}
(the Coulomb interaction acts like a color exchange force).
Solving the HF equation self-consistently is equivalent to
summing up nested rainbow graphs in a Schwinger-Dyson
equation.\cite{tHooft74b,Bars78}
On general grounds, the vacuum density matrix $\rho$ can be parametrized as
\begin{equation}
\rho(p) = \frac{1}{2} + \gamma^0 \rho_0(p)-{\rm i}\gamma^1 \rho_1(p)
+\gamma^5 \rho_5(p)
\end{equation}
where $\rho_1$ is only allowed if $m=0$.
Charge conjugation requires
$\rho_{0,1}$ to be even, $\rho_5$ to be odd functions of momenta.
The
Slater determinant condition $\rho^2(p)=\rho(p)$
characteristic for HF then translates into
\begin{equation}
\rho_0^2(p)+\rho_1^2(p)+\rho_5^2(p)=\frac{1}{4}  \ .
\end{equation}
Consider first the case of massive quarks.
For $m\neq 0$, $\rho_1$ vanishes and we can set
\begin{equation}
\rho_0= -\frac{1}{2}\sin \theta(p) \ , \qquad \rho_5=-\frac{1}{2}
\cos \theta(p) \ .
\end{equation}
Comparing
\begin{equation}
\rho(p) = v(p) v^{\dagger}(p)
\label{eq:densmat}
\end{equation}
with the HF spinors
of Sec. \ref{subsec:dirac} (Eq. \ref{eq:spinors}),
$\theta(p)$ is recognized as the Bogo\-liu\-bov angle. The
HF equation now becomes
\begin{equation}
\frac{2\pi}{L}(p+1/2)\sin \theta(p)-m \cos \theta(p) 
+ \frac{Ng^2 L}{16 \pi^2}
\sum_{n \neq 0} \frac{1}{n^2} \sin
\left( \theta(p)-\theta(p-n)\right) = 0 \ ,
\label{gap_tHooft}
\end{equation}
i.e., a nonlinear integral equation in the limit $L\to \infty$.
For $m=0$, $\rho_1$ can be non-zero.
As in the chiral GN model, there is now a continuum of vacua
parametrized by a chiral angle $\varphi$,
\begin{equation}
\left( \begin{array}{c}\rho_0(p) \\ \rho_1(p)  \\ \rho_5(p) \end{array}
\right) = -\frac{1}{2} \left( \begin{array}{c} \sin \theta(p) \cos \varphi
\\ \sin \theta(p) \sin  \varphi \\ \cos \theta(p) \end{array} \right)\ .
\end{equation}
The gap equation (\ref{gap_tHooft}) remains valid provided we set $m=0$.
The chiral condensates are
\begin{equation}
\langle \bar{q}q \rangle  =  - \frac{N}{L} \sum_p \sin \theta(p)
\cos \varphi \ , \quad
\langle \bar{q}\,{\rm i}\gamma^5 q \rangle  =
\frac{N}{L} \sum_p \sin \theta(p)
\sin \varphi \ .
\end{equation}
Although the gap equation (\ref{gap_tHooft}) for the 't~Hooft model
has to be
solved numerically,\cite{Li87,Salcedo91,Lenz91} the
value of the quark condensate
is known analytically, owing to an indirect determination via sum rules
and the 't~Hooft equation for mesons;\s\cite{Zhitnitsky85}
it is (for $\varphi=0$)
\begin{equation}
\langle \bar{q} q \rangle_v = - \frac{N}{\sqrt{12}}
\left( \frac{Ng^2}{2\pi} \right)^{1/2} \ .
\end{equation}
(This formula has been generalized to the case of finite bare quark
mass.\cite{Burkardt96}) In Fig.~\ref{fig:6},
we show a few typical results for the Bogoliubov angle for several
values of the quark mass and compare them to the free theory. For
larger quark masses, the interaction effects are very small
and can be treated perturbatively.

So far, there seems to be little difference between the GN and 't~Hooft
models --- superficially, the vacuum looks in both cases as if the
fermions had acquired a dynamical mass, although
momentum dependent in the 't~Hooft case.
How does confinement of quarks manifest itself? In QCD$_2$,
there are no massive quarks in the spectrum, and we have to understand
how their appearance can be avoided in an independent particle picture like the
HF approximation. As already suspected by 't~Hooft, the answer lies
in the IR divergence of the quark self-energy.

We first remark that the gap equation (\ref{gap_tHooft}) has no IR
problem in the limit $L\to \infty$: The Coulomb interaction term
has only a simple pole which can be regulated with a principal value
prescription (or symmetric cut-off). The IR divergence discussed in the
literature concerns the quark self-energies or, in our language,
the HF single particle energies which have to be interpreted
physically as
removal energies (Koopmans' theorem\s\cite{Koopmans33}).
We can deduce them by
varying the vacuum energy functional
\begin{eqnarray}
E_{HF}/N & = & -\sum_p \frac{2\pi}{L}(p+1/2)\cos \theta(p)
- m \sum_p \sin \theta(p) \nonumber \\
& & + \frac{Ng^2L}{32 \pi^2} \sum_{n \neq 0, p} \frac{1}{n^2}
\left(1-\cos (\theta(p)-\theta(p-n)) \right)
\label{E_HF}
\end{eqnarray}
with respect to the occupation number of level $p$ and
find
\begin{equation}
\omega(p) = \frac{2\pi}{L}(p+1/2) \cos \theta(p) + m \sin \theta(p)
+ \frac{Ng^2 L}{16\pi^2} \sum_{n \neq 0} \frac{1}{n^2}
\cos (\theta(p)-\theta(p-n))
\label{singlepom}
\end{equation}
or, in the thermodynamic limit $L\to \infty$ (setting $q=2\pi(p+1/2)/L$),
\begin{equation}
\omega(q) = q \cos \theta(q) + m \sin \theta(q)
+\frac{Ng^2}{4} \int \frac{{\rm d}q'}
{2\pi} \frac{\cos \left( \theta(q)-\theta(q') \right)}{(q-q')^2}\ .
\label{omegap}
\end{equation}
Due to the double pole, the integral is badly IR divergent.
To exhibit the divergence, let us isolate the divergent
part of the sum in Eq. (\ref{singlepom}) before taking the limit $L\to \infty$,
\begin{equation}
\omega(q) = q \cos \theta(q) + m \sin \theta(q)
+\frac{Ng^2}{4} \int \frac{{\rm d}q'}
{2\pi}\frac{\cos \left( \theta(q)-\theta(q') \right)-1}
{(q-q')^2}
+ \frac{Ng^2L}{48} \ .
\label{omega}
\end{equation}
The principal value prescription for the quadratic pole
advocated by Bars and Green\s\cite{Bars78} and Li {\em et al.}\s\cite{Li87}
amounts to throwing away the divergent constant $Ng^2L/48$. As a
consequence, one gets an awkward sign change (tachyonic behavior)
in the single particle energies at small $p$ and $m$.
The usual excuse is to say that these self-energies have no
physical meaning, being gauge dependent and non-covariant.
Indeed, this difficulty has no direct consequence for
the vacuum properties --- as pointed out above, the gap equation is free
of IR divergences. Moreover, the
calculation of color singlet mesons is also IR safe, as
pointed out by 't~Hooft (his
IR cutoff parameter $\lambda$ is related to our box size $L$ by
$\lambda=12/\pi L$)\s\cite{Lenz91}: The infinite self-energy term
is cancelled by an equally infinite piece in the Coulomb interaction.

Since the emergence of the constant $Ng^2 L/48$ in the single particle
energy (\ref{omega}) but not in the vacuum energy (\ref{E_HF}) is
a subtle but rather important point, let us briefly recall the
argument given in Ref. 45 (see also Ref. 74).
In many-body language, the
Coulomb interaction produces both one- and two-body terms in $H$.
If the
single particle energies are decomposed accordingly as
\begin{equation}
\omega(p) = \omega^{(1)}(p) + \omega^{(2)}(p) \ ,
\label{E_sp}
\end{equation}
then the HF ground state energy is
\begin{equation}
\langle 0 | H | 0 \rangle = -N \sum_p \left( \omega^{(1)}(p)
+ \frac{1}{2} \omega^{(2)}(p)  \right)\ .
\label{E_gs}
\end{equation}
As is well known, the factor 1/2 is necessary to avoid
double counting of the 2-body
interaction term.
In the 't~Hooft model, one finds
\begin{eqnarray}
\omega^{(1)}(p) & =& \frac{2\pi}{L}\left(p+\frac{1}{2}\right) \cos
\theta(p) +m\sin\theta(p)-\frac{N g^2 L}{48}  \ ,
\nonumber \\
\omega^{(2)}(p) & = &  \frac{N g^2 L}
{16 \pi^2} \sum_{n \neq 0} \frac{1}{n^2} \left( 1 + \cos (\theta(p)
-\theta(p+n)) \right) \ .
\end{eqnarray}
Remembering that $\sum_{n \neq 0} 1/n^2 = \pi^2/3$, we see  that
in the ground state energy (\ref{E_gs}) the IR tamed combination
$(\cos (\theta(p)-\theta(p+n))-1)/n^2$ appears whereas in the
quark self-energy (\ref{E_sp}), the 1-term in $\omega^{(2)}$ is cancelled
and an IR divergence (quadratic pole for $L\to \infty$) survives.

Why bother about this IR divergence at all? As one proceeds to other
applications of the
't~Hooft model than the meson spectrum,
it is no longer true that the quark removal energies
do not show up in any physical quantity. The prime example is finite temperature
field theory in the large $N$ expansion by means of the standard
generalization of HF to finite $T$.\cite{Fetter} Here, the thermal occupation
factors depend crucially on the single particle energies.\cite{Li86}
If these are made finite by some ad-hoc prescription, quarks will manifest
themselves through contributions to the pressure of order $N$, in
conflict with confinement. We shall return to this issue in more detail
in Sec. \ref{subsec:thooftcyl}.
Summarizing, we believe that the divergence of the
quark single particle energies simply reflects confinement. In
fact this may be the only way how confinement can be realized
in the independent
particle picture characteristic for the leading order in a $1/N$ expansion.

\subsubsection{Baryons}\label{subsubsec:baryonsth}

Since their structure is very different, we
distinguish between those baryons which become massless in the chiral limit
and those which stay massive. The first ones are closely related to
Goldstone bosons and will be deferred to Sec. \ref{subsec:chiral},
where they will be discussed for the chiral GN model as well.
Here, we briefly sketch what is known about the ``normal" type of baryons
in the 't~Hooft model.

Several papers deal with QCD$_2$ baryons for a small number of colors,
like $N=2$ (where they are degenerate with mesons\s\cite{Burkardt89})
or $N=3$.\cite{Durgut76} We concentrate on the large $N$
limit instead where baryons are described in relativistic HF approximation.
An analytic solution like in the GN model seems to be out of the
question. The work closest in spirit to our description of the GN model
baryons (cf. Sec. \ref{subsubsec:baryons})
is Ref. 44.
In this work, a numerical approach based on a lattice discretization
of space and the use of staggered fermions has
been employed. The HF equation is solved in
position space, both at strong ($Ng^2\gg m$) and weak ($Ng^2 \ll m$)
coupling. One valence level is taken into account in addition to the
Dirac sea. Unlike in the GN model where the filling fraction $n/N$
can be varied, here one has to fill the valence level completely with $N$ quarks
in order to form a color singlet baryon. For heavy quarks,
the HF equation reduces to the non-relativistic one with a linearly
rising potential and the sea becomes irrelevant; the HF equation
for the valence level then becomes
\begin{equation}
\left( -\frac{1}{2m}\frac{\partial^2}{\partial x^2} + \frac{Ng^2}{4} \int {\rm d}y
|x-y| |\varphi(y)|^2 \right) \varphi(x) =\epsilon \varphi(x)  \ .
\end{equation}
For moderately heavy quarks, significant relativistic effects
were found. The $B=2$ sector was also explored, with the result
that two $B=1$ baryons were observed interacting via an
exponentially decreasing repulsive potential.

\subsubsection{Mesons}\label{subsubsec:mesonsth}

Originally, 't~Hooft determined the meson spectrum by solving the
Bethe-Salpeter equation in light-cone gauge,
summing up all planar diagrams without quark loops. He found an
infinite tower of mesons with masses which behave asymptotically like
\begin{equation}
\mu^2(n) \to \pi^2 n + {\rm O}(\ln n) \ .
\end{equation}
This should be contrasted to the GN models with one or two mesons only;
the difference obviously reflects confinement (Regge behavior).
In the chiral limit, a massless meson appears (see Sec. \ref{subsec:chiral}).
These results have subsequently
been confirmed in normal coordinates, using the
axial gauge. The necessary equations were first derived by Bars and
Green\s\cite{Bars78}
and later solved numerically by several groups.\cite{Li87,Lenz91}
The equal-time approach turns out to be significantly more
involved than the light-cone approach. Nevertheless,
since we cannot use light-cone quantization for all the questions
which interest us and wish
to stay within one common framework, we have to outline this approach here.

As before, we consider small oscillations around the HF solution in
the space of Slater determinants. The quantized modes of these oscillations
are just the RPA modes. The RPA equation is identical to the
Bethe-Salpeter equation in a diagrammatic large $N$ approximation.\cite{Bars78}
To derive it, we
return to the equation of motion for the bilinear quark operator
$Q(p,p')$, Eq. (\ref{Q_eom}), insert the expansion (\ref{expandQ})
and focus onto the
next-to-leading order in $1/\sqrt{N}$,
\begin{eqnarray}
{\rm i} \partial_t \tilde{Q}(p',p)&=&h_{HF}(p')\tilde{Q}(p',p)
-\tilde{Q}(p',p) h_{HF}(p)
\label{Qtilde}  \\
& & \hspace{-0.3cm}+ \frac{Ng^2L}{8\pi^2} \sum_{n\neq 0} \frac{1}{n^2}
\{ \rho(p') \tilde{Q}(p'-n,p-n)
-\tilde{Q}(p'-n,p-n)\rho(p)\} \ .
\nonumber
\end{eqnarray}
Expanding $\tilde{Q}$ in the HF spinors, as in Sec. \ref{subsubsec:mesons}
Eq. (\ref{hole-part}) and (\ref{transMel}), and setting
\begin{eqnarray}
\langle n,K| \tilde{Q}_{21}(p',p)|0\rangle &=& \delta_{K,p-p'}\Phi_+^n(K,p)
\ , \nonumber\\
\langle n,K| \tilde{Q}_{12}(p',p)|0\rangle &=& \delta_{K,p-p'}\Phi_-^n(K,p)
\end{eqnarray}
then yields the two coupled equations
\begin{eqnarray}
\pm E(K) \Phi_{\pm}(K,p) & = & (\omega_0(p-K)+\omega_0(p))\Phi_{\pm}(K,p)
\nonumber
\\
& &- \frac{Ng^2L}{8\pi^2} \sum_{p'(\neq p)}\frac{1}{(p-p')^2}
\{ s(p,p',K) \Phi_{\mp}(K,p')\nonumber\\
& &\hspace{2cm}
+c(p,p',K)\Phi_{\pm}(K,p')\}\label{RPA_tHooft}
\end{eqnarray}
with the definitions
\begin{eqnarray}
s(p,p',K)&=&\sin \left(\frac{\theta(p')-\theta(p)}{2}\right)
\sin \left( \frac{\theta(p-K)-\theta(p'-K)}{2}\right) \ ,
\nonumber \\
c(p,p',K)&=&\cos \left(\frac{\theta(p')-\theta(p)}{2}\right)
\cos \left( \frac{\theta(p-K)-\theta(p'-K)}{2}\right) \nonumber\ .\\
&&
\end{eqnarray}
Eqs. (\ref{RPA_tHooft}) have the standard RPA form of Eq.
(\ref{matrixRPA}), with
\begin{eqnarray}
A_{p p'}(K) &=& \delta_{pp'}(\omega_0(p-K)+\omega_0(p))
-\frac{Ng^2L}{8\pi^2} \frac{(1-\delta_{pp'})}{(p-p')^2}
c(p,p',K)  \ ,
\nonumber \\
B_{pp'}(K) & = & - \frac{Ng^2L}{8\pi^2} \frac{(1-\delta_{pp'})}
{(p-p')^2} s(p,p',K)  \ .
\end{eqnarray}
The diagrammatic content of these equations is sketched in Fig. 5
(see Sec. \ref{subsubsec:vacuumth}).
Using a slightly different language,
Bars and Green\s\cite{Bars78} interpret
them in terms of forward and backward going strings. The covariance of
the spectrum is a non-trivial issue if one works canonically in
normal coordinates, but can be checked either numerically or
via the $1/N$ expansion of the Poincar\'{e} algebra. Meson
wavefunctions (or rather RPA amplitudes) can be found in the
literature for a few low-lying states.\cite{Li87} In the limit of heavy
quarks, the RPA reduces to the non-relativistic Schr\"odinger
equation with a linear potential. The light-cone TDA
('t~Hooft equation) on the other hand
can be recovered by going to the infinite momentum frame\s\cite{Bars78}
(backward going bubbles are suppressed, see also Sec. \ref{subsec:chiral}).

\subsection{Chiral limit and Goldstone modes}\label{subsec:chiral}

Since QCD$_2$ or the chiral GN model in the massless limit break chiral
symmetry spontaneously (at least at infinite $N$), the appearance of a
boson whose mass behaves like the square root of the bare quark mass
has to be expected. This has a lot in common with four
dimensional theories and makes 
these models good toy models for QCD.
However, the
chiral limit is somewhat obscured by a certain pathology of two dimensional
theories
which gives rise to a much greater variety of (decoupled) massless
states. In particular, non-interacting
massless mesons and baryons already appear at finite $N$ where the Coleman
theorem certainly forbids talking about Goldstone bosons. Our
main interest here is in the large $N$ limit where we will derive
both massless mesons and baryons. However, before doing so, we also
briefly explain the
more pathological aspects of two dimensional
theories leading to massless states at
finite $N$ as well.

The identification of massless excitations in two dimensional field theories
is one of the tasks where light-cone
quantization is particularly elegant.
Due to the light-cone dispersion relation $p_+=\frac{m^2}{2p_-}$,
a massless state in two dimensions has zero
energy at all momenta. Therefore its presence can be associated with
a {\em local} symmetry of the Hamiltonian. 
In the chiral limit the right-handed fermions drop out of the light-cone
Hamiltonian $P^+=H-P$.\cite{Amati81,Amati82,Lenz91}
Since the Coulomb interaction
admits local chiral transformations, $P^+$ is
invariant under local phase transformations of right-handed quarks.
Free, massless bosons then must appear
and it is sufficient to
identify the winding number of their field with baryon number to infer
the existence of free,
massless baryons as well. The IR trouble of Coleman's theorem is avoided by
the fact that the massless modes decouple.

The essence of the argument can of course also be phrased without recourse
to light-cone quantization.\cite{Witten78} Let us return to
the fermion currents discussed in Sec. \ref{subsec:dirac}.
Vector current conservation
allows us to set
\begin{equation}
j_V^{\mu} = \epsilon^{\mu \nu} \partial_{\nu} \phi
\end{equation}
with a scalar field $\phi$.
If the axial current is non-anomalous, we get in the limit $m \to 0$
(see $j_A^{\mu}$ above, Eq. (\ref{partialcons}))
\begin{equation}
\partial_{\mu}j_A^{\mu} = \Box \phi = 0    \ ,
\end{equation}
i.e., the announced massless boson.
This purely classical argument can easily be promoted  to the
quantum level within the canonical framework.\cite{Oderkerk92}
Define the right-handed current as
\begin{equation}
j_R(q,t)=\frac{1}{2}(j_V^0(q,t)+j_V^1(q,t))
\end{equation}
where Heisenberg operators, Fourier transformed with respect to the space
coordinate $x$ are used. In the chiral
limit, vector and axial vector current conservation imply
\begin{equation}
[j_R(q,t),H]  = q j_R(q,t) \ .
\end{equation}
Besides, if $P$ denotes the momentum operator, we have trivially
\begin{equation}
[j_R(q,t),P]  =  q j_R(q,t) \ .
\end{equation}
For any energy-momentum eigenstate $|i\rangle$ with eigenvalues ($E_i,P_i$),
we thus get
\begin{eqnarray}
H j_R(q,t)|i\rangle & = & (E_i -q) j_R(q,t)|i\rangle  \ , \nonumber \\
P j_R(q,t)|i\rangle & = & (P_i -q) j_R(q,t)|i\rangle \ .
\end{eqnarray}
Thereby decoupled massless bosons can be added to or removed from any stationary state.
Similarly, left moving massless bosons can be related
to the left-handed current.

Let us now turn to the large $N$ limit and discuss the
emergence of massless particles there.
Again we benefit from the experience in many-body theory.
The Goldstone
theorem has a close correspondence which we may invoke here (remember
that it also holds in non-relativistic theories):
If the HF ground
state breaks a continuous symmetry spontaneously, the RPA develops
a gapless mode. The basic argument can be given in two lines,
using the language of linear response theory
(without external force).\cite{Ring}
Take the equation of motion for the one-body density matrix,
\begin{equation}
{\rm i} \dot{\rho} = \left[h(\rho),\rho \right]\ .
\end{equation}
Expand $\rho$ around its ground state expectation value,
$\rho=\rho^{(0)}+\delta \rho$, and linearize
\begin{equation}
{\rm i} \delta \dot{\rho} = [h(\rho^{(0)}),\delta \rho]
+ \left[ \frac{\delta h}{\delta \rho} \delta \rho , \rho^{(0)}
\right]  \ .
\end{equation}
This equation has the form of the RPA, cf. the above derivation
in the equations of motion approach in Sec. \ref{subsubsec:mesons}.
If the ground state density matrix $\rho$
breaks a continuous symmetry, we find that the deformation
$\delta \rho$
corresponding to the symmetry transformation (e.g. a chiral rotation)
solves the RPA equation with zero energy.
The backward going ph-bubbles are crucial for getting the
zero modes; as is well known, the TDA does
not have this property.

These insights enable us to construct the Goldstone boson type
solution of the RPA equation explicitly. In the case of QCD$_2$,
the RPA amplitudes of the ``pion"
can easily be obtained by projecting the right-handed current
onto HF spinors. Since these depend only on the Bogoliubov angle, the
pion amplitudes are intimately related to the broken symmetry vacuum.
One finds\s\cite{Oderkerk92,Satzinger91}
\begin{eqnarray}
\Phi_{+}^\pi(K,p) & = & \sqrt{\frac{2\pi}{K}} \cos \left( \theta(p)/2\right)
\sin \left( \theta(p-K)/2 \right) \ , \nonumber \\
\Phi_{-}^\pi(K,p) & = & \sqrt{\frac{2\pi}{K}} \sin \left( \theta(p)/2\right)
\cos \left( \theta(p-K)/2 \right) \ .
\label{pion}
\end{eqnarray}
The chiral GN model can be analyzed in a similar manner.
Generalizing the derivation of the RPA equation, Sec. \ref{subsubsec:mesons},
to the chirally invariant model, one finds an integral equation with a
two-term separable kernel (instead of one-term in the model with
discrete chiral symmetry) and correspondingly two
bound states, the threshold bound state at 2$m$ (the ``$\sigma$")
and a massless one (the ``$\pi$").
The pion has the same
structure as in Eq. (\ref{pion}), except that the Bogoliubov angles now
refer to the GN  vacuum. They are just the Bogoliubov angles of the
free, massive fermions given in Eq. (\ref{freebog}).
In this particular case, we are in the rare position
of having complete analytical control over a meson (the $\pi$) in an
arbitrary Lorentz frame.
\begin{figure}[t]
    \null\begin{center}
    \parbox[t][4cm][t]{3.6cm}{
      \epsfig{file=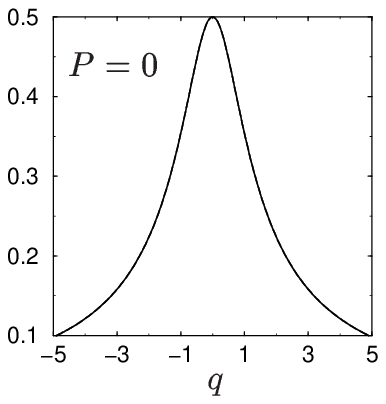,width=3.6cm}
      }\hfill
    \parbox[t][4cm][t]{3.6cm}{
      \epsfig{file=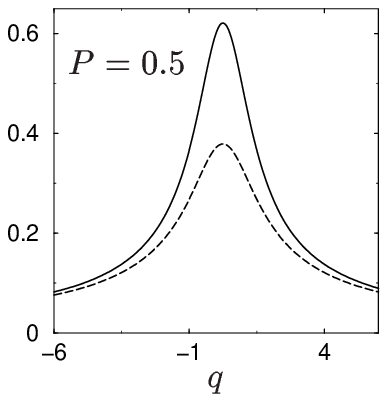,width=3.6cm}
      }\hfill
    \parbox[t][4cm][t]{3.6cm}{
      \epsfig{file=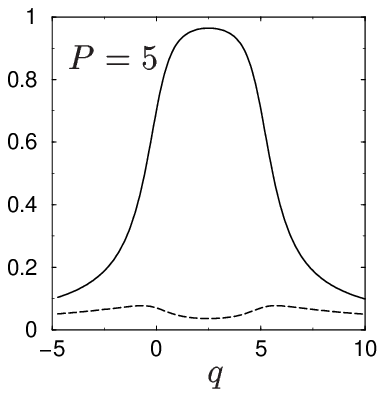,width=3.6cm}
      }\hfill
       \parbox[t]{7.8cm}{
    \parbox[t][4cm][t]{3.6cm}{
      \epsfig{file=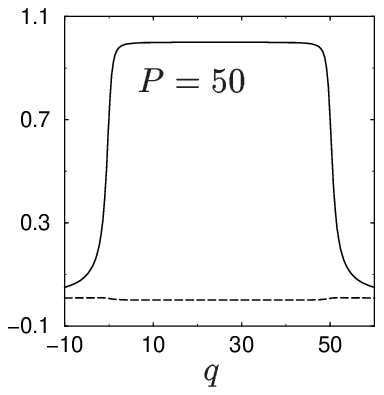,width=3.6cm}
      }\hfill
    \parbox[t][4cm][t]{3.6cm}{
      \epsfig{file=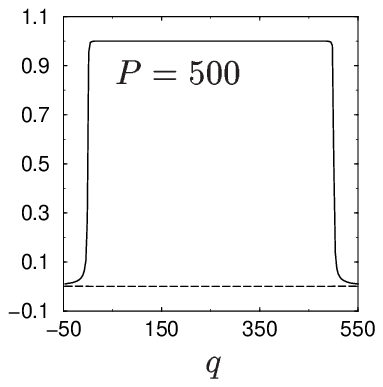,width=3.6cm}
      }\hfill}
      \caption{Forward (solid lines) and backward (dashed lines) RPA amplitudes
$\Phi_{\pm}^{\pi}(P,q)$ of the pion in the chiral GN model, for pion
momenta $P$ from 0 to 500 (in units of $m$). This illustrates the
evolution from the rest frame to the ``infinite momentum frame".\label{fig:7}}
  \end{center}
 \end{figure}
In Fig.~\ref{fig:7}, we
illustrate the disappearance of the backward components as one boosts
the pion to higher momenta (in the rest frame, both are exactly equal).
At very high momenta,
the constant wave function familiar from the light-cone meson equation
emerges.\cite{tHooft74b}
In the infinite momentum limit, the 't~Hooft and GN model
pions become identical, although their ``wave functions''
(\ref{pion}) differ in any other frame due to different Bogoliubov angles.
This illustrates nicely how the vacuum decouples in the infinite
momentum frame, one of the celebrated advantages of the light-cone
approach (triviality of the vacuum). Connected to this is the
disappearance of the backward amplitudes ---  the RPA reduces to the TDA in
the infinite momentum frame, the other characteristic benefit
on the light-cone.

We now turn to the issue of baryons which become massless in the
chiral limit, a phenomenon which does not happen in 3+1 dimensions. The best
way of thinking about them is in terms of coherent,
topological excitations of the pion field. Like the massive baryons
discussed in Secs. \ref{subsubsec:baryons}
and \ref{subsubsec:baryonsth}, they are accessible through the relativistic HF
approximation. In order to find these solutions in the first place, another
point of view has turned out to be very useful. It is
closely related to bosonization,\cite{Damgaard92}
but since only the chiral field
is bosonized, it is conceptually and technically much simpler.
It is also similar in spirit to the Skyrme picture of baryons in
3+1 dimensions.\cite{Skyrme62}

Following Salcedo {\em et al.},\cite{Salcedo91} we start from a
variational ansatz for the baryon near the chiral limit. The
interaction part of the ground state energy functional
\begin{eqnarray}
E[\rho]
& = &  N \int \rd x {\rm tr} \left[ \left( -{\rm i} \gamma^5
\vec{\partial}_x + m \gamma^0 \right) \rho(x,x) \right]
\nonumber \\
& & + \frac{N^2 g^2}{8} \int \rd x \rd y |x-y| {\rm tr}
\left( \rho(x,y) \rho(y,x) \right)
\label{Efunctional}
\end{eqnarray}
(where $\vec{\partial}_x$ acts only on the first variable of $\rho(x,x)$)
is invariant under any {\em local} chiral transformation. Let us perform a
local chiral rotation of the vacuum density matrix,
\begin{equation}
\rho'(x,y) = {\rm e}^{{\rm i}\chi(x)\gamma^5} \rho^v(x-y)
{\rm e}^{-{\rm i} \chi(y) \gamma^5} \ .
\label{ansatz}
\end{equation}
We can use the
following general decomposition
of the baryon density matrices $\rho^v$ and $\rho'$
\begin{equation}
\rho = \frac{1}{2}\rho_e + \gamma^0 \rho_0 -{\rm i} \gamma^1 \rho_1
+ \gamma^5 \rho_5 \ 
\end{equation}
with corresponding superscripts.
In the vacuum, $\rho_e= \delta(x-y)$, due to translational invariance
all components depend only on
$x-y$ and $\rho_1$ vanishes unless $m=0$. If the
vacuum breaks chiral symmetry, the ansatz (\ref{ansatz}) yields a new
density matrix which can be inserted into Eq. (\ref{Efunctional});
$\chi(x)$ is then determined by minimizing $E$. In order to understand
the emergence of baryon number out of winding number,
we need the short distance singularities of $\rho^v$ which are the same
as in the free theory,
\begin{eqnarray}
\rho_0(x) &=& - \frac{m}{2\pi}K_0(m|x|) \sim \frac{m}{2\pi}\ln(m|x|) \ ,
\nonumber \\
\rho_5(x)&=&\frac{m}{2\pi{\rm i}}{\rm sgn}(x) K_1(m|x|) \sim \frac{1}
{2\pi {\rm i}x} \ .
\end{eqnarray}
Using
\begin{eqnarray}
\rho_e'(x,y)\!&=&\!\cos (\chi(x)-\chi(y))\rho_e^v(x-y)
+ 2{\rm i}\sin (\chi(x)-\chi(y))\rho_5^v(x-y)\nonumber \ ,\\
\rho_5'(x,y)\!&=&\!\cos (\chi(x)-\chi(y))\rho_5^v(x-y)
+ \frac{\rm i}{2}\sin (\chi(x)-\chi(y))\rho_e^v(x-y)\ ,\nonumber\\
\end{eqnarray}
one then finds
\begin{eqnarray}
\lim_{y \to x} [ \rho_e'(x,y)-\rho_e^v(x-y)] &=& \frac{1}{\pi}\partial_x
\chi(x)\ , \nonumber \\
\lim_{y\to x} \partial_x [\rho_5'(y,x)-\rho_5^v(x-y)] &=&
\frac{\rm i}{4\pi}
(\partial_x \chi(x))^2 \ .
\end{eqnarray}
Similarly,
\begin{equation}
\rho_0'(x,x) \mp {\rm i} \rho_1'(x,x) = \frac{\langle \bar{q}q \rangle_v}
{2N} {\rm e}^{\mp {\rm i} 2 \chi(x)}
\end{equation}
where $\langle \bar{q}q \rangle_v$ refers to the vacuum.
If we invoke a finite spatial box of length $L$ for these topological considerations,
the baryon number gets identified with the winding number of the chiral field,
\begin{equation}
B= \int_0^L \rd x [ \rho_e'(x,x) -\rho_e^v(0)]
= \frac{1}{\pi} (\chi(L)-\chi(0)) \ .
\end{equation}
This is an integer if the baryon goes over into the vacuum at $0$ and $L$.
For small quark mass, find
\begin{equation}
E[\rho'] =E[\rho^v] + N \int \rd x \left\{ \frac{1}{2} (\partial_x \phi)^2
+ m \frac{\langle \bar{q}q \rangle_v}{N} \left[ \cos (\sqrt{4\pi}
\phi(x) ) -1 \right] \right\}
\label{Erhop}
\end{equation}
with the rescaled field
\begin{equation}
\phi(x)=-\frac{1}{\sqrt{\pi}}\chi(x)
\end{equation}
and
$\langle \bar{q} q\rangle_v$ referring to the vacuum in the chiral limit.
$E[\rho']$ is minimized if $\phi$ satisfies the sine-Gordon
equation\s\cite{Scott73}
\begin{equation}
\partial_x^2 \phi + \sqrt{4\pi} m \frac{\langle \bar{q}q \rangle_v}
{N} \sin (\sqrt{4\pi} \phi(x)) = 0       \ .\label{eq:SG}
\end{equation}
The solution with baryon number one is just the famous sine-Gordon kink,
\begin{equation}
\phi(x)=-\frac{2}{\sqrt{\pi}} \arctan \exp \left(  \frac{x}{x_B}\right) \ .
\end{equation}
It is localized with spatial extent $x_B$,
\begin{equation}
x_B = \left( - 4\pi m \frac{\langle \bar{q}q \rangle_v}{N}
\right)^{-1/2}
\end{equation}
and has mass
\begin{equation}
M_B=8N \left(-\frac{m}{4\pi} \frac{\langle \bar{q}q \rangle_v}{N}
\right)^{1/2} \ .
\end{equation}
If one linearizes the sine-Gordon equation (in the limit $m
\to 0$), one can immediately read off the Gell-Mann, Oakes, Renner
relation (GOR)\s\cite{Gell-Mann68}
\begin{equation}
m_{\pi}^2 = - 4 \pi m \frac{\langle \bar{q}q \rangle_v}{N}
\label{eq:GOR}
\end{equation}
for the Goldstone boson. We have thus derived an effective low
energy theory near the chiral limit. Since the potential energy does
not enter Eq. (\ref{Erhop}) but only the kinetic energy and mass
term, the resulting low energy effective theory
is universal for a whole class of such chiral models.

In the limit $m\to 0$, both the pion and the baryon become massless;
the baryon gets completely
delocalized. This could actually have been anticipated from current conservation:
$\partial_1 j_A^1=0$ holds for any stationary state and in two dimensions,
$j_A^1$ is just
the baryon density $j_V^0$ (cf. Sec. \ref{subsec:dirac}).

\begin{figure}[t]
  \begin{center}
    \epsfig{file=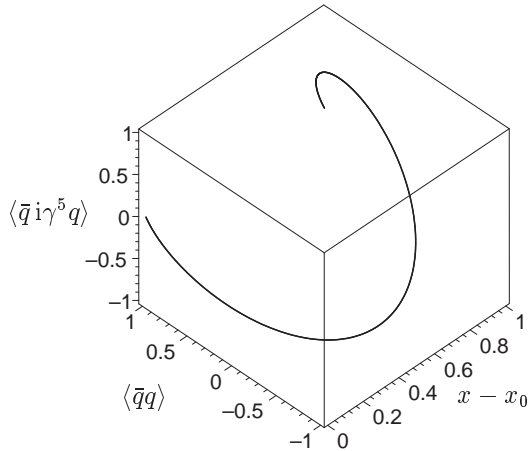,width=7cm}
    \caption{Attempt to visualize the topological structure of the
      massless baryon 
in terms of a ``chiral spiral" winding around once. Condensates 
in units of $\langle \bar{q}q\rangle_v$, $x$ in units of $L$.}
    \label{fig:8}
  \end{center}
\end{figure}

An attempt to illustrate the
baryon structure in the chiral limit has been made in Fig.~\ref{fig:8}.
For $m=0$, $\chi(x)$ becomes simply a linear function and the scalar and
pseudoscalar condensates
\begin{equation}
\langle \bar{q}q \rangle  =
\langle \bar{q}q \rangle_ v \cos ( 2\pi (x-x_0)/L)
\ ,\quad
\langle \bar{q}\,{\rm i} \gamma^5 q \rangle  = -\langle \bar{q}q \rangle_ v
\sin (2 \pi (x-x_0)/L)
\ ,
\end{equation}
can be regarded as projections of a spiral of radius $\langle \bar{q}q\rangle_v$
winding around once. This picture will be generalized later on
and turn out to be useful for characterizing
baryonic matter in two dimensional chiral models.

In Ref. 44
the Skyrme type picture of the baryon was compared
to the exact numerical solution for light quarks and good agreement was
found also for small, finite quark masses. It was pointed out that
these results support the Skyrme picture of the baryon more than the bag picture in two dimensions.

Finally, let us come back to the question of validity of the variational
calculation based on the ansatz (\ref{ansatz}).
In the massless limit,
it is easy to convince oneself that Eq. (\ref{ansatz})
with a linear function $\chi(x)$
solves the HF equation exactly. Thus for instance in
the 't~Hooft model,
the massless HF equation reads\s\cite{Salcedo91}
\begin{equation}
-{\rm i}(\gamma_5)_{\alpha \beta} \frac{\partial}{\partial x}
\varphi_{\beta}^{(n)}(x) + \frac{Ng^2}{4} \int {\rm d}y |x-y|
\rho'_{\alpha \beta}(x,y) \varphi_{\beta}^{(n)}(y) =
\omega_n \varphi_{\alpha}^{(n)}(x) \ .
\end{equation}
Upon substituting
\begin{equation}
\varphi_{\alpha}^{(n)}(x) = \left({\rm e}^{-{\rm i} \pi x \gamma_5/L}
\right)_{\alpha \beta}
\tilde{\varphi}^{(n)}_{\beta}(x)
\end{equation}
as we are instructed to do by Eq. (\ref{ansatz}), we discover
that $\tilde{\varphi}^{(n)}$ does indeed solve the HF equation,
the only change being that the single particle energy $\omega_n$
gets replaced by $\omega_n + \pi/L$. The same argument goes through
in the chiral GN model, or in any field theory where the
interaction term has a local chiral invariance for that matter.
This shows that the result becomes exact in
the chiral limit (to leading order in the $1/N$ expansion, of course).


\section{Finite temperature}\label{sec:finT}

\subsection{Finite temperature versus finite extension}\label{subsec:finTfinL}

\begin{figure}[t]
   \begin{center}
     \epsfig{file=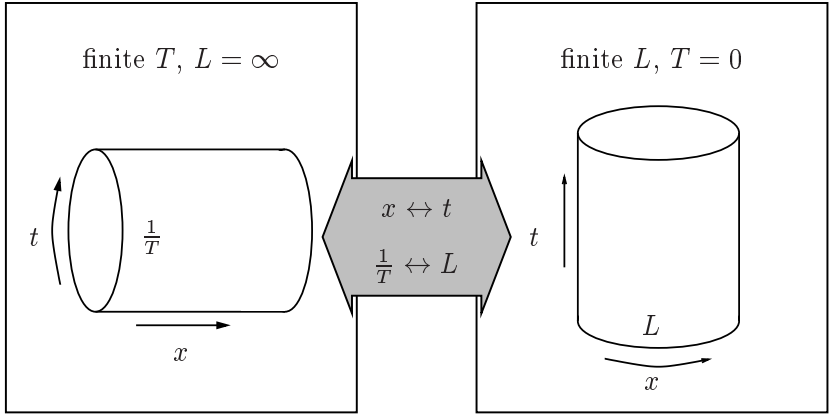,width=11cm}
     \caption{Correspondence between relativistic field
       theories at finite temperature
and finite extension.}
     \label{fig:9}
   \end{center}
\end{figure}

We briefly discuss the relation between field theories at finite extension
and finite temperature.\cite{Lenz98} Equivalence between relativistic
theories at finite
extension and finite temperature is formally almost trivial but gives
rise to some of
the most intriguing
consequences of covariance. By rotational invariance in the
Euclidean, the value of the partition function of a system with finite
extension $L$ in one space direction and $\beta$ in the time direction
is invariant under the exchange of these two parameters,
\begin{equation}
Z(\beta,L)=Z(L,\beta) \ ,
\label{interchange}
\end{equation}
provided bosonic (fermionic) fields satisfy periodic (antiperiodic)
boundary conditions in both time and compact space coordinate.
Sending one of the parameters ($\beta,L$) to infinity,
relativistic covariance connects the thermodynamic properties of a canonical
ensemble with the properties of the pure state of the vacuum corresponding to
the same physical system but at finite extension. In particular, as a
consequence of (\ref{interchange}), energy density and pressure are related by
\begin{equation}
\epsilon(\beta,L)=-p(L,\beta) \ .
\label{epsilon_p}
\end{equation}
For a system of non-interacting particles this equation can be used to relate
quantitatively  the
Stefan-Boltzmann law with the Casimir effect --- phenomena which at first
glance would not seem to be related at all.\cite{Toms80}
Another application of this discrete relativistic symmetry
worth mentioning are quark propagators and the interpretation
of lattice data in ``funny space".\cite{Koch92}
In four dimensional QCD, the
confinement-deconfinement phase transition and the chiral phase transition,
when quarks are present, appear as ``quantum phase transitions" if one swaps
space and Euclidean time. They are
driven by changes in quantum rather than thermal fluctuations which in turn are
induced by changes of a parameter of the system ($L$).\cite{Young95,Sondhi97}
We have found it useful to apply this technique to the two dimensional
models at hand
to see how it works in detail in the presence of interaction effects.
Consistency between finite extension and
thermal field theory calculations provides a non-trivial test of the approximations
used and will help us to clarify the role of confinement
in the 't Hooft model.
For the case of two dimensions
with one extension taken to be infinite, the symmetry
is illustrated in Fig. \ref{fig:9}.

To leading order in the large $N$ limit, it is easy to include finite temperature
and chemical potential since the required generalization of our main tool,
the HF approximation, is well known.\cite{Fetter} Let us focus on
finite temperature here since finite density and chemical
potential will be the subject of  Secs. \ref{sec:findens} and 
\ref{sec:phasediag}.
The main difference to the temperature
zero case is that for sytems in thermal equilibrium, one
is no longer dealing with a pure state but with a mixed state. The concepts of
independent particles and single
particle states remain valid. Whereas at $T=0$ each orbit is either empty or
occupied (sharp Fermi surface), at finite $T$ the occupation number is
given by the thermal occupation probability (Dirac-Fermi
distribution\s\cite{Kapusta}).
The one-body density matrix no longer satisfies the projector property $\rho^2=\rho$
characteristic for a single Slater determinant but assumes the form
\begin{equation}
\rho = N_+ u u^{\dagger} + N_- v v^{\dagger} \ ,
\end{equation}
with occupation numbers
\begin{equation}
N_{\pm} = \frac{1}{{\rm e}^{\pm \beta \epsilon}+1} \ .
\end{equation}
Using
\begin{equation}
N_+ + N_- = 1 \ , \qquad  N_+-N_-=-\tanh \frac{\beta \epsilon}{2} \ ,
\end{equation}
one can write
\begin{equation}
\rho = \frac{1}{2} + \left( \gamma^0 \rho_0 + \gamma^5 \rho_5
\right) \tanh \frac{\beta \epsilon}{2} \ .
\label{rho_T}
\end{equation}
where $\rho_0, \rho_5$ are given in terms of Bogoliubov angles in the same way as
at zero temperature (in the chiral limit, there could be a $\rho_1$-term in
addition.)
In contrast to the $T=0$ case now the single particle energies $\epsilon$ become
physically relevant and will appear explicitly in the self-consistency equation,
simply because excited states do play a role in the thermal equilibrium.
The alternative scenario, namely HF at $T=0$ but finite extension, does
not require any new tools, at least for the GN model. In the case of
the 't Hooft model, the situation will turn out to be more involved due
to the different gluon dynamics in a finite box
(cf. Sec. \ref{subsec:thooftcyl}).

\subsection{GN model on a cylinder}\label{subsec:GNcyl}

Before embarking on the details of the finite temperature GN model, we have to
come back once more to the issue of SSB in lower dimensional
systems.\cite{Barducci95}
Since the thermodynamic limit can be taken only with respect to one dimension
now, there are further restrictions as compared to those of Sec.
\ref{subsec:long}. According
to the Mermin-Wagner theorem, even a discrete symmetry
should be immediately restored at $T=0^+$. The restoration of chiral symmetry at finite
temperature is expected to be driven by the presence of kinks and
antikinks\s\cite{Dashen75a,Karsch97} for any large
but finite $N$. If however the limit $N\to \infty$ is taken before
the thermodynamic limit, these configurations are suppressed and
one is left with the mean field theory where
a second order phase transition at a critical temperature $T_c \neq 0$
becomes possible.
Similarly, for the continuous model, for large but finite $N$ at $T \neq 0$, the
almost long range order disappears. Only if $N\to \infty$ is taken
first, mean field theory applies and one finds the
same critical temperature as in the discrete model.
The non-analytic structure
revealed by the phase diagram of the infinite $N$ GN model depends
on the diverging number of fermion components like in reduced
models.\cite{Wolff85,Eguchi82}

Barducci {\em et al.}\s\cite{Barducci95} advocate to include a small bare
fermion mass,
thereby eliminating kink-antikink configurations in the thermodynamic limit.
Since their results go over rather smoothly into those obtained in mean
field theory at $m=0$, we see nothing wrong in considering the chiral
limit directly for some questions.

\begin{figure}[t]
  \begin{center}
    \epsfig{file=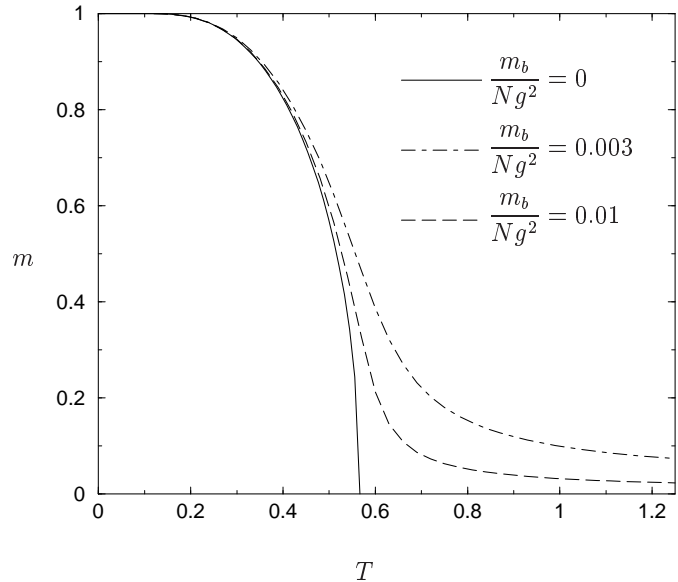, width=9cm}
    \caption{Temperature dependence of dynamical fermion mass for several
values of bare mass, in the GN model. In units of $m_0$.}
    \label{fig:10}
  \end{center}
\end{figure}

First consider finite extension which requires only minimal modifications
of the HF approach for the vacuum discussed in Sec. \ref{subsubsec:vacuum}.
Since the fermion momenta are discretized due to the antiperiodic
boundary conditions, the gap equation (\ref{eq:gapint}) (self-consistency
requirement for the scalar density) has to be replaced by
\begin{equation}
m = \frac{2Ng^2}{L} \sum_{n=0}^{N_{\Lambda}} \frac{m}{\sqrt{m^2+
k_n^2}}
\label{gap_L}
\end{equation}
with
\begin{equation}
N_{\Lambda}=\frac{L\Lambda}{4\pi} \ , \qquad \qquad k_n=\frac{\pi}{L}
(2n+1) \ .
\end{equation}
As the finite extension does not affect any UV properties, we can renormalize
the theory in the limit $L\to \infty$. Denoting the physical mass
on the infinite line by $m_0$ and using Eqs. (\ref{eq:gapint})
and (\ref{gap_L}),
we obtain
\begin{equation}
\lim_{\Lambda \to \infty} \left(
\int_0^{\Lambda/2}\frac{\rd k}{2\pi} \frac{1}{\sqrt{m_0^2+ k^2}}
- \frac{1}{L} \sum_{n=0}^{N_{\Lambda}} \frac{1}{\sqrt{m^2+k_n^2}}\right) = 0
\label{gapeqn}\ .
\end{equation}
This equation is now free of divergences but cannot be solved analytically
in general. To find
the critical length below which chiral symmetry gets restored
it is sufficient to set $m=0$ and solve for $L$,
\begin{equation}
\frac{1}{2} \ln  \frac{\Lambda}{m_0}= \sum_{n=0}^{N_{\Lambda}}
\frac{1}{2n+1}    \ .
\end{equation}
The sum can be performed,
\begin{equation}
\sum_{n=0}^{N_{\Lambda}}\frac{1}{2n+1} = \frac{1}{2}({\rm C}+\ln 4N_{\Lambda})
+ O(1/N_{\Lambda}^2)
\end{equation}
(${\rm C}\approx 0.5772$ Euler constant)
yielding the critical length
\begin{equation}
L_c = \frac{\pi}{m_0} e^{-{\rm C}} \ .
\label{Lcrit}
\end{equation}
A bare fermion mass $m_b$ can easily be taken into account if
one replaces the gap equation (\ref{gapeqn}) by
\begin{equation}
\left(\frac{m_b}{2Ng^2}\right)\left(\frac{1}{m_0}-\frac{1}{m}\right)
+\lim_{\Lambda \to \infty} \left(
\int_0^{\Lambda/2}\!\frac{\rd k}{2\pi} \frac{1}{\sqrt{m_0^2+ k^2}}
- \frac{1}{L} \sum_{n=0}^{N_{\Lambda}} \frac{1}{\sqrt{m^2+k_n^2}}\right)
\!\!=0 \label{gapbarmass}\ .
\end{equation}
Note the appearance of the new parameter
$\frac{m_b}{Ng^2}$ in addition to $m_0$.
Numerical solutions
of this gap equation for different
quark masses are shown in  Fig. \ref{fig:10},
exhibiting a second order phase transition
in the chiral limit and a cross-over phenomenon for non-zero bare masses.
Since it explicitly violates chiral symmetry, the bare quark mass plays a role
analogous to that of an external magnetic field in the ferromagnetic
phase transition.
Finally, we reemphasize that to this order in the $1/N$ expansion the
calculation
is identical for the GN models with discrete or continuous chiral symmetry.

We now switch to the equivalent finite temperature calculation to
familiarize ourselves
with the corresponding generalization of the HF approximation. As
indicated in Sec. \ref{subsec:finTfinL},
the gap equation of thermal HF has to be modified by $T$-dependent occupation
numbers which ``smear out" the Fermi surface,
\begin{equation}
m - m_b =  {Ng^2}\int \frac{\rd k}{2\pi}
\frac{m}{\epsilon(k)} \tanh \frac{\beta \epsilon(k)}{2} \ ,
\quad \epsilon(k)=\sqrt{m^2+k^2} \ .
\end{equation}
The $\tanh$ was derived in Eq. (\ref{rho_T}).
Renormalizing at $T=0$ then yields the following
equation for the $T$-dependence of the physical fermion mass,
\begin{equation}
0=m_b\left(\frac{1}{m_0}-\frac{1}{m}\right) +
Ng^2 \int \frac{\rd k}{2\pi} \left( \frac{\tanh \frac{1}{2}
\beta \sqrt{m^2+k^2}}{\sqrt{m^2+k^2}} - \frac{1}{\sqrt{m^2_0+k^2}}
\right) \ .
\label{gap_T}
\end{equation}
The integral in (\ref{gap_T}) converges and we recover the well known results
obtained by functional integration methods,\cite{Harrington75,Dashen75a}
in particular a second order phase transition at a critical
temperature $T_c$ which coincides
with $L_c^{-1}$ from Eq. (\ref{Lcrit}).
Bosonic excitations are
$1/N$ suppressed and play no role at this stage. Hence just like in
the finite extension case, there is no
distinction yet between GN models with continuous and discrete
chiral symmetry.

It is interesting to compare the two approaches
leading to the gap equations (\ref{gapbarmass}) and (\ref{gap_T}) respectively
at an intermediate stage to understand how covariance ``works".
The crucial identity which ensures the equivalence
of the finite $L$ and finite $T$ calculations is
\begin{equation}
\int_{-\Lambda/2}^{\Lambda/2} \frac{\rd k}{\sqrt{m^2+k^2}}
\tanh \left( \frac{1}{2} \beta \sqrt{m^2+k^2} \right)
\stackrel{!}{=} \frac{4\pi}{L} \sum_{n=0}^{N_{\Lambda}} \frac{1}{\sqrt{m^2+k_n^2}}
\label{identity}
\end{equation}
for $\beta = L$. Eq. (\ref{identity}) can be verified with the help
of Cauchy's
theorem.
The integrand has simple poles in the upper half plane at
\begin{equation}
k= {\rm i} \sqrt{m^2+((2n+1)\pi/\beta)^2}
\ , \quad (n=0...N_{\Lambda})
\end{equation}
with residues $-2{\rm i}/(\beta \sqrt{m^2+((2n+1)\pi/\beta)^2})$.
Thereby, a continuous average involving thermal occupation numbers
can be converted into a sum characteristic for a finite interval. This is of
course reminiscent of the imaginary time approach with Matsubara frequencies
although different in detail (in the canonical approach) due to the interchange
of frequencies and momenta. In the Euclidean path integral formalism,
we would expect no difference
whatsoever.

So far we have concentrated on the dynamical mass or equivalently
the chiral condensate.
Other bulk thermodynamic observables can easily be calculated from the
thermodynamic potential. Let us therefore consider the two functions
which are minimized with respect to $m$ in the two
approaches. At finite extension in the chiral limit,
it is the HF vacuum energy density
\begin{equation}
\frac{{\cal E}}{N} = \frac{m^2}{2Ng^2} - \frac{1}{L} \sum_n \sqrt{m^2+k_n^2}
\end{equation}
(cf. Sec. \ref{subsubsec:vacuum}).
The gap equation (\ref{gap_L}) follows from the condition
$\partial {\cal E}/\partial m=0$.
In the finite $T$ case on the other hand, we have to
minimize the thermodynamic potential $\phi$ or free energy density ${\cal F}$
related to the partition function via
\begin{equation}
\frac{\phi}{L}={\cal F} = - \frac{1}{\beta L} \ln {\rm tr}\, {\rm e}^{-\beta H}
\end{equation}
and equal to the negative of the pressure. In the GN model at the HF
level, the Hamiltonian
is given by that of the free, massive theory plus a $c$-number term
correcting for double counting
of the two-body interaction,
\begin{equation}
H = H^{(0)}_m + \frac{Lm^2}{2g^2} \ .
\end{equation}
Therefore the partition function can be evaluated following the textbook
approach
for a free Fermi gas,\cite{Kapusta}
\begin{equation}
\frac{{\cal F}}{N} = \frac{m^2}{2Ng^2} - \int \frac{{\rm d}k}{2\pi}
\left[\epsilon(k)
+ \frac{2}{\beta} \ln \left( 1+{\rm e}^{-\beta \epsilon(k)}
\right) \right] \ .
\label{freefermi}
\end{equation}
Differentiating with respect to $m$ reproduces Eq. ($\ref{gap_T}$)
(here for $m_b=0$).
Since we have already verified that the two gap equations are identical, it is
sufficient to check that ${\cal E}$ and ${\cal F}$ coincide for one value
of $m$
(upon identifying $L$ and $\beta$) to establish that the two functions agree
everywhere as
expected on general grounds (cf. Eq. (\ref{epsilon_p})). We choose $m=0$ where
all calculations can be done in closed form. For the energy density in the
finite interval
we find, using heat kernel regularization (as usually done in treating the
Casimir effect),
\begin{equation}
\frac{{\cal E}}{N} = - \frac{\pi}{6 L^2}\ .
\label{Casimir_m0}
\end{equation}
For finite temperature
Eq. (\ref{freefermi}) yields
\begin{equation}
\frac{{\cal F}}{N} = - \frac{\pi}{6\beta^2} \ .
\end{equation}
i.e., the Stefan-Boltzmann law. In both cases,
the same $L$- resp. $\beta$-independent quadratically
divergent term has been dropped. This illustrates the
above mentioned correspondence between the Casimir
effect and the Stefan-Boltzmann law and completes the demonstration
that compressed and hot relativistic systems are fully related by
covariance for the case of the GN model.

\begin{figure}[t]
  \begin{center}
    \epsfig{file=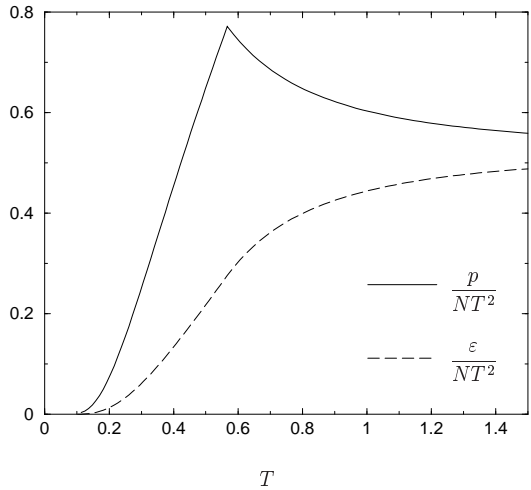,width=7cm}
    \caption{Pressure (solid line) and internal energy density
      (dashed line) per 
color in the GN model, dividing out the $T^2$ dependence of the
Stefan-Boltzmann law. In units of $m_0$; adapted from Ref. 92.}
    \label{fig:11}
  \end{center}
\end{figure}

In order to derive other bulk thermodynamic observables it
is useful to renormalize the
thermodynamic potential.\cite{Barducci95,Wolff85} Using the gap equation at
zero temperature one can write
\begin{equation}
\left. \frac{{\cal F}}{N} \right|_{\rm ren} = \frac{m^2}{4\pi}
\left( \ln \left( \frac{m^2}{m_0^2}\right)-1 \right) + \frac{m_0^2}{4\pi}-
\frac{2}{\beta} \int \frac{{\rm d}k}{2\pi} \ln \left( 1+ {\rm e}^{-\beta \epsilon(k)}
\right) \ .
\end{equation}
Following Barducci {\em et al.}\s\cite{Barducci95} we
have added a constant ``bag term" $m_0^2/4\pi$
in order to to normalize the (unobservable) vacuum pressure to zero.
The derivative $\partial {\cal F}_{\rm ren}/\partial m$
reproduces the renormalized gap equation (\ref{gap_T}).
From this, the bulk thermodynamical observables can be obtained.
For temperatures below the phase transition, pressure and internal energy density are
\begin{eqnarray}
\frac{p}{N} & = & -\frac{m^2}{4\pi}\left(\ln \left(\frac{m^2}{m_0^2}\right)
-1 \right)-\frac{m_0^2}{4\pi}
+\frac{2}{\beta \pi} \int_0^{\infty} {\rm d}k \ln
\left(1+{\rm e}^{-\beta\epsilon(k)}\right) \\
\frac{\varepsilon}{N} & = & \frac{m^2}{4\pi}
\left(\ln \left(\frac{m^2}{m_0^2}\right) -1 \right)
 + \frac{m_0^2}{4\pi}
+\frac{2}{\pi}
\int_0^{\infty}{\rm d}k \frac{\epsilon(k)}{1+{\rm e}^{\beta \epsilon(k)}}
\end{eqnarray}
where $m$ has to be evaluated at the minimum of ${\cal F}$ (solution of the gap equation).
In the region above $T_c$, the results are simply
\begin{eqnarray}
\frac{p}{N}&=& \frac{\pi T^2}{6}-\frac{m_0^2}{4\pi} \ ,\\
\frac{\varepsilon}{N} & = & \frac{\pi T^2}{6} + \frac{m_0^2}{4\pi} \ .
\end{eqnarray}
In  Fig. \ref{fig:11}, pressure and internal energy density have
been plotted after dividing out
the $T^2$ dependence of the Stefan-Boltzmann law, so as to enhance
interaction effects.

In conclusion, both finite extension and finite temperature
calculations agree perfectly for the GN model.
Covariance of the single particle energies is certainly the key
feature behind this result.

\subsection{'t Hooft model on a cylinder, decompactification}
\label{subsec:thooftcyl}

The behavior of the 't Hooft model at finite temperature is rather
controversial.
McLerran and Sen\s\cite{McLerran85} argue
that there is no deconfining phase transition, except possibly at
infinite temperature.
Ming Li,\cite{Li86} using standard finite temperature field theory
methods, concludes that chiral symmetry may get restored in
the limit $T\to \infty$.
In both of these studies, severe infrared problems were encountered,
either in the form of divergent diagrams or ambiguous quark
self energies. Hansson and Zahed\s\cite{Hansson93} address
the question of the $N$-dependence of thermodynamic quantities.
Several studies have considered
QCD$_2$ on a spatial circle at zero temperature. As explained above,
this can also be reinterpreted as finite
temperature calculations for a spatially extended system.
Lenz {\em et al.}\s\cite{Lenz91}
observe a chiral phase transition
in the massless 't~Hooft model
at some critical length, strongly reminiscent of the GN model.
Dhar {\em et al.}\s\cite{Dhar94a,Dhar94b,Dhar94c} treat the zero mode
gluons in a more ambitious way
than Ref. 45
using technology developed in the framework of matrix
models and string
theory, but are not able to fully solve the resulting complicated equations.
They propose that the gauge variables get decompactified by the
fermions in complete analogy with the Schwinger model,\cite{Manton85}
a claim which
has recently been disputed by Engelhardt.\cite{Engelhardt95a}

Let us first consider finite temperature HF. Here, the single
particle energies enter through the equilibrium thermal
occupation numbers. If we used the principal value prescription
with tachyonic behavior at small momenta\s\cite{Li86}
(Sec. \ref{subsubsec:vacuumth}), we
would necessarily get a non-trivial temperature dependence in leading
order of the $1/N$ expansion ($T$-dependent condensate and pressure of order
$N$). Physically, this would imply that the
quarks get deconfined. Color singlet mesons which might contribute
to thermodynamic observables
can show up only in next-to-leading order in $1/N$.
We have
argued before that the HF single particle energies (interpreted
as removal energies) should include the constant $Ng^2L/48$ diverging in the
thermodynamic limit (cf. Sec. \ref{subsubsec:vacuumth}).
If this is kept, ${\rm e}^{-\beta \epsilon}
\to 0$ and all thermal factors become independent of $T$.
As a consequence,
\begin{equation}
\frac{\partial}{\partial T} \lim_{N\to \infty}\frac{\langle \bar{q}q\rangle}
{N} = 0 \ ,
\quad  \lim_{N\to \infty}\frac{p}{N} = 0 \ .
\end{equation}
A system of independent quarks cannot be heated up due to confinement.
This leaves no room for a chiral phase transition,
not even in the limit $T\to \infty$.

Let us now turn to the finite extension alternative. The 't Hooft
model on a line reduces to a purely fermionic
theory with linear Coulomb interaction, see Sec. \ref{subsec:thooft}.
If we simply take this
theory and put it on a finite interval with antiperiodic boundary
conditions for the quarks, we can repeat the HF calculation of the
vacuum (Sec. \ref{subsubsec:vacuumth})
numerically, in analogy to what we did analytically
in the GN model (Sec. \ref{subsec:GNcyl}).
This calculation has in fact been done
long ago, although for reasons totally unrelated to
finite temperature.\cite{Lenz91,Thies92} A second order phase
transition to a chirally restored phase was found at a critical interval
length of $L_c\approx 19.4/\sqrt{Ng^2}$.
Here, the quark self-energies cannot be responsible since they
do not enter the
gap equation which is perfectly well behaved in the IR, see
Sec. \ref{subsubsec:vacuumth}.
So something must have gone wrong with
covariance --- the equivalence of finite extension and finite
temperature which we have verified in the GN model seems to
have been lost in the gauge theory.

The solution to this puzzle was found only recently.\cite{Schoen00a}
It turns out that
as soon as one considers finite $L$, it is not true
that the gluons in QCD$_2$ can be completely eliminated in favor of a
static Coulomb potential between quarks.
The large $N$ limit of the zero mode gluons (discussed for
pure Yang-Mills in Sec. \ref{subsec:gauge}) needs to be refined.
Eventually, the finite
interval calculation yields the same null-result as the thermal
HF approach. In effect, the zero mode gluons make it impossible
to compress the quark system, in the same way as the IR-divergent single
particle energies make it impossible to heat the system up
(to leading order in $1/N$). We are
familiar with the fact that gauge theories can hide their microscopic
degrees of freedom; it is perhaps less well known that they can even hide
the structure of space-time, making a cylinder look like an infinite plane
(``decompactification"\s\cite{Schoen00a}).
Since this is a novel and unexpected phenomenon, it may be
worth going briefly through the analysis.

\begin{figure}
\begin{center}
\epsfig{file=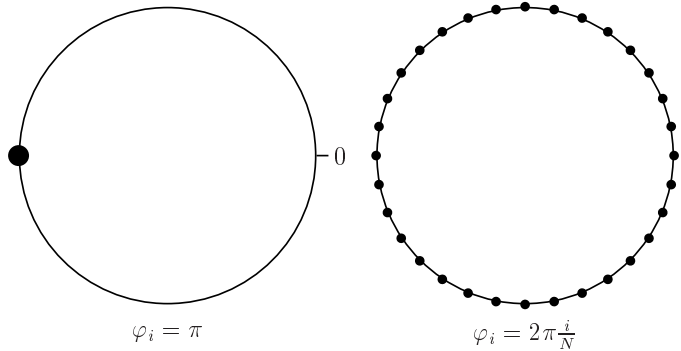,width=9cm}
\caption{Preferred gluon background field configuration, determining quark
boundary conditions for 't Hooft model on a spatial circle. 
Left: Neglecting the
Jacobian. Right: Taking into account the repulsion due to the
Vandermonde determinant (``pearl necklace").}
\label{fig:12}
\end{center}
\end{figure}

It is easy to include dynamical fermions into the pure Yang-Mills theory of
Sec. \ref{subsec:gauge}.
We again work canonically on a spatial circle of length $L$ in the gauge
$\partial_1 A_1 =0$, $(A_1)_{ij}=\delta_{ij}\frac{\varphi_i}{gL}$ diagonal in
color. The Hamiltonian reads\s\cite{Lenz91,Lenz94,Lenz95}
\begin{equation}
H=H_{\rm g} + H_{\rm f} + H_{\rm C} \ ,
\end{equation}
with the gauge field kinetic energy
\begin{equation}
H_{\rm g}=-\frac{g^2 L}{4} \sum_i \frac{\partial^2}{\partial \varphi_i^2}
\ ,
\end{equation}
the quark kinetic energy
\begin{eqnarray}
H_{\rm f} &=& \sum_{n,i}\frac{2\pi}{L}\left( n+\frac{\varphi_i}{2\pi}\right)
\left( a_i^{\dagger}(n)a_i(n)
- b_i^{\dagger}(n)b_i(n) \right)\nonumber\\
&&\phantom{\sum_{n,i}}
+m \sum_{n,i} \left(a_i^{\dagger}(n)b_i(n)+b_i^{\dagger}(n)
a_i(n)\right) \ ,
\end{eqnarray}
and the Coulomb interaction
\begin{equation}
H_{\rm C} = \frac{g^2 L}{4} \sum_{n,i,j} \frac{j_{ij}(n)j_{ji}(-n)}
{\left(2\pi n-\varphi_j+\varphi_i\right)^2} \ .
\end{equation}
The notation for the quark operators and currents is the same as in
Sec. \ref{subsec:thooft}.
As discussed in Sec. \ref{subsec:gauge} due to the curved
configuration space of the $\varphi_i$ and the SU($N$)
Haar measure originally appearing in $H_{\rm g}$, this Hamiltonian has to be
supplemented by the following boundary condition for the
wavefunctionals,
\begin{equation}
\Psi(\varphi_1,...,\varphi_N;{\rm fermions}) = 0 \quad {\rm if} \quad
\varphi_i=\varphi_j \ {\rm mod}\  2\pi \ .
\end{equation}
In Ref. 45 quantization was
performed after complete
classical gauge fixing.
Thereby, the fact that the $\varphi_i$ are curvilinear coordinates
was ignored.
This led to the assumption that all the
$\varphi_i$ are frozen at the value $\pi$ in the large $N$ limit. In the
resulting
purely fermionic theory, the only remnant of the gluons are
antiperiodic boundary conditions for the quarks in the compact space
direction. In the meantime,
this whole approach has been put on a more rigorous basis by first
quantizing in the Weyl gauge
and then resolving the Gauss law quantum mechanically.\cite{Lenz94} This
made it clear that the
$\varphi_i$ are parameters on the group manifold
with corresponding Jacobian, the SU($N$) reduced Haar measure.
When solving the theory, it is then possible to restrict
oneself to
the smallest region in field space bounded by zeros of the
Jacobian, see Ref. 27  where
the consequences for SU(2), SU(3) have been explored in the case of
adjoint fermions.
How can this be
generalized to the large $N$ limit?
A definite choice of ``fundamental
domain" obviously means that the $\varphi_i$ always remain ordered, say
$0 \leq \varphi_1 \leq \varphi_2 \leq ... \leq \varphi_N \leq 2\pi$.
If we think of the gluons as particles on a circle, they cannot
cross each other and become closely packed in the limit $N\to \infty$.
Their fluctuations are suppressed by $1/N$, simply due to lack
of space. The only
degree of freedom left, the collective rotation of this ``pearl
necklace", is a U(1) factor which anyway is not present
in the SU($N$) theory. This
suggests that the correct choice for the gluon background field
as seen by the fermions
is not $\varphi_i=\pi$, but rather
the continuum limit of the lattice points
\begin{equation}
\varphi_i = 2\pi \frac{i}{N} \ , \quad (i=1 \dots N)
\label{necklace}
\end{equation}
(see Fig. \ref{fig:12}).
Instead of antiperiodic boundary conditions, the fermions then acquire
color dependent boundary conditions which interpolate smoothly
between the phases 0 and $2\pi$,
\begin{equation}
\psi_k(L)={\rm e}^{{\rm i}2\pi k/N}\psi_k(0) \ , \quad (k=1 \dots N) \ .
\end{equation}
In the thermodynamic limit $L\to \infty$, both of these choices
of the gluon field configuration, $\varphi_i = \pi$ or Eq. (\ref{necklace}),
will become indistinguishable and yield
the well-known results.
We now show that at finite $L$ the effects of the gluons on the
quarks is radically different in these two cases, and that it is the
spread out
distribution (\ref{necklace}) which is in fact the correct one.

In such a gluonic background field,
the fermions can be treated in a relativistic HF approximation
along the lines explained in Sec. \ref{subsubsec:vacuumth} except that
the Bogoliubov angle acquires a color index.
The gap equation then becomes
\begin{equation}
\frac{2\pi}{L}(n+\alpha_i) \sin \theta_i(n) - m \cos \theta_i(n)
+ \frac{g^2L}{16 \pi^2} \sum_{n',j}\frac{\sin \left( \theta_i(n)-
\theta_j(n-n')\right)}{(n'-\alpha_j+\alpha_i)^2} = 0
\end{equation}
where we have switched to the slightly more convenient variable
$\alpha_i = \frac{\varphi_i}{2\pi}
\in [0,1]$ for the gluons.
If $\alpha_i=1/2$ as chosen in Ref. 45, $\theta_i(n)$ becomes
$i$-independent and we recover the old gap equation considered in that work.
Now, we assume $\alpha_i=i/N$ and perform the large $N$ limit before
solving the gap equation. Since $\alpha_i$ becomes a continuous variable,
we replace
$\theta_i(n)\to \theta_{\alpha}(n)$ and
$\sum_j \to N \int_0^1 {\rm d}\alpha'$, with the result
\begin{eqnarray}
&&\frac{2\pi}{L}(n+\alpha) \sin \theta_{\alpha}(n) -m\cos \theta_{\alpha}(n)
\nonumber\\
&&+ \frac{Ng^2L}{16\pi^2}
\sum_{n'} \int_0^1 {\rm d} \alpha' \frac{\sin \left( \theta_{\alpha}(n)
-\theta_{\alpha'}(n-n')\right)}{(n'-\alpha'+\alpha)^2} = 0 \ .
\end{eqnarray}
This infinite set of coupled integral equations collapses into a single,
one-dimensional
integral equation, if we set
\begin{equation}
\theta_{\alpha}(n)=\theta(n+\alpha) \ .
\end{equation}
Since $n$ is integer and $\alpha \in [0,1]$, this step in effect
decompactifies
the original spatial circle. With this ansatz, the notation
$n+ \alpha= \nu$, $n-n'+\alpha'=\nu'$ (where $\nu, \nu'$ are dimensionless,
continuous
variables) and the substitution $\sum_{n'} \int_0^1 {\rm d}\alpha'
\to \int_{-\infty}^
{\infty} {\rm d} \nu'$, we obtain
\begin{equation}
\frac{2\pi}{L}\nu \sin \theta(\nu) -m\cos\theta(\nu)
+ \frac{Ng^2L}{16\pi^2}\int
\!\!\!\!\!\!- {\rm d}\nu'
 \frac{\sin (\theta(\nu)-\theta(\nu'))}{(\nu
-\nu')^2} = 0 \ .
\end{equation}
The simple pole in the integral has been regularized by the principal value
prescription. After rescaling the variables via
\begin{equation}
\frac{2\pi}{L} \nu := p \ , \ \ \frac{2\pi}{L} \nu' :=p' \ ,
\end{equation}
where $p, p'$ have the dimension of momenta, we recover exactly
the continuum version
of the HF equation, namely
\begin{equation}
p \sin \theta \left( \frac{Lp}{2\pi} \right) -m \cos \theta
\left(\frac{Lp}{2\pi}\right)
+ \frac{Ng^2}{4}
\int\!\!\!\!\!\!- \frac{{\rm d} p'}{2\pi} \frac{\sin \left( \theta
\left( \frac{Lp}{2\pi} \right)
-\theta \left( \frac{Lp'}{2\pi} \right) \right)}{(p-p')^2} = 0 \ .
\end{equation}
Denoting the Bogoliubov angle of the continuum 't~Hooft model by
$\theta_{\rm cont}(p)$,
we conclude that
\begin{equation}
\theta(\nu) = \theta_{\rm cont}\left(\frac{2\pi}{L}\nu \right) \ ,
\end{equation}
or, in terms of the original, color dependent Bogoliubov angle,
\begin{equation}
\theta_i(n) \approx \theta_{\rm cont}\left(
\frac{2\pi}{L} \left(n+\frac{i}{N}\right) \right) \ , \quad (N\to \infty) \ .
\end{equation}
This last relation becomes exact in the limit $N \to \infty$ only. In this
limit, the color-
and $L$-dependent Bogoliubov angles for the 't~Hooft model on the circle
of length $L$
are all given by one universal function,
namely the momentum dependent Bogoliubov angle of the 't~Hooft model
on the infinite line. We emphasize that this universality only holds
for the ``pearl necklace" type distribution of gauge variables,
Eq. (\ref{necklace}).
If the $\varphi_i$
are all set equal to $\pi$, there is no analytic way known how to relate
$\theta(n)$ for different $L$ values, but one has to solve the gap
equation numerically for each $L$.\cite{Lenz91}

\begin{figure}[t]
\begin{center}
\epsfig{file=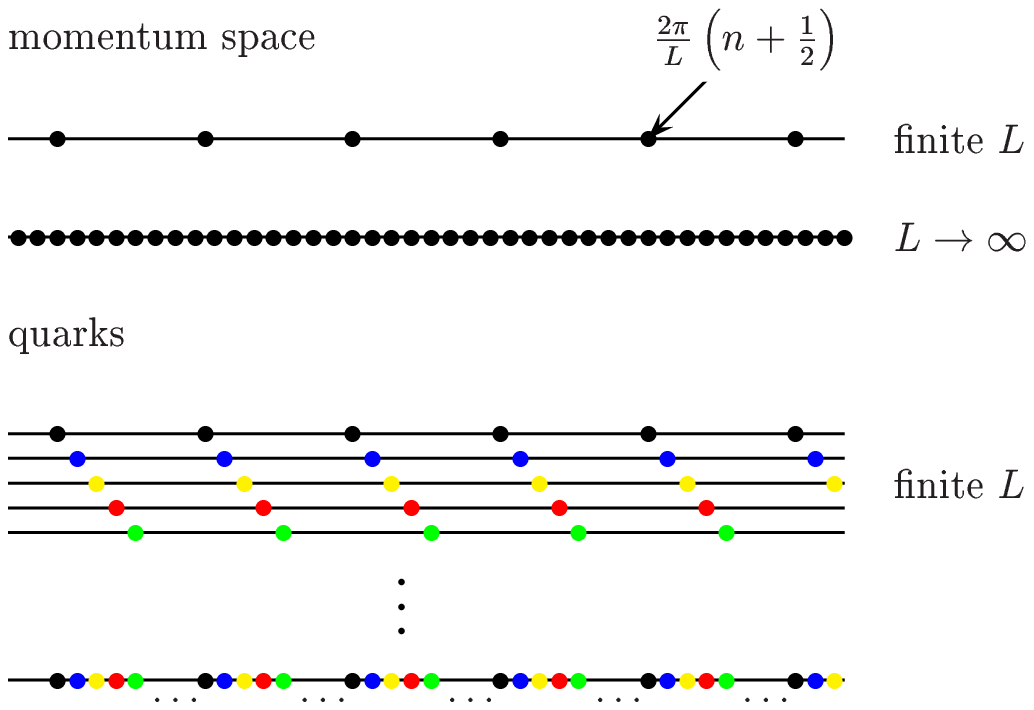}
\caption{Sketch (in momentum space) of how color-dependent quark
  boundary conditions
in a finite interval may simulate a theory on the infinite line,
in the limit $N\to \infty$
(``decompactification"). 
}
\label{fig:13}
\end{center}
\end{figure}

The upshot of this simple exercise is the following: In the large $N$ limit,
the gluon variables influence the fermion boundary conditions in such
a way that the circle gets replaced by a line --- they decompactify
space-time (see  Fig. \ref{fig:13}; this only works for gauge 
invariant variables,
where one sums over all colors). The length $L$ of the spatial
circle becomes an irrelevant
parameter.
To confirm this last point, let us evaluate the condensate as an
example of an observable,
\begin{equation}
\langle \bar{q}q \rangle = -\frac{1}{L} \sum_{n,i} \sin
\theta_i(n)
= - \frac{N}{L} \sum_n \int_0^1 {\rm d} \alpha \sin \theta (n+\alpha)
= -N \int \frac{{\rm d}p}{2\pi} \sin \theta_{\rm cont}(p) \ .
\end{equation}
The sum over the discrete momenta and the color sum in
the large $N$ limit conspire to produce the continuum result, independently
of the starting $L$ value. In the alternative thermodynamic interpretation,
the condensate becomes independent of temperature to leading order
in $1/N$, now in perfect agreement with the finite temperature HF result.

Notice that so far, we have discussed the influence of the gauge
fields on the
quarks which is indeed dramatic. Vice versa, we do not expect the quarks
to influence significantly the gluon zero point motion, again
due to the effects of the Jacobian. The kinetic energy
$H_{\rm g}$ will give the same result as in pure Yang Mills theory.
Since this contribution to the energy density is $L$-independent, it
again yields
zero pressure,
reflecting the absence of physical gluonic excitations in 1+1 dimensions.

\subsection{$1/N$ corrections at finite $T$, role of pions}
\label{subsec:1/N}

In the present section, we wish to discuss the role of the pions at
low temperature. Since this has to do with the next order in the $1/N$
expansion and cannot be covered systematically here, we will content
ourselves with briefly reviewing the state of the art and indicating
the issues which have attracted most attention recently.

The importance of pions for low temperature QCD was pointed out by
Leutwyler and collaborators\s\cite{Gasser87,Gerber89} some time ago.
In the vicinity of the chiral limit, pions are naturally the
lightest bosons and as such preferred degrees of freedom at low
temperature. Using chiral perturbation theory, interaction effects
in the pion gas can be computed reliably to rather high order.
An extrapolation of the temperature dependence of the chiral condensate
even gives a surprisingly good prediction for the
critical temperature of the QCD chiral phase transition. Glueballs
(if they exist in nature) are much too heavy to matter at low
temperatures, quarks are anyway ruled out due to confinement.

The situation is somewhat different in the case of the NJL model
which does not exhibit confinement. However also here thermal excitation
of a $q\bar{q}$ pair is strongly suppressed as compared to pions
at low $T$. It was soon recognized that the popular $1/N$ expansion
is somewhat misleading physically: To leading order, the pions do
not contribute at all to thermodynamic observables, they are dominated
by the
quarks due to their color weight factor (entropy).
For any finite $N$, one can
estimate a temperature below which pions dominate, but for $N\to \infty$
this temperature moves to $0$. Zhuang, Klevansky and
H\"ufner\s\cite{Zhuang94,Klevansky92}
and others have developed techniques to go to next-to-leading order in the
$1/N$ expansion, e.g. in the thermodynamic potential.
They take into account fluctuations about the mean field
which correspond diagrammatically to a
ring sum and physically to mesonic contributions. Both $q \bar{q}$ bound
states and
scattering phase shifts enter.
Since a
meson in the NJL model can decay into quarks, a
``mixed phase" arises at intermediate temperatures where
mesons as well as quarks play a role.
This problem was
also investigated by Barducci {\em et al.}\s\cite{Barducci96,Barducci99}
including the temperature dependence of meson masses.

The same issues are relevant for the lower dimensional
models which are the subject of the present work as well.
Since these are toy models
anyway, one usually takes the limit $N\to \infty$ seriously (not as an
approximation of $N=3$), at least this is the point of view which we have
adopted
here. In this case, one might argue that the pions
have no influence on the phase diagram of the chiral GN model. In the
strict chiral limit, this is a rather extreme picture:
$N$ has to be so large that
massive fermions win over massless pions in spite of their exponential
suppression through the Boltzmann factors, but as a
mathematical limit it is perhaps acceptable.

It is instructive to think about the $1/N$ corrections (meson effects)
in the ``rotated" picture of finite extension field theory.
There, in next-to-leading order one would have to
evaluate the RPA vacuum in a finite interval, in
clear correspondence
to the bubble sum of Zhuang {\em et al.}\s\cite{Zhuang94}
Unlike the thermodynamic calculation, this does not require any
new formalism since the RPA technique is well established.
Fermionic and mesonic degrees of freedom are then treated
consistently and it is also clear that scattering states as well as
bound states enter. Such a calculation has not yet been carried
out for the GN model. The thermodynamic treatment of the $1/N$
corrections has been presented by Barducci {\em et al.}\s\cite{Barducci97}
showing low temperature dominance of pion-like
excitations in the massive GN model at finite $N$.
Introducing a bare quark mass enables the authors to
vary the degree of explicit chiral symmetry breaking.

Let us finally come back to the 't Hooft model.
Here, due to confinement, the quarks do not contribute at all to
the pressure at order $N$ so that the mesonic contributions are in fact
the leading ones. Since in the large $N$ limit one has non-interacting
mesons, one can use the thermodynamics of the ideal Bose gas
to analyze this situation. Near the chiral limit, the pion will
naturally stick out at low temperatures.

A theoretical question of some interest is the corresponding finite
interval RPA calculation in the 't Hooft model. Here, we do not
expect that the mechanism which leads to ``decompactification"
in leading order remains intact, otherwise one would also not get any
meson contribution to the pressure. This must mean that the
fluctuations of the zero mode gluons come into the picture.
It would be interesting to study this in more detail since it would
teach us something about the gluon dynamics at large but finite $N$
and perhaps help to settle some of the difficult issues raised by Dhar
{\em et al.}\s\cite{Dhar94a,Dhar94b,Dhar94c} in connection with the
string picture of QCD$_2$.

\section{Finite density}\label{sec:findens}

In the present section, we address the problem of finite baryon density or
chemical potential in the GN and 't Hooft models. Since the physics
issues are quite different from those at finite temperature
(Sec. \ref{sec:finT}),
we restrict ourselves to the $T=0$ case here and try to clarify the 
structure of cold baryonic matter first. Ultimately, we are of course
interested in the phase diagram of our soluble models in the
whole ($T,\mu $)-plane;
this will be the subject of Sec. 6. Whereas the GN model
at finite density  has received a lot of attention in the
literature,\cite{Harrington75,Dashen75a,Wolff85,Treml89,Barducci95}
the corresponding  extension of the 't~Hooft model has remained
essentially unexplored 
until very recently.\cite{Li86,Schoen00b}
Throughout this section, our main focus will be on the relationship between
the structure of the single baryon and that of baryonic matter.

\subsection{GN model as a Fermi gas}\label{subsec:GNFermi}

To determine the ground state of the GN model at finite density, we 
can again use the HF method expected to become exact in the large $N$ limit.
As we shall see, this does not mean that one should trust the
results of a HF calculation blindly: If the ground state breaks
some symmetry, the HF equations will develop different
solutions and one may easily miss the solution with the lowest energy
due to erroneous assumptions.
Anyway, to start our investigation let us first assume translational invariance
and only offer the system the chance to break chiral symmetry. We thus simply
fill a number of positive energy plane wave states on top
of the Dirac sea until we reach the desired baryon density. This is the standard
approach to the finite density GN model, phrased in many-body language. 
Although we have in mind primarily the chiral  GN model,
the corresponding calculation would  in fact 
be identical for the model with discrete symmetry only. 

We denote the fermion density per color (or baryon density) by $\rho_B
= p_f/\pi$
($p_f$: Fermi momentum). At the mean field level, the fermions acquire
a physical mass $m$ which has to be determined self-consistently.
The ground state
energy density per color is then given by
\begin{equation}
\frac{{\cal E}}{N}=-2 \int_{p_f}^{\Lambda/2} \frac{{\rm d}k}{2\pi} \sqrt{
m^2+k^2} + \frac{m^2}{2N g^2}
\label{d1}
\end{equation}
where $\Lambda$ is the UV cutoff.
The first term sums up the single particle energies for
all occupied states (the Dirac sea plus all positive energy states
with $|p|<p_f$), the second term corrects for double
counting of the interaction. We renormalize the theory at $p_f=0$,
cf. Sec. \ref{subsubsec:vacuum}, and denote the physical fermion mass
in the vacuum by $m_0$.
Using the vacuum gap equation (\ref{eq:gap}) to renormalize the matter
ground state energy density, Eq. (\ref{d1}), we find (dropping an
irrelevant infinity)
\begin{equation}
\frac{{\cal E}}{N}= -\frac{m^2}{4\pi}
+ \frac{p_f}{2\pi} \sqrt{p_f^2+m^2}
+ \frac{m^2}{2\pi} \ln \left( \frac{p_f + \sqrt{m^2+p_f^2}}{m_0}\right) \ .
\label{d5}
\end{equation}
The energy is minimal provided $m$ satisfies
\begin{equation}
m \ln \left( \frac{p_f + \sqrt{m^2+p_f^2}}{m_0}\right)=0 \ ,
\label{d6}
\end{equation}
i.e., for
\begin{equation}
m=0 \ \quad \mbox{or} \quad m= m_0\sqrt{1-
\frac{2 p_f}{m_0}} \quad \left(p_f< \frac{m_0}{2}\right) \ .
\label{d7}
\end{equation}
The corresponding energy densities are
\begin{eqnarray}
\left. \frac{{\cal E}}{N}\right|_{m=0} &=& \frac{p_f^2}{2\pi} \ ,
\nonumber \\
\left. \frac{{\cal E}}{N} \right|_{m\neq 0}& = & -\frac{m_0^2}{4\pi} + \frac{p_f m_0}
{\pi} -\frac{p_f^2}{2\pi} \qquad \left(p_f < \frac{m_0}{2}\right)\ .
\label{d8}
\end{eqnarray}

\begin{figure}[t]
\parbox[t]{5.6cm}{
    \epsfig{file=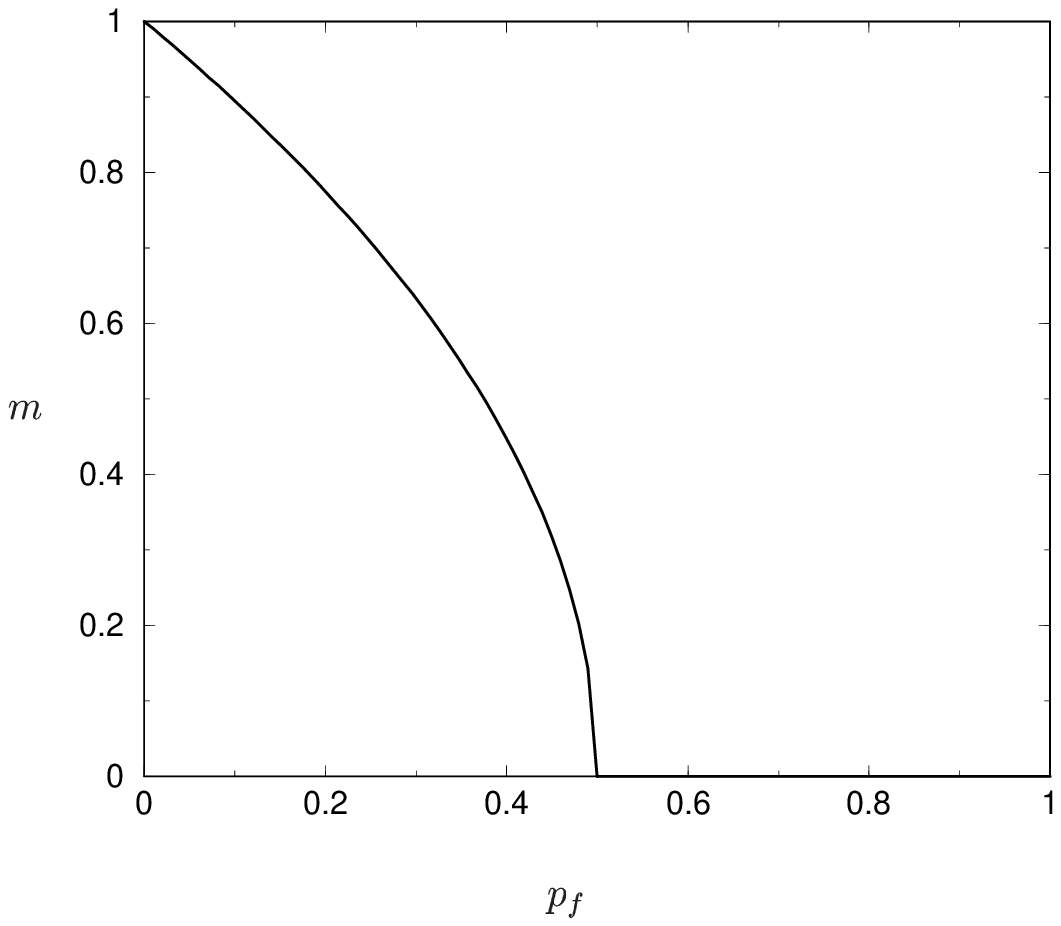,width=5.6cm}
    \caption{Physical fermion mass as a function of Fermi momentum
in the GN model. In units of $m_0$.}
    \label{fig:20}}
  \hfill
  \parbox[t]{5.6cm}{
    \epsfig{file=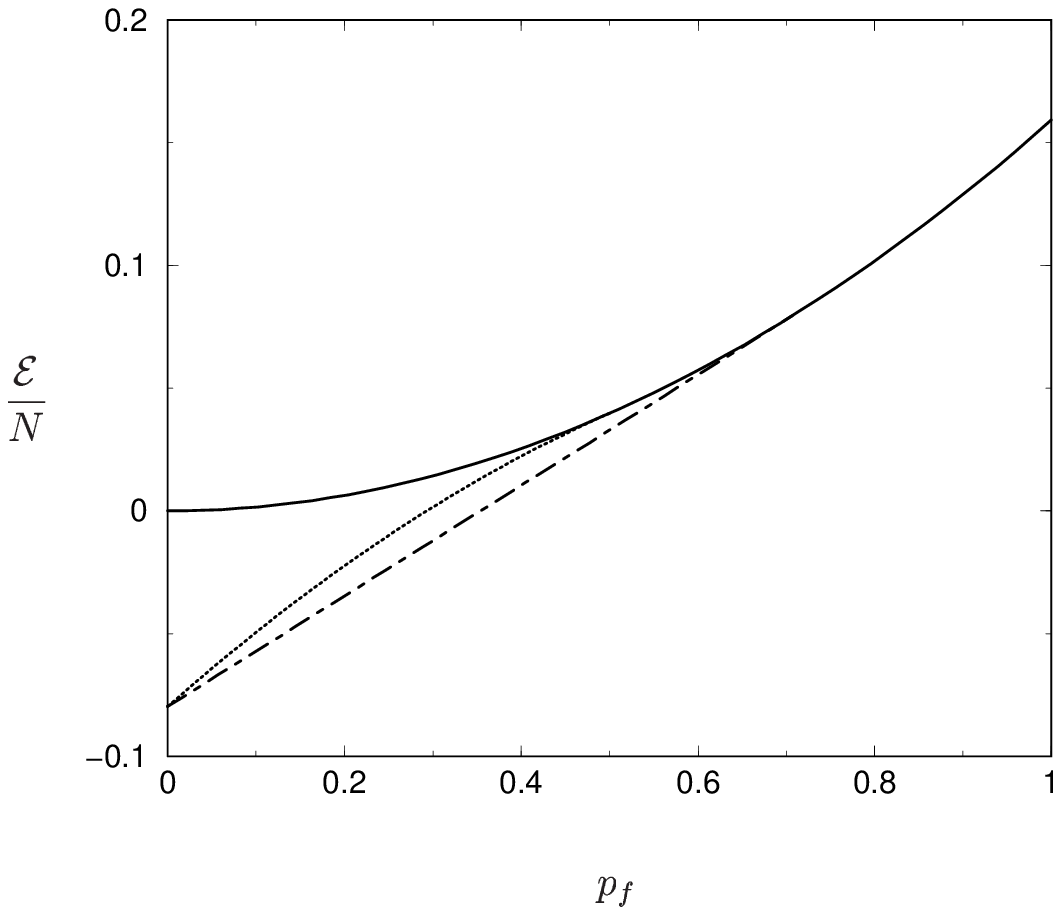,width=5.6cm}
    \caption{Energy density per color as a function of Fermi momentum in the 
GN model. Solid curve: Chirally symmetric solution ($m=0$).
Dotted curve: Broken chiral symmetry ($m$ according to Fig. 14).
Dash-dotted line: Mixed phase. In units of $m_0$.}
    \label{fig:21}}
\end{figure}

The physical quark masses (\ref{d7}) and the energy densities (\ref{d8})
are
plotted in Figs. \ref{fig:20} and \ref{fig:21}.
From these figures one might be tempted to
conclude that chiral symmetry is broken at low densities and
gets restored in a second order
phase transition at $p_f=m_0/2$. As is well known,
this does not occur, rather there is a first order chiral phase transition
at $p_f=m_0/\sqrt{2}$. This can easily be inferred by inspection of the
thermodynamic potential of the GN model.\cite{Wolff85,Barducci95} 
At zero temperature, it is possible to understand it also
more directly in terms of the following instability.
Let us compare the energy densities (\ref{d8}) with the energy
density for a system of size $L$ divided into two homogeneous regions I
(size $\ell$) and II (size $L-\ell$). In region I chiral symmetry is
restored; it contains the extra fermions needed to get the prescribed
average density (the ``MIT bag"\s\cite{Chodos74}). Region II
consists of the physical vacuum
with broken chiral symmetry, void of excess fermions. The mean energy
density obtained in this way is
\begin{equation}
\frac{{\cal E}}{N} = -\left(\frac{L-\ell}{L}\right) \frac{m_0^2}{4\pi}
+ \frac{L p_f^2}
{2\pi \ell} \ .
\label{d9}
\end{equation}
Minimization with respect to $\ell$ yields
\begin{equation}
\ell = \frac{\sqrt{2} p_f L}{m_0}
\label{d10}
\end{equation}
valid for $p_f<m_0/\sqrt{2}$, and hence the optimal energy density
\begin{equation}
\frac{{\cal E}}{N} = - \frac{m_0^2}{4\pi} + \frac{p_f m_0}{\sqrt{2}\pi} \qquad
\left(p_f < \frac{m_0}{\sqrt{2}} \right) \ .
\label{d11}
\end{equation}
As shown in Fig. \ref{fig:21}, this solution is lower in energy than the homogeneous one;
moreover, it yields the convex hull of ${\cal E}$.
It ends exactly at the first order phase transition point $p_f=
m_0/\sqrt{2}$ where all space is filled with one big bag.
This should be contrasted to the scenario underlying Fig. \ref{fig:20}
where the
fermion mass decreases continuously. We thus recover
the well known mixed phase interpretation of the
GN model at finite density, in the zero temperature limit.
Notice also that only the total size of regions I and II matters, not how they
are subdivided; there could be baryon ``droplets" as well.\cite{Alford98} 

One important point to which we would like to draw the attention
of the reader is
the behavior of ${\cal E}$ near $\rho_B=p_f/\pi=0$. Since ultimately, at
very low density, the fermionic matter problem must reduce
to the problem of a single baryon, one would expect
\begin{equation}
\left. \frac{\partial {\cal E}}{\partial \rho_B}\right|_{\rho_B=0} =
M_B
\label{d12}
\end{equation}
where $M_B$ is the baryon mass.
In the present calculation,
$M_B$ is not the physical baryon mass, but the mass of an alleged
``delocalized" baryon. This is inherent in the translationally invariant
HF approach, i.e., the
assumption that the single particle orbitals are momentum eigenstates.
Using Eq.~(\ref{d12}) we obtain in
the homogeneous, single phase calculation, Eq.~(\ref{d8}),
$M_B=Nm_0$, consistent with a short range force and a delocalized baryon.
The (physically more viable) mixed phase approach, Eq.~(\ref{d11}),
predicts a baryon mass lower by a factor of $1/\sqrt{2}$
which can be understood in terms of the bag model for the
baryon.\cite{Schoen00b}

However, the GN model possesses bound baryons with lowest mass
$Nm_0/\pi$ (kink solution for the model with discrete chiral symmetry,
cf. Sec. \ref{subsubsec:baryons}),
or even massless baryons (model with continuous chiral
symmetry, cf. Sec. \ref{subsec:chiral}).
These binding effects are not $1/N$ suppressed
and should be
correctly reproduced in a HF approach, in the low density limit.
They have obviously been missed here due to our tacit assumption of
translational invariance.
There is no good reason why such effects should not play a role at higher
densities as well. Moreover,
differences between the continuous and discrete chirally
symmetric GN models based on their different baryon structure
and spectra are not at all captured by the Fermi gas approach.
In Sec. \ref{subsec:breakdown}
we shall present a cure for this disease. Before that however,
let us first repeat the naive calculation for the 't~Hooft model  where we can 
also investigate the role of confinement. 

\subsection{'t Hooft model as a Fermi gas}\label{subsec:thooftfermi}

We have discussed the vacuum and baryons in the 't~Hooft model in Sec.
\ref{subsec:thooft}. Assuming translational invariance, it is again
straightforward to include a finite baryon density into the HF calculation.
If $p_f$ denotes the Fermi momentum we only need to replace the density matrix
(\ref{eq:densmat}) by
\begin{eqnarray}
\rho(p)&=&\Theta(p_f-|p|)u(p)u^{\dagger}(p) + v(p)v^{\dagger}(p)
\nonumber \\
& = & \Theta(p_f-|p|) + \Theta(|p|-p_f) v(p)v^{\dagger}(p)
\label{d25}
\end{eqnarray}
where we have used the completeness relation for the spinors in
the second step. In the expression for the HF
ground state energy density at $m=0$ (\ref{E_HF}),
according to the second line of
Eq.~(\ref{d25}), we must exclude the region
$[-p_f,p_f]$ from the momentum integrations and pick up an additional
term due to the change in the baryon density ${\rm tr} \rho$,
\begin{eqnarray}
\frac{{\cal E}}{N} &=& - \int \frac{{\rm d}p}{2\pi} \Theta(|p|-p_f) p \cos \theta(p)
\nonumber \\
& & - \frac{Ng^2}{8} \int \frac{{\rm d}p}{2\pi} \int \frac{{\rm d}p'}{2\pi}
\Theta(|p|-p_f)
\Theta(|p'|-p_f) \frac{\cos (\theta(p)-\theta(p'))
-1}{(p-p')^2}
\nonumber \\
& & +\frac{Ng^2}{4}\int \frac{{\rm d}p}{2\pi}\int \frac{{\rm d}p'}{2\pi}
\Theta(p_f-|p|)\Theta(|p'|-p_f) \frac{1}{(p-p')^2} \ .
\label{d26}
\end{eqnarray}
This yields at once the following finite density generalization
of the gap equation,
\begin{equation}
p \sin \theta(p) + \frac{Ng^2}{4} \int \!\!\!\!\!\!- \frac{{\rm d}p'}
{2\pi}
\Theta(|p'|-p_f)\frac{
\sin \left(\theta(p)-\theta(p')\right)}{(p-p')^2} = 0 \ ,
\quad (|p|> p_f) \ ,
\label{d27}
\end{equation}
whereas the condensate now becomes
\begin{equation}
\langle \bar{q}q \rangle = -N \int \frac{{\rm d}p}{2\pi} \theta(|p|-p_f)
\sin \theta(p) \ .
\label{d28}
\end{equation}
The gap equation (\ref{d27}) can easily be solved numerically for
various $p_f$.
\begin{figure}[t]
  \begin{center}
    \epsfig{file=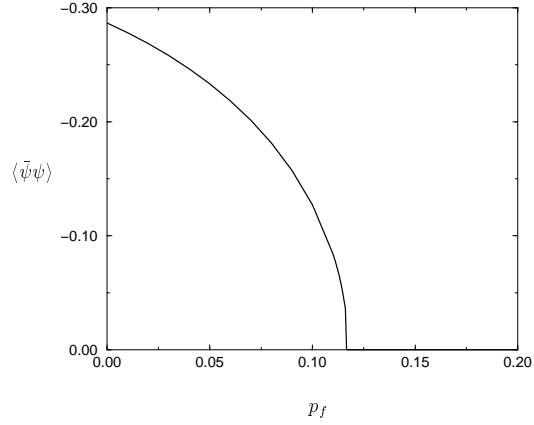,width=7cm}
    \caption{Quark condensate as a function of Fermi momentum in the 't Hooft 
model, in units where $Ng^2=2\pi$.}
    \label{fig:22}
  \end{center}
\end{figure}
The resulting condensate is shown in Fig. \ref{fig:22}.
We find that it decreases monotonically with increasing
density, disappearing at a critical Fermi momentum
\begin{equation}
p_f^c \approx 0.117 \left( \frac{Ng^2}{2\pi} \right)^{1/2} \ .
\label{d28a}
\end{equation}
This behavior is strikingly similar to the
corresponding result for the GN model depicted in
Fig. \ref{fig:20}, again
suggesting
some phase transition with restoration of chiral symmetry at high
density. Unfortunately, here we are not able to assess
whether we are
dealing with a first or second order phase transition. The reason
lies in the following problem:
If we compute the energy density (\ref{d26}) for the 't~Hooft model we
discover that subtraction of the value at $p_f=0$ is not sufficient to
give a finite result. Unlike in the GN model,
the difference is still IR divergent.
To be able to proceed, we enclose the system in a box of length
$L$. We then find that the divergence is due to the last term
in Eq.~(\ref{d26}) (the one which does not involve the Bogoliubov angles)
which now contributes the following double sum to the energy per color,
\begin{equation}
\left. \frac{E}{N}\right|_{\rm div} = \frac{Ng^2 L}{16 \pi^2} \sum_{p\in I}
\sum_{n\neq 0,(p-n)\not\in I} \frac{1}{n^2} \ .
\label{d29}
\end{equation}
Here antiperiodic boundary conditions for fermions have been
employed in the box regularization, and correspondingly
the interval $I$ is defined in the following way,
\begin{equation}
I = [-n_f, n_f] \ \ \ \mbox{for} \ \ B=2 n_f+1 \ \mbox{odd} \ , \ \ \
I = [-n_f-1, n_f] \ \ \ \mbox{for} \ \ B=2n_f+2\  \mbox{even} \ .
\label{d29a}
\end{equation}
The result (\ref{d29}) is even more alarming than the non-convex behavior of
${\cal E}$ in the GN model, Fig. \ref{fig:21},
due to its $L$-dependence.
Adding quarks to the
vacuum causes the energy to increase by an infinite amount in the limit
$L \to \infty$. Evaluating the double sums in Eq.~(\ref{d29}) for
low values of $B$, we obtain information on the origin of this divergent
behavior. For $B=1$
($I=[0,0]$) in particular, the calculated baryon mass (to leading order in $L$) is
\begin{equation}
M_B= N \left(\frac{N g^2 L}{48}\right) \ .
\label{d30}
\end{equation}
This is the same relation as $M_B=N m_0$ in the GN model
except that the physical fermion mass is replaced by the infinite
constant $Ng^2 L/48$ characteristic of confinement, cf. Eq.~(\ref{omega}).

Summarizing,
the problems encountered in the GN model with translationally
invariant baryonic matter again show up in the 't~Hooft model, although in
aggravated form.
The physics reason is clear: In the GN
model the cost of distributing $N$ fermions over the whole space is
governed by their physical mass; in the 't~Hooft model, due to confinement
of quarks, the corresponding quark effective mass diverges with the
volume. On the other
hand, it is known that both models
do possess massless, delocalized baryons in the chiral limit. Evidently, this
has to be accounted for, and we conclude that the naive, translationally
invariant HF approximation cannot be trusted.

\subsection{Breakdown of translational invariance}\label{subsec:breakdown}

\subsubsection{Strict chiral limit}\label{subsubsec:strict}

\begin{figure}[t]
  \begin{center}
    \epsfig{file=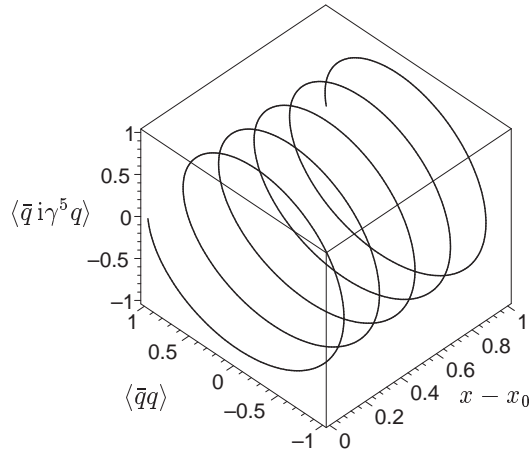,width=7cm}
    \caption{Same as Fig. \ref{fig:8},
      but for baryonic matter in the chiral GN or 't Hooft model
(``chiral spiral"). The wavenumber of the oscillations in the condensates
is $2p_f$. }
    \label{fig:23}
  \end{center}
\end{figure}

We now return to the massless baryons introduced in Sec. \ref{subsec:chiral}
in the chiral limit of
two-dimensional field theories. By using literally the same techniques
(except for a different value 
of the baryon number) we can easily find the
ground state of the system for any baryon density.
As discussed in Sec. \ref{subsec:chiral} and illustrated in
Fig. \ref{fig:8},
the single baryon can be visualized in terms of  
a ``chiral spiral" parametrized by $\chi(x)$ with one single turn
in the whole volume.
A finite density $\rho_B=B/L=p_f/\pi$ on the other hand implies that
\begin{equation}
\chi(x)=p_f (x-x_0) \ ,
\label{d42}
\end{equation}
i.e., one full rotation over a physical distance which now has a well
defined
limit for $L\to \infty$, namely 2/$\rho_B$. The baryon density remains
constant in space for the reasons discussed in Sec. \ref{subsec:chiral},
but the condensates are modulated as
\begin{eqnarray}
\langle \bar{q} q\rangle  & = & \langle \bar{q} q \rangle_v
\cos 2 p_f (x-x_0) \ , \nonumber \\
\langle \bar{q}\,{\rm i}\gamma_5 q\rangle
& = & -\langle \bar{q} q \rangle_v
\sin 2 p_f (x-x_0) \ .
\label{d43}
\end{eqnarray}
They can again be viewed as projections of a chiral spiral of radius
$|\langle \bar{q}q \rangle_v |$ onto two
orthogonal planes, see Fig. \ref{fig:23}.

\begin{figure}[t]
  \begin{center}
    \epsfig{file=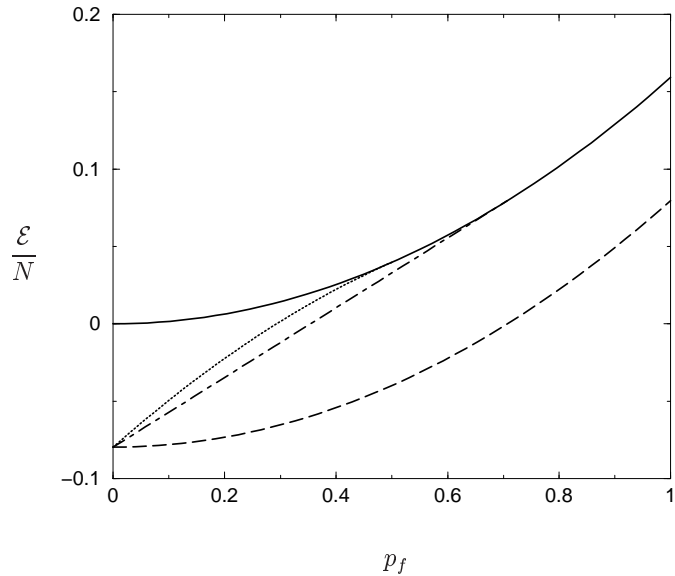,width=9cm}
    \caption{Same as Fig. \ref{fig:21}
      (GN model), except that the energy density of the 
Skyrme crystal type of state (the true ground state) has been included
as the dashed line.}
    \label{fig:24}
  \end{center}
\end{figure}

This state breaks translational symmetry; it is a crystal. In fact, it
may be viewed as the simplest possible realization of the old idea
of a Skyrme crystal,\cite{Klebanov85} here in the context of large $N$
two-dimensional field
theories. It is meaningless to ask
where one baryon begins and ends; each full turn
of the spiral adds one unit to the baryon number. Only the condensates reveal that
translational symmetry has been broken down to a discrete subgroup. The
energy density
of this unusual kind of ``nuclear matter" is simply (after subtracting
the vacuum energy density)
\begin{equation}
\frac{{\cal E}}{N}=\frac{p_f^2}{2\pi} \ .
\label{d44}
\end{equation}
Surprisingly,
this is exactly what one would get for a free Fermi gas of
massless quarks although Eq.~(\ref{d44}) holds for interacting
theories where the vacuum has lower energy due to chiral symmetry
breaking. In Fig. \ref{fig:24} we compare the energy density for
this state to the
ones discussed
above for the GN model, where translational symmetry had been
assumed. The crystal is always energetically favored,
the dependence on $p_f$ is now
convex, and there is no trace of a phase transition, neither first
nor second order, at any density. The hori\-zon\-tal slope at $p_f=0$
correctly
signals the presence of massless baryons and eliminates the
above-mentioned problems with the spurious massive, delocalized baryons.
We cannot even draw the corresponding picture for the 't~Hooft model,
simply because in this case the quark Fermi gas is infinitely
higher in energy than the Skyrme crystal for $L\to \infty$.
Nevertheless, all the results for baryonic matter discussed
in this section apply to the 't~Hooft model as well.

In the high density limit the oscillations
of the condensates become more and more rapid. If we are interested only
in length scales large as compared to $1/p_f$, the
condensates average to zero. In this sense, one might argue that
chiral symmetry gets restored at high density,
although not in the naive way suggested by Figs. \ref{fig:20} or \ref{fig:21}.

Finally, we remark that the ``chiral spiral" ground
state for fixed baryon density still preserves one continuous, unbroken
symmetry,
namely the combination of translation and chiral rotation generated by
$P+p_f Q_5$ ($P$: momentum operator, $Q_5$: axial charge).
One would therefore expect that RPA excitations on this ground
state\s\cite{Salcedo91,Lenz91} (or mesons in nuclear matter)
will have only one
collective, gapless mode, a hybrid of a ``phonon" and a ``pion".

\subsubsection{Non-vanishing bare quark masses}\label{subsubsec:nonvanishing}

As discussed in Sec. \ref{subsec:chiral}
following Ref. 44, a finite bare quark
mass can be included in the Skyrme type approach to the baryon; it changes
the effective action for the chiral phase
$\chi(x)$ from that of a free massless field into a sine-Gordon model. The baryon was identified with
the kink solution. It is then clear which type of solution is likely to be a good 
candidate for baryonic matter: The sine-Gordon kink crystal.
It goes over into the results of the previous section in the limit $m\to 0$
and is an approximate solution to the HF equations
(presumably a very accurate one, judging from the results of
Ref. 44 for the single baryon).
Luckily, the sine-Gordon kink crystal has already been studied thoroughly
in the literature, first in solid state
physics\s\cite{McMillan77,Theodorou78}
and more recently as a toy model for the Skyrme crystal,\cite{Takayama93}
in terms of Jacobi elliptic functions and elliptic
integrals.\cite{Abramowitz}
We take over the results from Ref. 116
which is close in spirit to the present study although the authors
apparently did not
have in mind two-dimensional large $N$ field theories.
Adapting the formulae of their work to our notation, the following
steps allow us to generalize the Skyrme crystal of the previous
section to small, finite bare quark masses: Let $m_{\pi}$ denote the
mass of the Goldstone boson, Eq.~(\ref{eq:GOR}), and $\bar{\rho}_B=
p_f/\pi$ the average baryon density (this is our definition of $p_f$
for the case of broken translational symmetry).
We then first have to solve the
transcendental equation
\begin{equation}
\frac{\pi m_{\pi}}{p_f} = 2 k {\bf K}(k)
\label{d50}
\end{equation}
for $k$ where ${\bf K}(k)$ is the complete elliptic integral of the
first kind. The sine-Gordon kink crystal is then given by the following
solution of Eq.~(\ref{eq:SG}),
\begin{equation}
\chi(x) = \frac{\pi}{2} + {\rm am}(\xi,k) \ , \qquad \xi=
\frac{m_{\pi}}{k}(x-x_0) \ ,
\label{d51}
\end{equation}
(${\rm am}(\xi,k)$ is the Jacobian elliptic amplitude function). From this,
we can
express the baryon density and the various condensates in terms of
further Jacobian elliptic
functions (${\rm dn}, {\rm sn}, {\rm cn}$) as
follows,
\begin{eqnarray}
\rho_B(x)&=& \frac{1}{\pi} \partial_x \chi(x) \ = \
\frac{m_{\pi}}{\pi k} {\rm dn}(\xi,k) \ ,
\nonumber \\
\langle \bar{q}q \rangle & = & \langle \bar{q}q \rangle_v
\cos 2 \chi(x) \ = \
- \langle \bar{q}q \rangle_v
\left( {\rm cn}^2(\xi,k)-{\rm sn}^2(\xi,k)\right) \ ,
\nonumber \\
\langle \bar{q}\,{\rm i}\gamma_5 q \rangle & = &
-\langle \bar{q}q \rangle_v \sin 2 \chi(x) \ = \
+ \langle \bar{q}q \rangle_v
2 {\rm sn}(\xi,k) {\rm cn}(\xi,k)\ .
\label{d52}
\end{eqnarray}
Here $\xi$ is as defined in Eq.~(\ref{d51}). Finally, the energy per volume of
this kind of matter is given by
\begin{equation}
\frac{{\cal E}}{N}= \frac{m_{\pi} p_f}{4 \pi^2} \left\{ \frac{8}{k} {\bf E}
(k) + 4 k \left(1-\frac{1}{k^2}\right) {\bf K}(k) \right\} \ ,
\label{d53}
\end{equation}
${\bf E}(k)$ denoting the complete elliptic integral of the second kind.

\begin{figure}[t]
  \begin{center}
    \epsfig{file=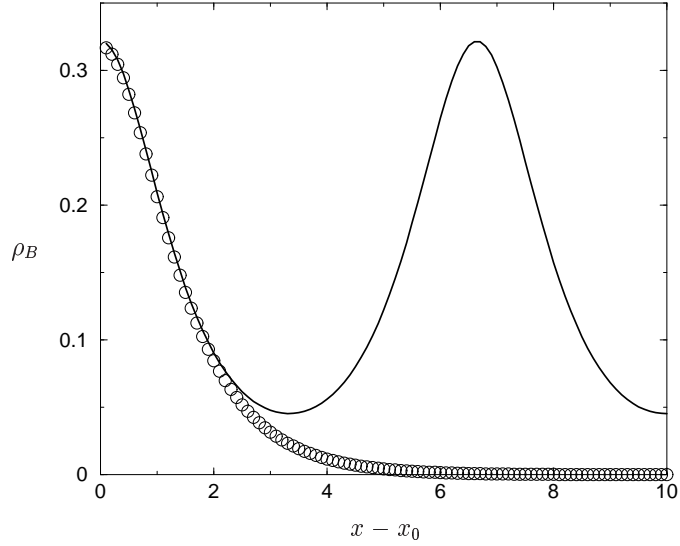,width=9cm}
    \caption{Solid curve: Spatial oscillation of the baryon density
      in the regime
$p_f \ll m_{\pi}$. Circles: Baryon density for a single baryon for
comparison.}
    \label{fig:25}
  \end{center}
\end{figure}
\begin{figure}
\parbox[t]{5.7cm}{
    \epsfig{file=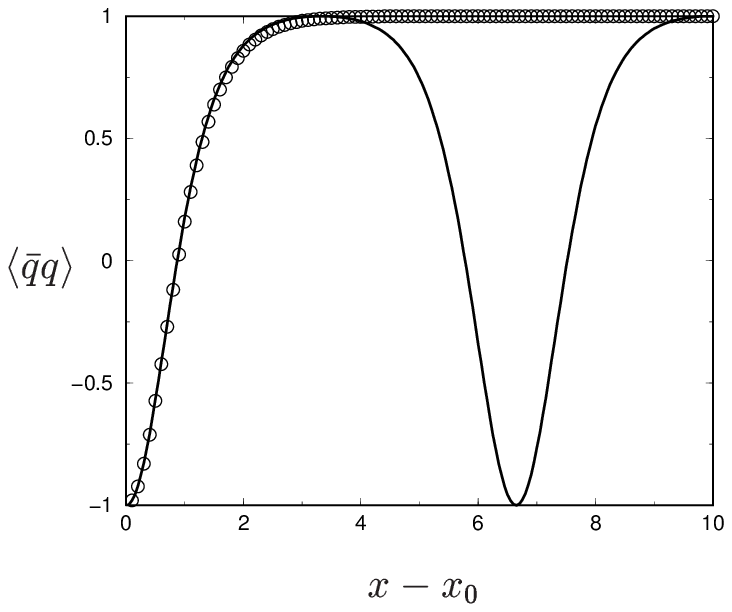,height=4.3cm}
    \caption{Solid curve: Spatial oscillation of the scalar
      chiral condensate in the 
regime $p_f \ll m_{\pi}$. Circles: Scalar chiral condensate
for a single baryon.}
    \label{fig:26}}
  \hfill
\parbox[t]{5.7cm}{
    \epsfig{file=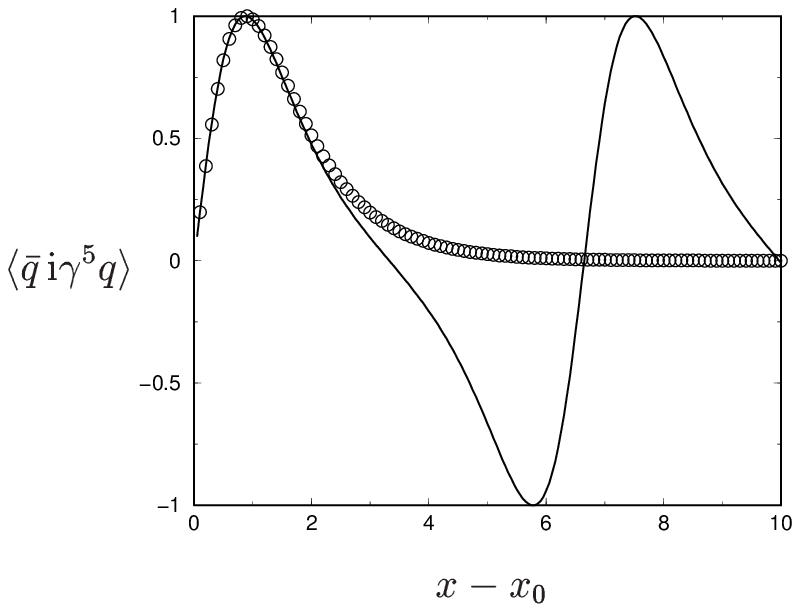,height=4.3cm}
    \caption{Same as Fig. 20, but for the pseudo-scalar chiral condensate.}
    \label{fig:27}}
  \end{figure}

  \begin{figure}[t]
    \begin{center}
      \epsfig{file=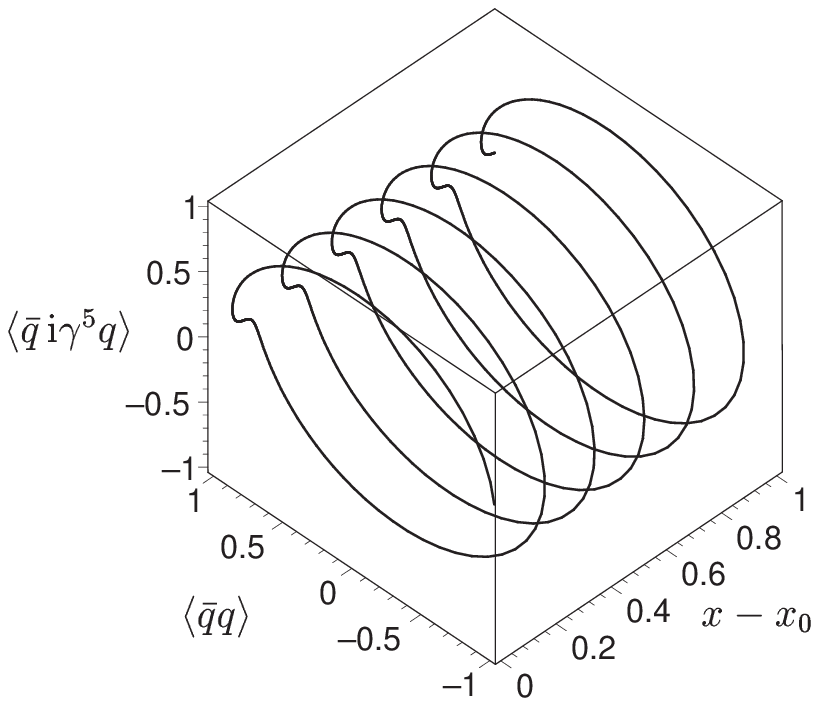}
      \caption{Illustration of the distorted ``chiral spiral"
        for baryonic matter at non-zero bare 
quark mass (sine-Gordon kink crystal).}
      \label{fig:28}
    \end{center}
  \end{figure}
\begin{figure}[t]
  \begin{center}
    \epsfig{file=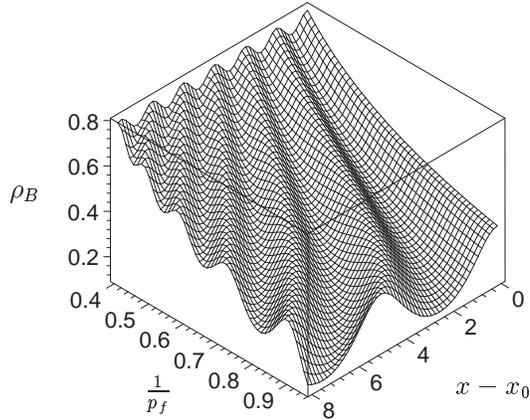,width=7cm}
    \caption{Spatial dependence of the baryon density as
      it evolves with increasing
average density (or Fermi momentum), in units of $m_{\pi}$.}
    \label{fig:29}
  \end{center}
\end{figure}

Let us now illustrate these results in two regimes of interest, namely
at low and high density. At low density ($p_f\ll m_{\pi}$), $k$ in
Eq.~(\ref{d50}) approaches 1 exponentially, and the baryon density
features a chain of well resolved lumps whose shape is determined by
the single kink solution (Fig. \ref{fig:25}). Likewise, the condensates behave
like those of a
single baryon: $\langle \bar{q}q \rangle$ changes
from the vacuum value outside the baryons to its negative in their center
whereas $\langle \bar{q}\,{\rm i}\gamma_5 q   \rangle $ is peaked in the
surface region of each baryon (Figs. \ref{fig:26}, \ref{fig:27}).
These condensates
are projections of the distorted ``chiral spiral" shown in Fig. \ref{fig:28}.
The energy (\ref{d53}) for low densities behaves as
\begin{equation}
{\cal E} \approx N \frac{2 m_{\pi}p_f}{\pi^2} = M_B \rho_B \ ,
\label{d54}
\end{equation}
showing the expected connection to the baryon mass. At high
densities ($p_f \gg m_{\pi}$), $k$ approaches 0 like
\begin{equation}
k \approx \frac{m_{\pi}}{p_f} \ .
\label{d55}
\end{equation}
Thus $\xi$ in Eq.~(\ref{d51}) becomes $p_f (x-x_0)$. Moreover, for $k\to 0$,
the
Jacobian elliptic functions ${\rm am}(\xi,k), {\rm sn}(\xi,k),{\rm  cn}(\xi,k)$
are known to reduce to the argument $\xi$ and the ordinary trigonometric
functions $\sin \xi$ and $\cos \xi$, respectively. We thus recover the
results for the simple chiral spiral in Sec. \ref{subsec:chiral}
(the parameter $x_0$ has to be readjusted to take care of the shift by $\pi/2$
in Eq.~(\ref{d51})). The energy in this case is approximately
\begin{equation}
\frac{{\cal E}}{N} \approx \frac{p_f^2}{2\pi} + \frac{m_{\pi}^2}{8\pi} \ .
\label{d56}
\end{equation}
The condensates look very much like the sin- and cos-functions of
the massless case and need
not be plotted. The baryon density oscillates around a constant value,
reflecting the strong overlap of the baryons, and are well approximated
at high density by
\begin{equation}
\rho_B(x) \approx \frac{p_f}{\pi} \left( 1-\frac{1}{2} \left( \frac{m_{\pi}}
{p_f}\right)^2 \sin^2p_f(x-x_0)\right) \ .
\label{d57}
\end{equation}
The behavior of the baryon density $\rho_B(x)$ as one increases
$p_f$ (i.e., the mean density) is
illustrated in Fig. \ref{fig:29}. In the chiral or high-density
limit ($m_{\pi}/p_f \to 0$) $\rho_B(x)$ eventually becomes $x$-independent.
This provides us with another way of understanding the structure of
matter described
in the previous section, namely as arising from a chain
of very extended, strongly overlapping lumps.

\subsubsection{Comparison to other works}\label{subsubsec:comparison}

It is noteworthy that a similar chiral structure of fermionic matter
has been reported previously in a variety of models different from
the ones discussed here.
This indicates that the basic results are more generally
valid than our derivation might suggest. 
Let us briefly go through these works to see what is common and what
is different from our case.

The first mention of nonuniform structures we are aware of is
Ref. 118 on the massive Schwinger model. Since in the
standard Schwinger model baryon charge is confined, one either has to 
add an inert, uniform background charge to neutralize the
system or introduce flavor. In the first case, Fisch\-ler {\em et al.}
find that
a spatially varying, periodic charge density is induced which they
interpret as 1+1 dimensional Wigner crystal.\cite{Wigner34,Schulz93}
More recently, the massless Schwinger model at finite 
chemical potential has also been 
examined and oscillating chiral condensates with wavenumber
$k=2\mu$ were found, by
bosonization\s\cite{Kao94} as well as by functional 
methods.\cite{Christiansen96a} 
In the massless case, the charge density becomes uniform. 
This phenomenon has obviously a lot in
common with our findings, although it is different in detail. In 
particular, the way to avoid conflict with the no-go theorem for
spontaneous symmetry breaking in two dimensions must be different
since one cannot invoke large
$N$ arguments; presumably, in the U(1) case, the anomaly and the
long range Coulomb interaction play a crucial role.
In the generalized Schwinger model with two flavors, one can have
baryonic charge
for neutral systems. Here, Fischler {\em et al.}\s\cite{Fischler79}
also arrive at a sine-Gordon
equation for the light meson but point out that a possible crystal
solution would be destroyed immediately by quantum fluctuations.
This is of course
avoided in our case by the large $N$ limit.

QCD$_2$ with flavor at finite chemical potential has been discussed 
by Christiansen {\em et al.}\s\cite{Christiansen96b} in the chiral limit. 
Oscillatory condensates where found within the path-integral approach.
Since these authors work at finite $N$, it is not clear how the no-go
theorem is avoided. On the other hand, since we know that QCD$_2$
does have massless mesons and baryons at finite $N$, one cannot
rule out that there is yet another way of bypassing the
Coleman theorem.

Even more surprising are perhaps quite a number of reports about
spatially inhomogeneous chiral
condensates with exactly the same wave number as in our case, but
in 3+1 dimensions. 
Kutschera {\em et al.}\s\cite{Kutschera90} have studied quark matter with pion 
condensates within the $\sigma$-model, using a mean-field approximation. Whereas their
quark matter is closely analogous to our Fermi gas,
the pion condensed state resembles the chiral spiral, especially
as far as the condensates are concerned. 
Another important work is the one by Deryagin
{\em et al.}\s\cite{Deryagin92} 
on large $N$ QCD in 3+1 dimensions. With the help of
a variational calculation, these
authors find an instability of cold and dense matter with respect to
formation of an inhomogenous, anisotropic condensate, a standing
wave with wave number $2\mu$. They point out that
unlike color superconductivity, this
condensation survives in the $N\to \infty$ limit.
Physically, they interpret their results as condensation of pairs
of particles and holes on opposite points of the Fermi surface
$|\vec{p}\,|=\mu$ (``Overhauser effect"\s\cite{Overhauser78}).
This scenario was
subsequently taken up by Shuster {\em et al.}\s\cite{Shuster99} and
Park {\em et al.}\s\cite{Park99} who elaborated
on the competition between BCS (pp or hh) pairing and Overhauser (ph) pairing.
They confirm that the large $N$ limit favors the chiral wave, although
it seems very unlikely to be relevant for $N$ values as low as 3.

Let us see how this pairing picture fits into the 1+1 dimensional models
considered here. Since relativistic HF is very similar to BCS theory, we
can interpret the vacuum result (as encoded in the Bogoliubov angles, cf.
Fig. \ref{fig:6}, Sec. \ref{subsubsec:vacuumth})
as a ph-pairing in a momentum region with a width of order
$m$ or $\sqrt{Ng^2}$ (for the GN and 't Hooft models, respectively)
and centered  around $p=0$.
The crystal solution is obtained by substituting
\begin{equation}
q(x) \to {\rm e}^{{\rm i} p_f x\gamma^5} q(x) \ ,
\label{d58}
\end{equation} 
thereby splitting momenta between right-handed and left-handed
fermion components by $\pm p_f$. Then,
provided that $p_f \gg (m, \sqrt{Ng^2})$, we can indeed describe
the result as ph 
pairing on opposite sides of the Fermi ``sphere" (here an interval).
However our results also hold for small $p_f$ where this picture is not
really applicable.

\section{Phase diagram in the ($T,\mu$)-plane}\label{sec:phasediag}

Most of the work done so far on the phase diagram for the two dimensional large $N$
models has
been devoted to the original, non-chiral GN model. Its amazingly
rich phase 
diagram has been 
discussed comprehensively by Wolff\s\cite{Wolff85} (for $m_b=0$)
and by Barducci {\em et al.}\s\cite{Barducci95}
(for finite bare quark masses). 
It has also been invoked for understanding some questions
about the statistical
physics of polyacetylene,\cite{Chodos94} a linear polymer. We are primarily
interested in the role
played by chiral symmetry and will therefore concentrate on the chiral
GN
model. Actually, as far as the conventional Fermi gas approach is concerned,
both variants of the GN model yield identical results to leading
order in
$1/N$.\cite{Barducci95}
Apart from briefly reviewing the common lore about the GN
model, we will also investigate the novel ``crystalline" phase (chiral
spiral, cf. Sec. \ref{subsec:breakdown})
and determine which phase is thermodynamically stable. 
In order to keep things technically as simple as possible, we consider
only the $m_b=0$ case
and work to leading order in the $1/N$ expansion throughout this section.
Incidentally, the 't Hooft model will be no issue here: Since in the
limit $N\to \infty$
confinement suppresses any $T$-dependence
(cf. Sec. \ref{subsec:thooftcyl}), there is
nothing we could add to what has already been said about $T=0$ and finite
density in Sec. \ref{sec:findens}.

\subsection{Chiral GN model with translational invariance}
\label{subsec:chiralGN}

Let us assume unbroken translational invariance and, without loss of
generality,
$\langle \bar{q}\,{\rm i}\gamma^5q\rangle=0$. If we are interested
in equilibrium thermodynamics at finite temperature and density, it
is convenient to work
with the grand canonical ensemble where the baryon density is adjusted
via a
Lagrange multiplier, the chemical potential $\mu$. The HF method can be
generalized without
any difficulty to finite $\mu$.\cite{Fetter} Thereby, the HF equation 
remains unchanged and equal to the free Dirac equation,
\begin{equation}
\left( \gamma^5 \frac{1}{{\rm i}}\frac{\partial}{\partial x} +
  m \gamma^0 \right)
\varphi_n(x) =\epsilon_n \varphi_n(x) \ ,
\label{freeDirac}
\end{equation} 
whereas the chemical potential only enters the gap equation
through the standard Fermion occupation numbers
\begin{equation}
-\frac{m}{Ng^2} = \frac{\langle \bar{q}q\rangle}{N} =
\sum_n \bar{\varphi}_n(x)\varphi_n(x)\frac{1}
{{\rm e}^{\beta(\epsilon_n-\mu)}+1} \ .
\label{gap}
\end{equation}
The solution of the HF equation is once more given by the free,
massive Dirac theory with
$\epsilon_n=\pm\epsilon(k)=\sqrt{k^2+m^2}$. The relevant thermodynamic 
function here is the grand canonical potential often referred to as   
``effective potential" in this context,    
\begin{equation}
V_{\rm eff} = - \frac{1}{\beta}\ln  Z = - \frac{1}{\beta}\ln
{\rm tr}\, {\rm e}^{-\beta (H-\mu N)} \ .
\end{equation}
Recalling
that the Hamiltonian of the GN model differs from that of the free
massive Dirac theory only by a $c$-number (in leading order in $1/N$) and
using the partition function for a free Fermi gas as follows,
\begin{eqnarray}
\ln Z & = & \ln \prod_n \left( 1+{\rm e}^{-\beta(\epsilon_n-\mu)}\right)\\
& = & L\int \frac{{\rm d}k}{2\pi}\ln \left[
{\rm e}^{\beta(\epsilon(k)+\mu)}
\left(1+{\rm e}^{-\beta(\epsilon(k)-\mu)}\right)
\left(1+{\rm e}^{-\beta(\epsilon(k)+\mu)}\right) \right] \ ,\nonumber
\end{eqnarray}
the effective potential density (${\cal V}_{\rm eff}=V_{\rm eff}/L$)
per color becomes
\begin{equation}
\frac{{\cal V}_{\rm eff}}{N} = \frac{m^2}{2Ng^2} -
\!\int \frac{{\rm d}k}{2\pi} \left\{ \epsilon(k) + \mu
  + \frac{1}{\beta}\ln \left[ \left( 1+{\rm e}^
      {-\beta(\epsilon(k)-\mu)}\right)\!\!
\left( 1 + {\rm e}^{-\beta (\epsilon(k)+\mu)} \right) \right]
\right\}\label{free}.
\end{equation}
In this way, the ``vacuum part" ($T=\mu=0$) can easily be identified
and separated
from the ``matter part" which in turn is made up of particle and anti-particle
contributions.
Incidentally, the term
\begin{equation}
- \int \frac{{\rm d}k}{2\pi} \mu
\label{term1}
\end{equation}
in eq. (\ref{free}) is a pure vacuum term which is usually omitted
in the literature; it arises from the infinite fermion density of the
Dirac sea.
The gap equation (\ref{gap}) can be further worked out to yield
\begin{equation}
m = \frac{Ng^2}{2\pi} \int {\rm d}k \frac{m}{\epsilon(k)}
\left(1- \frac{1}{{\rm e}^{\beta (\epsilon(k)-\mu)}+1}
- \frac{1}{{\rm e}^{\beta (\epsilon(k)+\mu)}+1}\right) \ ,
\label{gapeq}
\end{equation}
where we have used
\begin{equation}
\bar{u}(k)u(k)=-\bar{v}(k)v(k)= \frac{m}{\epsilon(k)} \ .
\end{equation}
Eq. (\ref{gapeq}) is equivalent to the
condition that the effective potential (\ref{free}) 
is stationary with respect to $m$,
\begin{equation}
\frac{\partial  {\cal V}_{\rm eff}}{\partial m}  = 0 \ .
\end{equation}
Renormalization can be performed at 
$T=0,\mu=0$ (physical fermion mass $m_0$) resulting in   
the renormalized potential
\begin{eqnarray}
\left. \frac{{\cal V}_{\rm eff}}{N}\right|_{\rm ren} &=&  \frac{m^2}{4\pi}
 \ln \left(\frac{m^2}{m_0^2}\right)+ \frac{m_0^2-m^2}{4\pi}\label{free_ren}\\
&&- \frac{1}{\beta} \int \frac{{\rm d}k}{2\pi}
\ln \left[ \left( 1 + {\rm e}^{-\beta(\epsilon(k)-\mu)}\right)
\left( 1 + {\rm e}^{-\beta(\epsilon(k)+\mu)}\right)\right] \ \nonumber.
\end{eqnarray}
By adding the ``bag pressure" $m_0^2/(4\pi)$ and dropping the 
term in Eq. (\ref{term1}), it has been adjusted to vanish in the vacuum.
The effective potential (\ref{free_ren}) contains
all the information about the conventional phase
diagram of the GN model. (For the path
integral derivation and the representation of ${\cal V}_{\rm eff}$
in terms of sums over Matsubara frequencies instead of
integrals over thermal occupation numbers,
we refer to Refs. 95, 110.)
For the sake of completeness, we record the expressions for standard
thermodynamical bulk observables: The free energy density ${\cal F}$ (or,
equivalently, minus the
pressure $p$) is equal to ${\cal V}_{\rm eff}$, Eq. (\ref{free_ren}).
The quark density $\rho$
and the entropy density $s$ are obtained by 
differentiating $p$ with respect to $\mu$ and $T$,
\begin{eqnarray}
\frac{\rho}{N} & = & \frac{1}{\pi} \int_0^{\infty}{\rm d}k
\left(\frac{1}{{\rm e}^{\beta(\epsilon(k)-\mu)}+1}
- \frac{1}{{\rm e}^{\beta(\epsilon(k)+\mu)}+1}\right) \ ,  \\
\frac{s}{N} & = & \frac{1}{\pi} \int_0^{\infty} {\rm d}k \ln \left[ 
\left( 1+ {\rm e}^{-\beta(\epsilon(k)-\mu)} \right)
\left( 1+ {\rm e}^{-\beta(\epsilon(k)+\mu)} \right) \right] \nonumber \\
& & + \frac{\beta}{\pi} \int_0^{\infty} {\rm d}k
\left( \frac{\epsilon(k)-\mu}{{\rm e}^{\beta (\epsilon(k)-\mu)}+1}
+ \frac{\epsilon(k)+\mu}{{\rm e}^{\beta (\epsilon(k)+\mu)}+1}\right)  \ ,
\end{eqnarray}
whereas the internal energy density can now be deduced via
\begin{equation}
\varepsilon =   Ts-p+\mu \rho \label{eq:inten}\ .
\end{equation}
Needless to say, all of these observables have to be evaluated
in the absolute minimum
of the effective potential with respect to $m$, i.e., by solving the gap 
equation.

\begin{figure}[t]
  \begin{center}
    \epsfig{file=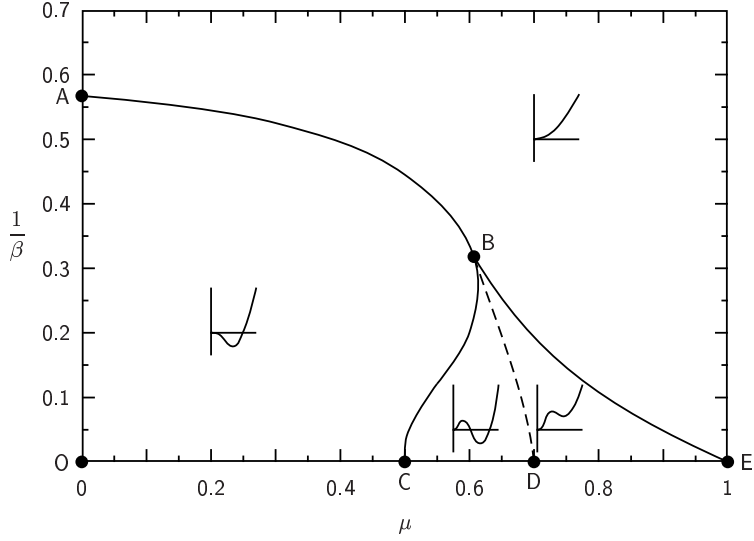,width=10cm}
    \caption{Phase diagram of the (chiral or non-chiral) GN models, assuming
unbroken translational symmetry. For a discussion, see text below
Eq. (\ref{eq:inten}).
Units of $m_0$; adapted from Ref. 95.}
    \label{fig:30}
  \end{center}
\end{figure}

The phase diagram which one obtains in this way is depicted in
Fig. \ref{fig:30}, adapted from Ref. 95. 
The line {\sffamily AB} is a critical line of second order transitions
(${\cal V}_{\rm eff}$ changes from one minimum to a
maximum and a minimum).
The point {\sffamily B} is a tricritical point located at
$1/\beta = 0.3183,\ \mu=0.6082$ (all numbers here 
are in units of $m_0$); it separates the second order line from a
first order line {\sffamily BD}
along which ${\cal V}_{\rm eff}$ has two degenerate minima and a maximum.
The endpoint {\sffamily D} lies at $\mu=1/\sqrt{2}$ where the $T=0$ phase
transition occurs, cf. Sec. \ref{subsec:GNFermi}.
Lines {\sffamily BC} and {\sffamily BE} are boundaries
of metastability; when crossed,
the potential acquires or loses a second minimum;
point {\sffamily C} is at $\mu=1/2$.
In region {\sffamily OABD},
chiral symmetry is broken and the quarks are massive;
the outside region has unbroken chiral symmetry
and massless 
fermions.

\begin{figure}[t]
  \parbox[t]{5.6cm}{
    \epsfig{file=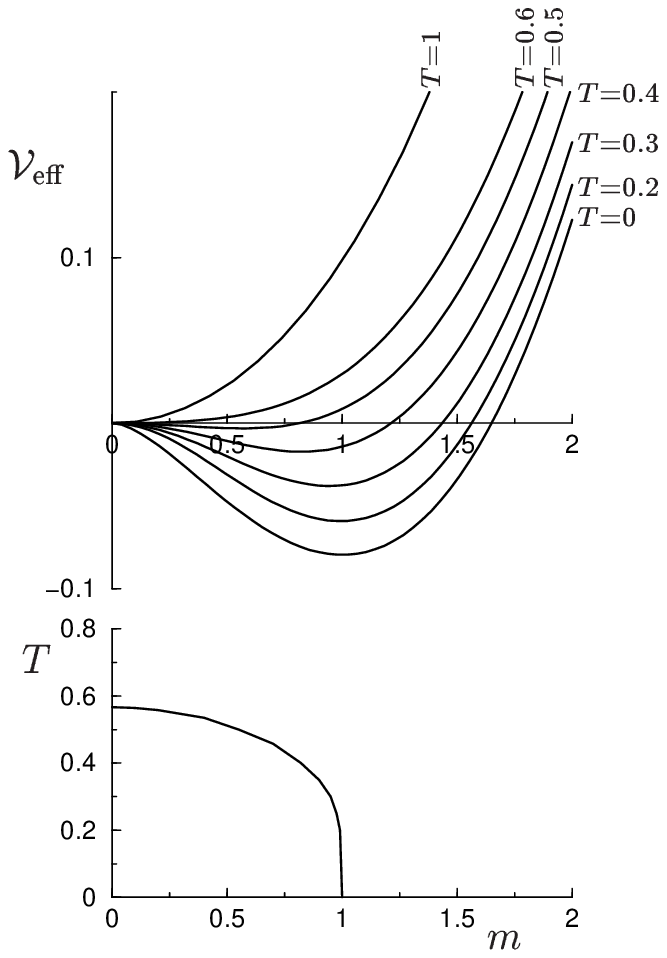,width=5.6cm}
    \caption{Upper graph: Effective potential density as a function of dynamical
fermion mass $m$ at $\mu=0$ for various temperatures below and
above $T_c \approx 0.567$. Lower graph: $m$ as
a function of $T$ (i.e., position of minimum in upper curves).
Second order phase transition.}
    \label{fig:31}}
  \hfill
  \parbox[t]{5.6cm}{
    \epsfig{file=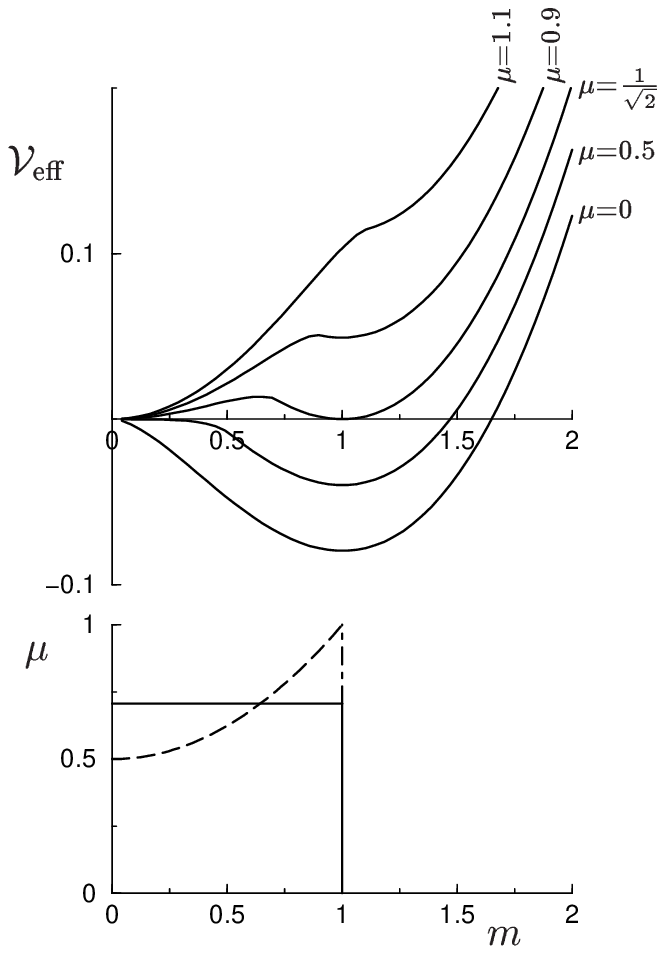,width=5.6cm}
    \caption{Upper graph: Effective potential density as a function of dynamical
fermion mass $m$ at $T=0$ for various chemical potentials
below and above $\mu_c=1/\sqrt{2}$. Lower graph: $m$ as a
function of $\mu$ (dashed line: position of
maximum, corresponding to Fig. 14; solid line: position of minimum in upper
curves). First order phase transition.}
    \label{fig:32}}  
\end{figure}

In order to highlight the different behavior of the system when
going through a
first and second order phase transition, respectively, we have
illustrated in Figs. \ref{fig:31} and \ref{fig:32} the evolution of
${\cal V}_{\rm eff}$ along the $\mu=0$ and
$T=0$ axes (this
corresponds to the special cases discussed in Secs. \ref{sec:finT}
and \ref{sec:findens}, respectively).
Also
shown is the behavior of the physical fermion mass. In this way, the unstable
HF solution which we have found at finite density (with continuously
decreasing mass,
see Sec. \ref{subsec:GNFermi}) can be identified as corrsponding to a local
maximum of
${\cal V}_{\rm eff}$.
In preparing Fig. \ref{fig:32}, we have taken advantage of the fact that the
integration in 
Eq. (\ref{free_ren}) can be carried out at $T=0$ with the result
\begin{eqnarray}
\frac{{\cal V}_{\rm eff}}{N} &=&  \frac{m^2}{4\pi}
 \ln \left( \frac{m^2}{m_0^2}\right)  + \frac{m_0^2-m^2}{4\pi}\\
&&+\frac{1}{2\pi}\theta(\mu-m)\left[m^2 \ln 
\left(\frac{\mu + \sqrt{\mu^2-m^2}}{m}\right)-\mu \sqrt{\mu^2-m^2}\right]
\nonumber\ .
\end{eqnarray}
\subsection{Chiral spiral at finite temperature}\label{subsec:chiralspiral}

We have seen before that the translationally non-invariant finite density
solution (constant baryon
density, but oscillating condensates in the chiral limit) of the HF equation
at $T=0$ can
be obtained from the vacuum solution by a space dependent chiral rotation of
the quarks,
cf. Sec. \ref{subsec:chiral}.
As already noted by Kao {\em et al.}\s\cite{Kao94} for the
massless Schwinger model and by 
Sadzikowski {\em et al.}\s\cite{Sadzikowski00} for the $\sigma$-
and NJL models, the same
procedure works at finite
temperature as well. We show that the Fermi gas solution at zero chemical
potential and finite temperature can be mapped onto the chiral spiral
type solution at 
finite $\mu$ and the same $T$. This will allow
us to follow the fate of the crystalline solution as we heat up
baryonic matter.

We start from the HF equation at finite $T,\mu$, but neither
invoking translational invariance
nor assuming that $\langle \bar{q}\,{\rm i}\gamma^5 q\rangle$ vanishes,  
\begin{equation}
\left\{ \gamma^5 \frac{1}{\rm i}\frac{\partial}{\partial x} -
g^2 \left( \langle \bar{q}(x)q(x)\rangle \gamma^0 + \langle
\bar{q}(x) {\rm i}\gamma^5 q(x)\rangle {\rm i}\gamma^1 \right) \right\}
\varphi_n(x) = \omega_n \varphi_n(x) \ .
\end{equation}
As before,
temperature and chemical potential only show up in the
self-con\-sis\-ten\-cy relations
which now read
\begin{eqnarray}
\langle \bar{q}(x)q(x)\rangle &=& N \sum_n \bar{\varphi}_n(x) \varphi_n(x)
\frac{1}{{\rm e}^{\beta(\omega_n-\mu)}+1} \ , \nonumber \\
\langle \bar{q}(x){\rm i}\gamma^5 q(x)\rangle &=& N \sum_n \bar{\varphi}_n(x)
{\rm i} \gamma^5 \varphi_n(x)
\frac{1}{{\rm e}^{\beta(\omega_n-\mu)}+1} \ .
\label{eq2}
\end{eqnarray}
Guided by the $T=0$ results (cf. Sec. \ref{subsubsec:strict}),
we make the following ansatz for the $x$-dependence of the
scalar and pseudo-scalar
condensates,
\begin{equation}
\langle \bar{q}(x)q(x)\rangle = -\frac{m}{g^2}\cos (2 \mu x)
\ ,\;
\langle \bar{q}(x){\rm i}\gamma^5 q(x)\rangle = \frac{m}{g^2} \sin
(2 \mu x) \ .
\label{cond1}
\end{equation}
The HF equation for this particular potential can be turned into
\begin{equation}
\left\{ \gamma^5 \frac{1}{\rm i}\frac{\partial}{\partial x} +
m {\rm e}^{{\rm i}\mu x \gamma^5}\gamma^0 {\rm e}^{-{\rm i}\mu
x \gamma^5}\right\}\varphi_n(x) = \omega_n \varphi_n(x) 
\end{equation}
where we have used the fact that
\begin{equation}
{\rm e}^{{\rm i}\mu x \gamma^5} \left( \begin{array}{c}
\gamma^0 \\ {\rm i} \gamma^1 \end{array} \right)
{\rm e}^{-{\rm i}\mu x \gamma^5}=
\left( \begin{array}{rr} \cos 2 \mu x & -\sin 2 \mu x \\
\sin 2 \mu x & \cos 2 \mu x \end{array} \right)
\left( \begin{array}{c} \gamma^0 \\ {\rm i} \gamma^1 \end{array}\right) \ .
\end{equation}
A local chiral rotation of the form
\begin{equation}
\varphi_n(x)= {\rm e}^{{\rm i} \mu x \gamma^5}\chi_n(x) 
\end{equation}
then eliminates the $x$-dependence from the HF potential,
\begin{equation}
\left\{ \gamma^5 \frac{1}{\rm i}\frac{\partial}{\partial x} + \mu
+ m \gamma^0 \right) \chi_n(x)=\omega_n \chi_n(x)  \ ,
\end{equation}
and enables us to solve the HF equation trivially,
\begin{eqnarray}
\chi_n(x) & = & {\rm e}^{{\rm i}kx} u (k) \quad \quad (\omega_n=
\epsilon(k)+\mu) \ , \nonumber \\
\chi_n(x) & = & {\rm e}^{{\rm i}kx} v (k) \quad \quad (\omega_n=
-\epsilon(k)+\mu) \ ,
\label{solveHF}
\end{eqnarray}
with $\epsilon(k) = \sqrt{k^2+m^2}$ 
and $u(k),v(k)$ free, massive spinors.
Since 
\begin{equation}
\chi_n^{\dagger}(x) {\rm i}\gamma^1 \chi_n(x) = 0 \ ,
\end{equation}
the two self-consistency conditions (\ref{eq2})
collapse into the single condition
\begin{equation}
- \frac{m}{Ng^2} = \sum_n \chi_n^{\dagger}(x) \chi_n(x) \frac{1}
{{\rm e}^{\beta (\omega_n- \mu)}+1} \ .
\end{equation}
The additive term $\mu$ in $\omega_n$, Eq. (\ref{solveHF}),
just cancels the chemical potential and we have mapped
our problem exactly onto the $\mu=0$, translational invariant HF
problem at finite $T$,
\begin{equation}
- \frac{m}{Ng^2} = \sum_n \chi_n^{\dagger}(x) \chi_n(x) \frac{1}
{{\rm e}^{\beta \epsilon_n}+1} \ , \qquad \epsilon_n=
\omega_n-\mu=\pm \sqrt{k^2+m^2} \ . 
\end{equation}
The parameter $m$ therefore has to be identified with the
physical fermion mass of the translationally
invariant vacuum at $\mu=0$ and the same finite $T$.
We can thereby construct a translationally broken solution of the HF
problem for any $(\beta, \mu)$; it can be pictured as a chiral
spiral whose radius
shrinks with increasing temperature until it disappears at
$T_c=m_0 {\rm e}^{\rm C} /\pi$, cf. Sec. \ref{subsec:GNcyl}.

Let us determine the free energy of this new phase.
The density of the effective potential 
per color is still given by eq. (\ref{free}), or equivalently by
\begin{equation}
\frac{{\cal V_{\rm eff}}}{N} =  \frac{m^2}{2Ng^2} -\frac{1}{L\beta}
\sum_n \ln \left( 1 + {\rm e}^{-\beta(\omega_n-\mu)}\right) \ .
\end{equation}
The double counting term depends only on the radius of the chiral
circle and is therefore unchanged (cf. Eq. (\ref{cond1})),
\begin{equation}
\langle \bar{q} q \rangle^2 + \langle \bar{q} \, {\rm i} \gamma^5
q \rangle^2 = \frac{m^2}{g^4} \ .
\end{equation}
In view of the relation  $\omega_n-\mu = \epsilon_n$,
it appears superficially as if nothing depended on $\mu$ and
we got the same result as for the translationally invariant
system
at zero chemical potential. As we know from the $T=0$
case, this argument is too rough --- the $\mu$-dependence is intimately
related to the UV regularization. Let us impose an UV-cutoff
on the single particle energies at the bottom of the Dirac sea,
\begin{equation}
 \omega_n > -\Lambda/2 \  \qquad \Leftrightarrow \qquad 
|k|< \Lambda/2 + \mu \ .
\end{equation}
The entire $\mu$-dependence of ${\cal V}_{\rm eff}$ arises from this
shift in the cutoff. We can then easily evaluate the difference
between the free energy density of the ``chiral spiral" at $(\beta,\mu)$
and the standard solution at $(\beta,0)$,
\begin{eqnarray}
\left. \frac{{\cal V}_{\rm eff}(\beta,\mu)}{N} \right|_{\rm spir} \left.
-\frac{{\cal V}_{\rm eff}(\beta,0)}{N}\right|_{\rm stand} & = &
- \int_{-(\Lambda/2+\mu)}^{\Lambda/2+\mu} \frac{{\rm d}k}{2\pi}\sqrt{k^2+m^2}\nonumber\\
&&+ \int_{-\Lambda/2}^{\Lambda/2} \frac{{\rm d}k}{2\pi} \sqrt{k^2+m^2}
\nonumber \\
& \approx & -\frac{(\Lambda+\mu)\mu}{2\pi} \ .
\end{eqnarray}
This result does not depend on temperature, a feature characteristic
for an UV effect. The fermion density per color can be
obtained by differentiation with respect to $\mu$; we find the same result
as at zero temperature,
\begin{equation}
\frac{\rho}{N} = - \frac{\partial}{\partial \mu} \frac{{\cal V}_{\rm eff}}{N}
=  \frac{\Lambda}{2\pi} + \frac{\mu}{\pi} \ ,
\label{rhoN}
\end{equation}
hence, with the definition of the Fermi momentum adopted in Sec.
\ref{subsubsec:strict},
\begin{equation}
\mu = p_f
\end{equation}
(the term $\sim \Lambda$ in Eq. (\ref{rhoN}) is the fermion
density of the Dirac sea).
Summarizing, we can now compare the effective potentials for two
solutions of the HF equation at finite $(\beta, \mu)$. The
Fermi gas solution with unbroken translational invariance has already been
given in Eq. (\ref{free_ren}). The crystalline phase on the other hand yields
\begin{equation}
\left. \frac{{\cal V}_{\rm eff}}{N} \right|_{\rm spir}  = 
\frac{m^2}{4\pi}\ln  \left(  \frac{m^2}{m_0^2}\right)+ \frac{m_0^2-m^2}{4\pi}
- \frac{2}{\beta} \int \frac{{\rm d}k}{2\pi}
\ln  \left( 1 + {\rm e}^{-\beta \epsilon(k)} \right) -\frac{\mu^2}{2\pi} \ ,
\end{equation}
where we have again dropped the term $-\Lambda \mu/(2\pi)$.
In both cases, one still has to minimize the effective potential with respect
to $m$. The depth of the respective minima (or physically minus
the pressure) allows us to decide which phase is the stable one.
The chiral spiral is favored
everywhere over the homogeneous phase, below the critical
temperature.  This can be shown by a straightforward numerical
evaluation, but also deduced analytically from the simple
relation
\begin{equation}
\left. {\cal V}_{\rm eff}(\beta,\mu) \right|_{\rm spir} =
{\cal V}_{\rm eff}(\beta, 0)
-N\frac{\mu^2}{2\pi}\ .\label{Vspir}
\end{equation}
As the additional term in Eq. (\ref{Vspir}) does not depend on
$m$, both $\left. {\cal V}_{\rm eff}(\beta,\mu) \right|_{\rm spir}$
and ${\cal V}_{\rm eff}(\beta, 0)$ are minimized by the same $m$
\begin{equation}
m_{\rm spir}(\beta,\mu)=m(\beta,0)\ .
\end{equation}
The free energy density of the spiral at $\beta$ and $\mu$
evaluated at its minimum $m_{\rm spir}(\beta,\mu)$ is
thus equal to
\begin{equation}
\left. {\cal V}_{\rm eff}(\beta,\mu) \right|_{{\rm spir},m=m_{\rm spir}}
=\left.{\cal V}_{\rm eff}(\beta, 0)\right|_{m=m(\beta,0)}
-N\frac{\mu^2}{2\pi}\ .\label{Vspirmin}
\end{equation}
$\left.{\cal V}_{\rm eff}(\beta, \mu)\right|_{m=m(\beta,\mu)}$
is a monotonously decreasing function of $\mu$ at fixed $\beta$
(its derivative with respect to $\mu$ at fixed $\beta$ is
just $-\rho$).
\begin{figure}[t]
  \begin{center}
    \epsfig{file=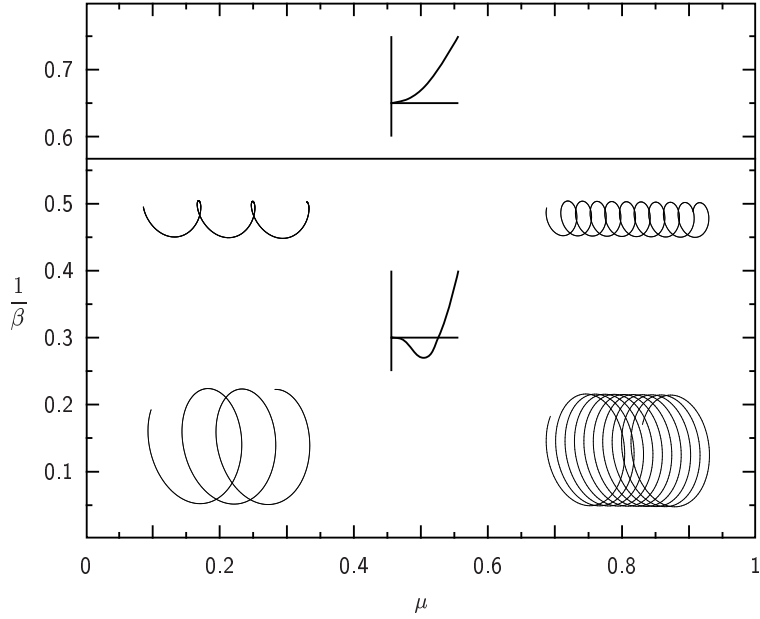,width=10cm}
    \caption{Phase diagram of chiral GN model, based on spatially inhomogeneous
solution. The horizontal line at $T_c \approx 0.567$ is a second order
line of chiral symmetry restoration. The dependence of radius and 
helix angle of the chiral spiral on $\mu$ and $T$ is
indicated qualitatively by means of the inserted drawings.}
    \label{fig:33}
  \end{center}
\end{figure}
Furthermore from the phase diagram (Fig. \ref{fig:30}) it
is clear that for any fixed $\beta$ there is a critical
chemical potential $\mu_c$, where chiral symmetry gets restored.
Above this critical chemical potential,
${\cal V}_{\rm eff}(\beta, \mu)$ evaluated at its minimum
just behaves like the energy density of free fermions in
1+1 dimensions
\begin{equation}
\left.{\cal V}_{\rm eff}(\beta, \mu)\right|_{m=m(\beta,\mu)}
= - N \frac{\mu^2}{2\pi}\; {\rm for}\; \mu \ge \mu_c\ .\label{Vfree}
\end{equation}
Comparing Eq. (\ref{Vspirmin}) to
$\left.{\cal V}_{\rm eff}(\beta, \mu)\right|_{m=m(\beta,\mu)}$
it is clear that the chiral spiral is always lower in free energy than
the homogeneous phase.

Eq. (\ref{Vspir}) is also useful if one wants to evaluate
the thermodynamic observables for the crystal phase
\begin{eqnarray}
\left. \rho(\beta,\mu)\right|_{\rm spir} & = &N \frac{\mu}{\pi} \ ,\nonumber \\
\left. p(\beta,\mu)\right|_{\rm spir} & = & p(\beta,0)+N\frac{\mu^2}
{2\pi}\ ,  \nonumber \\
\left. s(\beta,\mu)\right|_{\rm spir} & = & s(\beta,0) \ , \nonumber \\
\left. \varepsilon(\beta,\mu)\right|_{\rm spir} & =
& \varepsilon(\beta,0)+N\frac{\mu^2}{2\pi}\ .
\end{eqnarray}
(At $\mu=0$, there is no distinction between the Fermi gas
and the other phase.)
Since the translationally broken phase is the thermodynamically stable phase
wherever it exists, the phase diagram of the 
chiral GN model has only one horizontal second order line at $T=T_c$
where chiral symmetry is restored and the radius of the chiral spiral 
shrinks to zero (Fig. \ref{fig:33}); no phase transition occurs if we
increase the chemical potential.
Note that our results are rather different from a recent study of a non-uniform
chiral phase in 3+1 dimensional effective field theories ($\sigma$-model, 
NJL model\s\cite{Sadzikowski00}). In our language, these authors
use the radius and the
helix angle of the chiral spiral as independent variational parameters.
Since in 1+1 dimensions baryon
number is topologically connected to winding number, we do not seem
to have this
option but can vary only with respect to the radius (or $m$).
 
As a last remark, we wish to comment on the original GN model
with only discrete chiral symmetry. The chiral spiral does not exist here and
one might think that our analysis does not have anything to say about it.
However, the criticism of Sec. \ref{subsec:GNFermi} also
applies here. As the non-chiral
GN model has (massive) bound baryons, the low density behavior 
of the energy obtained in standard HF approximation cannot be correct and 
the phase diagram has to be reconsidered. This is a harder task than for
the chiral model since here dynamics is not completely dominated by
chiral symmetry.
It is quite possible that the first order phase transition survives, but this
still has to be worked out in detail.

\section{Closing remarks}\label{sec:outlook}
 
In this paper, we have illustrated the kind of insights one
gets from studies of exactly
soluble field theories by means of two classic examples, the
GN and the 't Hooft model.
The undeniable success of these models stems from a fortunate
combination of two very
bold substitutions: Replacing the number of spatial dimensions
by 1 and the number 
of colors by $\infty$ (ironically, both are 3 in real QCD).
Without the first choice,
there is no solvability; without the 2nd, there are hardly
enough interesting phenomena
due to rigorous no-go theorems.
If one is willing to swallow both of these bitter pills,
one is awarded with a   
striking number of analogies between these toy models and
the real world which have led
to a long list of QCD-motivated applications. These
analogies include such central 
features of QCD as asymptotic freedom, dimensional transmutation,
chiral symmetry breaking,
and confinement. In view of the fact that these models have the simplest
Lagrangians of certain types which one can write down and  are not
contrived in any sense,
this is quite amazing and underlines their pedagogical value for
QCD related questions.
Nowadays, it may well be that the study of lower dimensional field
theories is a better
training ground for students interested in QCD than a standard
field theory course based
on Feynman diagrams and perturbative QED.

Throughout this work we have tried to emphasize the fact that
the solution of these models
can be found by rather elementary means --- we have deliberately
chosen the relativistic HF
approach as our main tool. This does not mean though that there
are no subtleties:
We only mention the emergence of baryon number from the UV regularization, or 
how zero mode gluons affect the thermodynamics of quarks. Such features 
make the use of analytical methods particularly important and set
the scale for
the even more subtle properties to be expected from 4-dimensional QCD. 

As far as our more specific subject \  --- finite temperature and
density behavior --- \  is concerned,
we have found that even after more than 25 years these models are
still good for some
surprises. The strongest impression for us is the way in which chiral symmetry
governs not only the hadron spectrum, but also the phase structure at
finite density
and temperature. Confinement is most clearly seen in the temperature
dependence
of observables, whereas the dependence on the chemical potential is dominated
by chiral symmetry. Other non-trivial insights are related to the realization
of the Skyrme picture in a precise fashion, a triple point in the
phase diagram
of the non-chiral GN model and inhomogeneous chiral condensates. 
As a by-product, we have resolved some puzzles in the literature
and identified issues
where further work is needed, for instance the phase diagram of
the non-chiral 
GN model.

In conclusion, we see no indication that models such as the GN
or 't Hooft model
are anywhere close to having exhausted their potential as a
theoretical laboratory for QCD.

\section*{Acknowledgements}
 
We are grateful to Misha Shifman for the invitation to contribute to this
Festschrift in honor of Boris Ioffe, thereby forcing us to clarify some issues
before a certain deadline.
M.T. thanks his collaborators, colleagues and students
for most enjoyable and useful interactions on 2d field theory, in particular
M. Burkardt, V. Dolman, M. Engelhardt, O. Jahn, M. Kurz, F. Lenz, S. Levit,
R. Oderkerk, K. Ohta, R. Pausch, C. Satzinger, B. Schreiber,
M. Seeger, A. Smilga,
D. Stoll, R. Weing\"artner, S. W\"orlen, K. Yazaki, and I. Zahed.

\section*{References}

\bibliographystyle{unsrt}

\begin{thebibliography}{99}

\bibitem{Seiberg94}
N. Seiberg and E. Witten, {\em Nucl. Phys.} B {\bf 426}, 19 (1994).

\bibitem{Schwinger62}
J. Schwinger, {\em Phys. Rev.} {\bf 128}, 2425 (1962).

\bibitem{Gross74}
D.J. Gross and A. Neveu, {\em Phys. Rev.} D {\bf 10}, 3235 (1974).

\bibitem{tHooft74b}
G. 't Hooft, {\em Nucl. Phys.} B {\bf 75}, 461 (1974).

\bibitem{Hartree28}
D.R. Hartree, {\em Proc. Camb. Phil. Soc.} {\bf 24}, 89 (1928).

\bibitem{Fock30}
V.A. Fock, {\em Z. Phys.} {\bf 61}, 126 (1930).

\bibitem{Blau93}
M. Blau and G. Thompson, {\em Trieste HEP Cosmol.}, 0175-244 (1993).

\bibitem{Abdalla96}
E. Abdalla and M.C.B. Abdalla, {\em Phys. Rep.} {\bf 265}, 253 (1996).

\bibitem{Brodsky98}
S.J. Brodsky, H.-C. Pauli, and S.S. Pinsky, {\em Phys. Rep.} {\bf 301}, 299 (1998).

\bibitem{Nakanishi}
N. Nakanishi and I. Ojima, {\em Covariant operator formalism of gauge theories and quantum gravity}, World Scientific 1991.

\bibitem{Coleman73}
S. Coleman, {\em Comm. Math. Phys.} {\bf 31}, 259 (1973).

\bibitem{Itzykson}
C. Itzykson and J.-B. Zuber, {\em Quantum field theory}, McGraw-Hill 1980. 

\bibitem{Mermin66}
N.D. Mermin and H. Wagner, {\em Phys. Rev. Lett.} {\bf 17}, 1133 (1966).

\bibitem{Huang}
K. Huang, {\em Quantum Field Theory. From Operators to Path Integrals}, John Wiley 1998, ch. 18.

\bibitem{Witten78}
E. Witten, {\em Nucl. Phys.} B {\bf 145}, 110 (1978).

\bibitem{Affleck86}
I. Affleck, {\em Nucl. Phys.} B {\bf 265}, 448 (1986).

\bibitem{Kosterlitz73}
J.M. Kosterlitz and D.J. Thouless, {\em J. Phys.} C {\bf 6}, 1181 (1973).

\bibitem{Kosterlitz74}
J.M. Kosterlitz, {\em J. Phys.} C {\bf 7}, 1046 (1974).

\bibitem{Berezinsky71}
V.L. Berezinsky, {\em Sov. Phys. JETP} {\bf 32}, 493 (1971).

\bibitem{Manton85}
N.S. Manton, {\em Ann. Phys. (N.Y.)} {\bf 159}, 220 (1985).

\bibitem{Rajeev88}
S.G. Rajeev, {\em Phys. Lett.} B {\bf 212}, 203 (1988).

\bibitem{Hetrick89}
J.E. Hetrick and Y. Hosotani, {\em Phys. Lett.} B {\bf 230}, 88 (1989).

\bibitem{Hetrick93}
J.E. Hetrick, {\em Int. J. Mod. Phys.} A {\bf 9}, 3153 (1994).

\bibitem{Lenz94}
F. Lenz, H.W.L. Naus, and M. Thies, {\em Ann. Phys. (N.Y.)} {\bf 233}, 51 (1994).

\bibitem{Minahan93a}
J.A. Minahan and A.P. Polychronakos, {\em Phys. Lett.} B {\bf 326}, 288 (1994).

\bibitem{Svetitsky86}
B. Svetitsky, {\em Phys. Rep.} {\bf 132}, 1 (1986).

\bibitem{Lenz95}
F. Lenz, M. Shifman, and M. Thies, {\em Phys. Rev.} D {\bf 51}, 7060 (1995).

\bibitem{Engelhardt95b}
M. Engelhardt and B. Schreiber, {\em Z. Phys.} A {\bf 351}, 71 (1995).

\bibitem{Dhar94a}
A. Dhar, G. Mandal, and S.R. Wadia, {\em Nucl. Phys.} B {\bf 436}, 487 (1995).

\bibitem{Seeger98}
M. Seeger and M. Thies, {\em Phys. Rev.} D {\bf 58}, 027701 (1998).

\bibitem{Dhar94b}
A. Dhar, G. Mandal, and S.R. Wadia, {\em Phys. Lett.} B {\bf 329}, 15 (1994).

\bibitem{Dhar94c}
A. Dhar, P. Lakdawala, G. Mandal, and S.R. Wadia, {\em Int. J. Mod. Phys.} A {\bf 10}, 2189 (1995).

\bibitem{Thirring58}
W.E. Thirring, {\em Ann. Phys. (N.Y.)} {\bf 3}, 91 (1958).

\bibitem{tHooft74a}
G. 't Hooft, {\em Nucl. Phys.} B {\bf 72}, 461 (1974).

\bibitem{Coleman79}
S. Coleman, Erice lecture 1979, in: {\em Aspects of symmetry}, Cambridge University press 1985, p. 351.

\bibitem{Witten79}
E. Witten, {\em Nucl. Phys.} B {\bf 160}, 57 (1979).

\bibitem{Manohar98}
A.V. Manohar, in: {\em Probing the Standard Model of Particle Interactions}, F. David and R. Gupta, eds. (Les Houches 1997).

\bibitem{Nambu60}
Y. Nambu and G. Jona-Lasinio, {\em Phys. Rev.} {\bf 122}, 345 (1960); {\bf 124}, 246 (1961).

\bibitem{Zamolodchikov78}
A.B. Zamolodchikov and A.B. Zamolodchikov, {\em Phys. Lett.} B {\bf 72}, 481 (1978).

\bibitem{Andrei79}
N. Andrei and J.H. Lowenstein, {\em Phys. Rev. Lett.} {\bf 43}, 1698 (1979).

\bibitem{Andrei80}
N. Andrei and J.H. Lowenstein, {\em Phys. Lett.} B {\bf 90}, 106 (1980); B {\bf 91}, 401 (1980).

\bibitem{Dashen75b}
R.F. Dashen, B. Hasslacher, and A. Neveu, {\em Phys. Rev.} D {\bf 12}, 2443 (1975).

\bibitem{Feinberg94}
J. Feinberg, {\em Phys. Rev.} D {\bf 51}, 4503 (1995).

\bibitem{Salcedo91}
L.L. Salcedo, S. Levit, and J.W. Negele, {\em Nucl. Phys.} B {\bf 361}, 585 (1991).

\bibitem{Lenz91}
F. Lenz, M. Thies, S. Levit, and K. Yazaki, {\em Ann. Phys. (N.Y.)} {\bf 208}, 1 (1991).

\bibitem{Pausch91}
R. Pausch, M. Thies, and V.L. Dolman, {\em Z. Phys.} A {\bf 338}, 441 (1991).

\bibitem{Gross73}
D.J. Gross and F. Wilczek, {\em Phys. Rev. Lett.} {\bf 30}, 1343 (1973).

\bibitem{Politzer73}
H.D. Politzer, {\em Phys. Rev. Lett.} {\bf 30}, 1346 (1973).

\bibitem{Wetzel85}
W. Wetzel, {\em Phys. Lett.} B {\bf 153}, 297 (1985).

\bibitem{Luperini91}
C. Luperini and P. Rossi, {\em Ann. Phys. (N.Y.)} {\bf 212}, 371 (1991).

\bibitem{Shei76}
S. Shei, {\em Phys. Rev.} D {\bf 14}, 535 (1976).

\bibitem{Feinberg96a}
J. Feinberg and A. Zee, {\em Phys. Lett.} B {\bf 411}, 134 (1997).

\bibitem{Feinberg96b}
J. Feinberg and A. Zee, {\em Phys. Rev.} D {\bf 56}, 5050 (1997).

\bibitem{Scott73}
A.C. Scott, F.Y.F. Chu, and D.W. McLaughlin, {\em Proc. of the IEEE} {\bf 61}, 1443 (1973).

\bibitem{CCGZ}
C.G. Callan, S. Coleman, D.J. Gross, and A. Zee, unpublished (referred to by Ref. 42).
  
\bibitem{Thies93}
M. Thies and K. Ohta, {\em Phys. Rev.} D {\bf 48}, 5883 (1993).

\bibitem{Einhorn76}
M.B. Einhorn, {\em Phys. Rev.} D {\bf 14}, 3451 (1976).

\bibitem{Callan76}
C.G. Callan, N. Coote, and D.J. Gross, {\em Phys. Rev.} D {\bf 13}, 1649 (1976).

\bibitem{Einhorn77}
M.B. Einhorn, S. Nussinov, and E. Rabinovici, {\em Phys. Rev.} D {\bf 15}, 2282 (1977).

\bibitem{Ellis77}
J. Ellis, {\em Acta Phys. Polon.} B {\bf 8}, 1019 (1977).

\bibitem{Weis78}
J.H. Weis, {\em Acta Phys. Polon.} B {\bf 9}, 1051 (1978).

\bibitem{Wu77}
T.T. Wu, {\em Phys. Lett.} B {\bf 71}, 142 (1977).

\bibitem{Bars78}
I. Bars and M.B. Green, {\em Phys. Rev.} D {\bf 17}, 537 (1978).

\bibitem{Li86}
M. Li, {\em Phys. Rev.} D {\bf 34}, 3888 (1986).

\bibitem{Li87}
M. Li, L. Wilets, and M.C. Birse, {\em J. Phys.} G {\bf 13}, 915 (1987).

\bibitem{Dalley93}
S. Dalley and I. Klebanov, {\em Phys. Rev.} D {\bf 47}, 2517 (1993).

\bibitem{Bhanot93}
G. Bhanot, K. Demeterfi, and I. Klebanov, {\em Phys. Rev.} D {\bf 48}, 4980 (1993).

\bibitem{Bigi99}
I. Bigi, M. Shifman, N. Uraltsev, and A. Vainshtein, {\em Phys. Rev.} D {\bf 59}, 054011 (1999).

\bibitem{Grinstein99}
B. Grinstein and R.F. Lebed, {\em Phys. Rev.} D {\bf 59}, 054022 (1999).

\bibitem{Burkardt00}
M. Burkardt and N. Uraltsev, hep-ph/0005278.
  
\bibitem{Zhitnitsky85}
A.R. Zhitnitsky, {\em Phys. Lett.} B {\bf 165}, 405 (1985).

\bibitem{Burkardt96}
M. Burkardt, {\em Phys. Rev.} D {\bf 53}, 933 (1996).

\bibitem{Koopmans33}
T.H. Koopmans, {\em Physica} {\bf 1}, 104 (1933).

\bibitem{Schoen00b}
V. Sch\"on and M. Thies, hep-th/0003195.
  
\bibitem{Fetter}
A.L. Fetter and J.D. Walecka, {\em Quantum theory of many-particle systems}, McGraw-Hill 1971.

\bibitem{Burkardt89}
M. Burkardt, {\em Multiquarksysteme in der 1+1 dimensionalen QCD}, phd thesis, Universit\"at Erlangen-N\"urnberg 1989.

\bibitem{Durgut76}
M. Durgut, {\em Nucl. Phys.} B {\bf 116}, 233 (1976).

\bibitem{Amati81}
D. Amati and E. Rabinovici, {\em Phys. Lett.} B {\bf 101}, 407 (1981).

\bibitem{Amati82}
D. Amati, K.-C. Chou, and S. Yankielowicz, {\em Phys. Lett.} B {\bf 110}, 309 (1982).

\bibitem{Oderkerk92}
R.P. Oderkerk, {\em Relativistic hadrons in an external field}, phd thesis, Free University of Amsterdam (VU) 1992.

\bibitem{Ring}
P. Ring and P. Schuck, {\em The nuclear many-body problem}, Springer, New York 1980. 

\bibitem{Satzinger91}
C. Satzinger, {\em Wiederherstellung der chiralen Symmetrie in QCD$_2$}, Diplomarbeit, Universit\"at Erlangen-N\"urnberg 1991.

\bibitem{Damgaard92}
P.H. Damgaard, H.B. Nielsen, and R. Sollacher, {\em Nucl. Phys.} B {\bf 385}, 227 (1992).

\bibitem{Skyrme62}
T.H.R. Skyrme, {\em Proc. Roy. Soc. Lond.} A {\bf 260}, 127 (1961).

\bibitem{Gell-Mann68}
M. Gell-Mann, R.J. Oakes, and B. Renner, {\em Phys. Rev.} {\bf 175}, 2195 (1968).

\bibitem{Lenz98}
F. Lenz and M. Thies, {\em Ann. Phys. (N.Y.)} {\bf 268}, 308 (1998).

\bibitem{Toms80}
D.J. Toms, {\em Phys. Rev.} D {\bf 21}, 928 (1980).

\bibitem{Koch92}
V. Koch, E.V. Shuryak, G.E. Brown, and A.D. Jackson, {\em Phys. Rev.} D {\bf 46}, 3169 (1992).

\bibitem{Young95}
A.P. Young, {\em Nucl. Phys.} B {\bf 42} (Proc. Suppl.), {\bf 201} (1995).

\bibitem{Sondhi97}
S.L. Sondhi, S.M. Girvin, J.P. Carini, and D. Shahar, {\em Rev. Mod. Phys.} {\bf 69}, 317 (1997).

\bibitem{Kapusta}
J.I. Kapusta, {\em Finite-temperature field theory}, Cambridge Univesity Press 1989.

\bibitem{Barducci95}
A. Barducci, R. Casalbuoni, M. Modugno, G. Pettini, and R. Gatto, {\em Phys. Rev.} D {\bf 51}, 3042 (1995).

\bibitem{Dashen75a}
R.F. Dashen, S.K. Ma, and R. Rajaraman, {\em Phys. Rev.} D {\bf 11}, 1499 (1975).

\bibitem{Karsch97}
F. Karsch, J. Kogut, and H.W. Wyld, {\em Nucl. Phys.} B {\bf 280}, 289 (1987).

\bibitem{Wolff85}
U. Wolff, {\em Phys. Lett.} B {\bf 157}, 303 (1985).

\bibitem{Eguchi82}
T. Eguchi and H. Kawai, {\em Phys. Rev. Lett.} {\bf 48}, 1063 (1982).

\bibitem{Harrington75}
B.J. Harrington and A. Yildiz, {\em Phys. Rev.} D {\bf 11}, 779 (1975).

\bibitem{McLerran85}
L.D. McLerran and A. Sen, {\em Phys. Rev.} D {\bf 32}, 2794 (1985).

\bibitem{Hansson93}
T.H. Hansson and I. Zahed, {\em Phys. Lett.} B {\bf 309}, 385 (1993).

\bibitem{Engelhardt95a}
M. Engelhardt, {\em Phys. Lett.} B {\bf 355}, 507 (1995).

\bibitem{Thies92}
M. Thies, {\em Nucl. Phys.} A {\bf 546}, 233c (1992).

\bibitem{Schoen00a}
V. Sch\"on and M. Thies, {\em Phys. Lett.} B {\bf 481}, 299 (2000).

\bibitem{Gasser87}
J. Gasser and H. Leutwyler, {\em Phys. Lett.} B {\bf 184}, 83 (1987).

\bibitem{Gerber89}
P. Gerber and H. Leutwyler, {\em Nucl. Phys.} B {\bf 321}, 387 (1989).

\bibitem{Zhuang94}
P. Zhuang, J. H\"ufner, and S.P. Klevansky, {\em Nucl. Phys.} A {\bf 576}, 525 (1994).

\bibitem{Klevansky92}
S.P. Klevansky, {\em Rev. Mod. Phys.} {\bf 64}, 649 (1992).

\bibitem{Barducci96}
A. Barducci, R. Casalbuoni, M. Modugno, G. Pettini, and R. Gatto, {\em Mod. Phys. Lett.} A {\bf 11}, 1579 (1996).

\bibitem{Barducci99}
A. Barducci, M. Modugno, G. Pettini, R. Casalbuoni, and R. Gatto, {\em Phys. Rev.} D {\bf 59}, 114024 (1999).

\bibitem{Barducci97}
A. Barducci, R. Casalbuoni, M. Modugno, G. Pettini, and R. Gatto, {\em Phys. Rev.} D {\bf 55}, 2247 (1997).

\bibitem{Treml89}
T.F. Treml, {\em Phys. Rev.} D {\bf 39}, 679 (1989).

\bibitem{Chodos74}
A. Chodos, R.L. Jaffe, K. Johnson, C.B. Thorn, and V.F. Weisskopf, {\em Phys. Rev.} D {\bf 9}, 3471 (1974).

\bibitem{Alford98}
M. Alford, K. Rajagopal, and F. Wilczek, {\em Phys. Lett.} B {\bf 422}, 247 (1998).

\bibitem{Klebanov85}
I. Klebanov, {\em Nucl. Phys.} B {\bf 262}, 133 (1985).

\bibitem{McMillan77}
W.L. McMillan, {\em Phys. Rev.} B {\bf 16}, 4655 (1977).

\bibitem{Theodorou78}
G. Theodorou and T.M. Rice, {\em Phys. Rev.} B {\bf 18}, 2840 (1978).

\bibitem{Takayama93}
K. Takayama and M. Oka, {\em Nucl. Phys.} A {\bf 551}, 637 (1993).

\bibitem{Abramowitz}
{\em Handbook of Mathematical Functions}, M. Abramowitz and I.A. Stegun, eds., Dover, New York 1970.

\bibitem{Fischler79}
W. Fischler, J. Kogut, and L. Susskind, {\em Phys. Rev.} D {\bf 19}, 1188 (1979).

\bibitem{Wigner34}
E. Wigner, {\em Phys. Rev.} {\bf 46}, 1002 (1934).

\bibitem{Schulz93}
H.J. Schulz, {\em Phys. Rev. Lett.} {\bf 71}, 1864 (1993).

\bibitem{Kao94}
Y.-C. Kao and Y.-W. Lee, {\em Phys. Rev.} D {\bf 50}, 1165 (1994).

\bibitem{Christiansen96a}
H.R. Christiansen and F.A. Schaposnik, {\em Phys. Rev.} D {\bf 53}, 3260 (1996).

\bibitem{Christiansen96b}
H.R. Christiansen and F.A. Schaposnik, {\em Phys. Rev.} D {\bf 55}, 4920 (1997).

\bibitem{Kutschera90}
M. Kutschera, W. Broniowski, and A. Kotlorz, {\em Nucl. Phys.} A {\bf 516}, 566 (1990).

\bibitem{Deryagin92}
D.V. Deryagin, D.Yu. Grigoriev, and V.A. Rubakov, {\em Int. J. Mod. Phys.} A {\bf 7}, 659 (1992).

\bibitem{Overhauser78}
A.W. Overhauser, {\em Adv. in Phys.} {\bf 27}, 343 (1978).

\bibitem{Shuster99}
E. Shuster and D.T. Son, {\em Nucl. Phys.} B {\bf 573}, 434  (2000).

\bibitem{Park99}
B.-Y. Park, M. Rho, A. Wirzba, and I. Zahed, {\em Phys. Rev.} D {\bf 62}, 034015 (2000).

\bibitem{Chodos94}
A. Chodos and H. Minakata, {\em Phys. Lett.} A {\bf 191}, 39 (1994).

\bibitem{Sadzikowski00}
M. Sadzikowski and W. Broniowski, hep-ph/0003282.
  
\end{thebibliography}

\end{document}